%% file: 3soK-NRG-CFT.tex
\documentclass[twocolumn, 10pt, aps, superscriptaddress, showpacs, prx, citeautoscript, longbibliography]{revtex4-1}

\usepackage[ngerman, english]{babel}   
\usepackage[utf8]{inputenc}
\usepackage{lmodern}
\usepackage{amssymb}
\usepackage{amsmath}
\usepackage{graphicx}
\usepackage{diagbox} 
\usepackage{hhline}  

\usepackage{xr} 
\usepackage{xcite} 
\externalcitedocument{3soK-NRG-CFT_supp} 
  
\usepackage[usenames,dvipsnames]{xcolor} 
\usepackage[colorlinks, citecolor={blue!50!black}, urlcolor={blue!50!black}, linkcolor={red!50!black},hypertexnames=false]{hyperref}

\usepackage{subfigure}   
\graphicspath{{figures/}} 
\usepackage{tikz}             

\usepackage{tabularx}                    
\usepackage{multirow, bigdelim}  
\usepackage{longtable}
\usepackage{makecell}                   
\usepackage{cellspace}

\usepackage{enumitem}  

\usepackage{mdwlist}      

\usepackage{verbatim}       
\usepackage{bbm}              
\usepackage{nicefrac}        
\usepackage[vcentermath]{youngtab} 
\usepackage{cancel}

\usepackage{etoolbox} 

\newcolumntype{L}[1]{>{\raggedright\arraybackslash}p{#1}} 
\newcolumntype{C}[1]{>{\centering\arraybackslash}p{#1}} 
\newcolumntype{R}[1]{>{\raggedleft\arraybackslash}p{#1}} 
\definecolor{schwarz}{RGB}{0,0,0}
\definecolor{braun}{RGB}{102,51,0}
\definecolor{blau}{RGB}{0,84,159}
\definecolor{tiefblau}{RGB}{0,0,255}
\definecolor{maigruen}{RGB}{189,205,0}
\definecolor{rot}{RGB}{204,7,30}
\definecolor{tiefrot}{RGB}{255,0,0}
\definecolor{bordeaux}{RGB}{161,16,53}
\definecolor{violett}{RGB}{97,33,88}
\definecolor{lila}{RGB}{122,111,172}
\definecolor{tieflila}{RGB}{204,0,204}
\definecolor{magenta}{RGB}{255,0,255}
\definecolor{orange}{RGB}{255,100,0}    
\definecolor{gelb}{RGB}{246,168,0}       
\definecolor{gruen}{RGB}{87,171,39}      
\definecolor{petrol}{RGB}{0,97,101}
\definecolor{rot2}{RGB}{205,2,37}        
\definecolor{blau2}{RGB}{0,86,153}      
\definecolor{darkpetrol}{RGB}{0,73,76}

\definecolor{darkgreen}{rgb}{0,0.5,0}
\definecolor{purple}{rgb}{0.6,0,0.5}
\definecolor{orange}{rgb}{1,0.5,0}
\definecolor{darkred}{rgb}{.7,0,0}
\definecolor{darkblue}{rgb}{0,0,.3}
\definecolor{grey}{rgb}{.6,.6,.6}
\definecolor{dimgreen}{rgb}{0.2,0.6,0.1}
\hypersetup{colorlinks,linkcolor=blue,urlcolor=blue,citecolor=blue}

\usepackage[normalem]{ulem}

\newcommand{\y}[2]{%
  \ifstrequal{#1}{0}%
   {\ifstrequal{#2}{0}{$\bullet$}%
    {\ifstrequal{#2}{1}{{\tiny$\yng(1,1)$}}%
     {\ifstrequal{#2}{2}{{\tiny$\yng(2,2)$}}%
      {{\tiny$\yng(#1,#2)$}}%
     }%
    }%
   }%
   {\ifstrequal{#1}{1}%
    {\ifstrequal{#2}{0}{{\tiny$\yng(1)$}}%
     {\ifstrequal{#2}{1}{{\tiny$\yng(2,1)$}}%
      {{\tiny$\yng(#1,#2)$}}%
     }%
    }%
    {\ifstrequal{#1}{2}%
     {\ifstrequal{#2}{0}{{\tiny$\yng(2)$}}%
      {{\tiny$\yng(#1,#2)$}}%
     }%
    }%
   }%
}%

\newcommand{\f}[3]{$\frac{#1}{#2#3}$}
\newcommand{\fr}[2]{$\frac{#1}{#2}$}

\newcommand{\m}[1]{\multirow{2}{*}{#1}}
\newcommand{\mt}[1]{\multirow{3}{*}{#1}}
\newcommand{\mfour}[1]{\multirow{4}{*}{#1}}

\newcommand{\msix}[1]{\multirow{6}{*}{#1}}
\newcommand{\mseven}[1]{\multirow{7}{*}{#1}}
\newcommand{\meight}[1]{\multirow{8}{*}{#1}}

\newcommand{\lb}{\ldelim\{{2}{*}}

\newcommand{\lbt}{\ldelim\{{3}{*}}
\newcommand{\lbfour}{\ldelim\{{4}{*}}

\newcommand{\lbsix}{\ldelim\{{6}{*}}
\newcommand{\lbseven}{\ldelim\{{7}{*}}
\newcommand{\lbeight}{\ldelim\{{8}{*}}
\newcommand{\minimum}{{\rm{min}}}
\newcommand{\nfl}{{\rm{NFL}}}

\newcommand{\fl}{{\rm{FL}}}

\newcommand{\Qbar}{\bar{Q}}
\newcommand{\imp}{{\rm{imp}}}
\newcommand{\bath}{{\rm{bath}}} 

\newcommand{\FI}{{\rm{FI}}}
\newcommand{\threeoAH}{{\rm{3oAH}}}
\newcommand{\threesoK}{{\rm{3soK}}}
\newcommand{\twosoK}{{\rm{2soK}}} 

\newcommand{\Tspin}{T_\spin}
\newcommand{\Torb}{T_\orb}
\newcommand{\Tss}{T_{\rm{ss}}}
   
\newcommand{\spin}{{\rm{sp}}}   
\newcommand{\Fermi}{{\rm{F}}}   
\newcommand{\orb}{{\rm{orb}}}
\newcommand{\ch}{{\rm{ch}}}
\newcommand{\flavor}{{\rm{fl}}}
\newcommand{\keep}{{\rm{keep}}}
\newcommand{\bulk}{{\;\mathrm{bulk}}}
 
\newcommand{\reference}{{\rm{ref}}} 
\newcommand{\interaction}{{\rm{int}}}
\newcommand{\bPhi}{\mathbf{\Phi}}

\newcommand{\tPsi}{\tilde \Psi}
\newcommand{\tDelta}{\tilde \Delta}
\newcommand{\bJ}{\mathbf{J}}
\newcommand{\bcJ}{\boldsymbol{\mathcal{J}\!}}
\newcommand{\cE}{{\mathcal{E}}}
\newcommand{\spinorb}{{\operatorname{sp-orb}}} 
\newcommand{\bS}{\mathbf{S}}
\newcommand{\bU}{\mathbf{U}}
\newcommand{\bT}{\mathbf{T}}
\newcommand{\bc}{\mathbf{c}}
\newcommand{\uone}{\text{U}(1)} 
\newcommand{\uonecharge}{\text{U}(1)_\text{ch}}
\newcommand{\sutwo}{\text{SU}(2)} 
\newcommand{\suN}{\mbox{{\text{SU}($N$)}}} 
\newcommand{\sutwospin}{\text{SU}(2)_\text{sp}}
\newcommand{\suthree}{\text{SU}(3)}
\newcommand{\suk}{\mbox{{\text{SU}($k$)}}}
\newcommand{\suthreeorb}{\text{SU}(3)_\text{orb}}
\newcommand{\susix}{\text{SU}(6)}
\newcommand{\sufour}{\text{SU}(4)}
\newcommand{\soplus}[1]{\oplus}

\newcommand{\pnfl}{$\bc_\nfl^\ast$}
\newcommand{\pfl}{$\bc_\fl^\ast$}
\newcommand{\NRG}{\mathrm{NRG}}
\newcommand{\md}{\mathrm{d}}

\newcommand{\step}[1]{(C#1)}

\begin{document}

\title{Uncovering Non-Fermi-Liquid Behavior in Hund Metals: \\ 
Conformal Field Theory Analysis of an $\sutwo\times \suthree$ Spin-Orbital Kondo Model}

\author{E. Walter}
\affiliation{Arnold Sommerfeld Center for Theoretical Physics, 
Center for NanoScience,\looseness=-1\,  and Munich 
Center for \\ Quantum Science and Technology,\looseness=-2\, Ludwig-Maximilians-Universität München, 80333 Munich, Germany}
\author{K. M. Stadler}
\affiliation{Arnold Sommerfeld Center for Theoretical Physics, 
Center for NanoScience,\looseness=-1\,  and Munich 
Center for \\ Quantum Science and Technology,\looseness=-2\, Ludwig-Maximilians-Universität München, 80333 Munich, Germany}
\author{S.-S. B. Lee}
\affiliation{Arnold Sommerfeld Center for Theoretical Physics, 
Center for NanoScience,\looseness=-1\,  and Munich 
Center for \\ Quantum Science and Technology,\looseness=-2\, Ludwig-Maximilians-Universität München, 80333 Munich, Germany}
\author{Y. Wang}
\affiliation{Condensed Matter Physics and Materials Science Department, Brookhaven National Laboratory, Upton, New York 11973, USA}
\author{G. Kotliar}
\affiliation{Condensed Matter Physics and Materials Science Department, Brookhaven National Laboratory, Upton, New York 11973, USA}
\affiliation{Department of Physics and Astronomy, Rutgers University, Piscataway, New Jersey 08854, USA}
\author{A. Weichselbaum}
\affiliation{Condensed Matter Physics and Materials Science Department, Brookhaven National Laboratory, Upton, New York 11973, USA}
\affiliation{Arnold Sommerfeld Center for Theoretical Physics, 
Center for NanoScience,\looseness=-1\,  and Munich 
Center for \\ Quantum Science and Technology,\looseness=-2\, Ludwig-Maximilians-Universität München, 80333 Munich, Germany}
\author{J. von Delft}
\affiliation{Arnold Sommerfeld Center for Theoretical Physics, 
Center for NanoScience,\looseness=-1\,  and Munich 
Center for \\ Quantum Science and Technology,\looseness=-2\, Ludwig-Maximilians-Universität München, 80333 Munich, Germany}
\date{\today}

\begin{abstract}
Hund metals have attracted attention in recent years due to their
unconventional superconductivity, which supposedly originates from
non-Fermi-liquid (NFL) properties of the normal state.  When studying
Hund metals using dynamical mean-field theory, one arrives at a
self-consistent ``Hund impurity problem'' involving a multiorbital
quantum impurity with nonzero Hund coupling interacting with a
metallic bath. If its spin and orbital degrees of freedom are screened
at different energy scales, $T_\mathrm{sp} < T_\mathrm{orb}$, the
intermediate energy window is governed by a novel NFL fixed point,
whose nature had not yet been clarified.  We resolve this problem by
providing an analytical solution of a paradigmatic example of a Hund
impurity problem, involving two spin and three orbital degrees of
freedom.  To this end, we combine a state-of-the-art implementation of
the numerical renormalization group, capable of exploiting
non-Abelian symmetries, with a generalization of Affleck and Ludwig's
conformal field theory (CFT) approach for multichannel Kondo models.
We characterize the NFL fixed point of Hund metals in detail for a
Kondo model with an impurity forming an $\sutwo \times \suthree$
spin-orbital multiplet, tuned such that the NFL energy window is very
wide. The impurity’s spin and orbital susceptibilities then exhibit
striking power-law behavior, which we explain using CFT arguments. We
find excellent agreement between CFT predictions and numerical renormalization group results.
Our main physical conclusion is that the regime of spin-orbital
  separation, where orbital degrees of freedom have been screened but
  spin degrees of freedom have not, features anomalously strong local
  spin fluctuations: the impurity susceptibility
  increases as $\chi_\spin^\imp \sim \omega^{-\gamma}$, with $\gamma > 1$.
\end{abstract}

\maketitle

\section{Introduction}
\label{sec:Introduction}

\subsection{Motivation: Hund metals}

Hund metals are multiorbital materials with broad bands which are
correlated via the ferromagnetic Hund coupling $J_{\rm H}$, rather than the
Hubbard interaction $U$. The coupling $J_{\rm H}$ implements Hund’s rule,
favoring electronic states with maximal spin, which causes Hund metals
to be fundamentally different from Mott insulators.  
This is a new exciting area of condensed matter physics;
for a recent  review with numerous references, see Ref.\ \cite{Georges2013}.
Hund metals are a very diverse class of materials, including transition metal oxides with partially filled $d$ shells, such as the iron-based pnictide and selenide superconductors, the ruthenates, and many others
\cite{Georges2013,Werner2008,Haule2009,Yin2011a,Yin2011b,Medici2011,Mravlje2011,Yin2012,Aron2015,Hoshino2015,Dang2015,Mravlje2016,Zingl2019}.

The iron-based superconductors, in particular, raised much interest in recent years because of the unconventional nature of their superconductivity. It has been argued that the Hund
nature of their normal state is essential for the onset of
superconductivity \cite{Lee2018a}.  In particular, spin fluctuations
with a power-law divergent susceptibility $\propto \omega^{-\gamma}$,
with $\gamma>1$, have been evoked in an explanation for the
anomalously large ratio of $2\Delta_\text{max}/T_\text{c}$ observed
experimentally, where $\Delta_\text{max}$ is the maximum
superconducting gap and $T_\text{c}$ the critical temperature
\cite{Lee2018a}.  The normal state of Hund metals is of great interest
on its own, since it typically shows bad-metal behavior
\cite{Yi2013,Hardy2013,Medici2011}.  Motivated by these
considerations, computational and experimental studies of Hund metals
have begun to uncover their rich physics in recent years
\cite{Yin2011a,Yin2011b,Yin2012,Dang2015,Horvat2016,Mravlje2016,Tamai2019,Werner2019,
  Coleman2019,Chen2019}.

When studying Hund metals in the context of dynamical mean-field theory (DMFT), the problem of a crystal lattice with many strongly interacting lattice sites is mapped onto a ``Hund impurity,'' coupled self-consistently to an effective noninteracting metallic bath. 
A Hund impurity has both spin \textit{and} orbital degrees of freedom and a finite Hund coupling, favoring a large local spin.

A particularly fascinating consequence of the interplay between spin and orbital degrees of freedom is the phenomenon of spin-orbital separation (SOS): Kondo screening of Hund impurity models occurs in two stages, and the energy scales below which free spin and orbital degrees are screened differ, $\Tspin < \Torb$ \cite{Yin2012,Aron2015,Stadler2015,Stadler2018,Deng2019}. The low-energy regime below $\Tspin$ shows Fermi-liquid (FL) behavior. The intermediate SOS window $[\Tspin, \Torb]$, by contrast, shows incoherent behavior, featuring almost fully screened orbital degrees of freedom coupled to almost free spin degrees of freedom. The incoherent regime has been conjectured to have non-Fermi-liquid (NFL) properties and argued to be relevant for the bad-metal behavior of Hund metals \cite{Yin2012,Akhanjee2013}. However, the nature of the putative underlying NFL state has not yet been clarified. 

A major obstacle for analyzing the conjectured NFL regime of Hund metals has been a lack of detailed, analytical understanding of the basic properties of Hund impurity models, since theoretical work has overwhelmingly focused on Kondo models without orbital degrees of freedom. In this work, we overcome this obstacle in the context of an instructive case study of a specific Hund impurity model. 

Before specifying the latter in detail, though, let us put our study into perspective by providing a brief historical overview of Hund impurity models.

\subsection{Brief history of Hund impurity models}

Hund impurity models are natural multiorbital generalizations of single-orbital magnetic impurity models such as the Kondo model used by Kondo in 1964 to explain the resistance minimum in magnetic alloys \cite{Kondo1964}. 
The search for a detailed understanding of the Kondo model beyond Kondo's perturbative calculation was a cornerstone toward the development of renormalization group techniques, starting with Anderson's poor man's scaling approach \cite{Anderson1970} and culminating in Wilson's numerical renormalization group (NRG) \cite{Wilson1975}. 
These methods confirmed that below a characteristic Kondo temperature the metallic bath screens the impurity spin, leading to the formation of a singlet state between impurity and conduction electrons.

Following these findings, naturally the question arises: What happens
if the impurity has multiple orbitals? In particular, electrons on a
multiorbital impurity experience not only a Coulomb interaction
stabilizing a magnetic moment on the impurity, but also a Hund
coupling, enforcing the effect of Hund's rule to maximize the total
impurity spin. These two interactions lead to an intricate interplay,
crucially depending on the number of electrons on the impurity.  
Indeed, it had been observed already in the 1960s that the Kondo
scale for impurities in transition metal alloys with partially filled
$d$shells decreases exponentially as the shell filling approaches
1/2 \cite{Schrieffer1967,Daybell1968}, drawing attention to the
question of understanding Kondo screening in the presence of multiple
orbitals.  Coqblin and Schrieffer \cite{Coqblin1969} developed a
generalization of the Kondo model for multiorbital impurities, yet
only involving the spin degree of freedom.  Okada and Yosida
\cite{Okada1973} included orbital degrees of freedom and in particular
pointed out the importance of a finite Hund coupling, enforcing the
effect of Hund's rule in such multiorbital systems.  However,
theoretical tools for analyzing a model with non-zero Hund coupling
away from half filling were lacking at the time.

Later, Nozi\`eres and Blandin \cite{Nozieres1980} studied a spin Kondo impurity immersed in a metallic bath with multiple orbital channels. A major conclusion of their work was that such models lead to overscreening of the impurity spin and NFL behavior, if the number of channels exceeds twice the impurity spin ($k>2S$). This generated great theoretical interest in multichannel Kondo models, including exact Bethe solutions
providing information on thermodynamical properties \cite{Andrei1984,Tsvelik1984,Tsvelik1985a,Tsvelik1985,Andrei1995,Jerez1998},
and NRG studies~\cite{Cragg1980,Pang1991}.
Affleck and Ludwig (AL) \cite{Affleck1990,Affleck1991a,Affleck1991,Affleck1993,Ludwig1994a} developed a powerful conformal field theory (CFT) approach for studying the strong-coupling fixed points of such multiband Kondo models, providing analytical results for finite-size spectra and the scaling behavior of correlation functions.
However, their work was restricted to pure spin impurities without nontrivial orbital structure. Thus, their methods have not yet been applied to Hund impurity models, including orbital degrees of freedom and a finite Hund coupling.

In this work, we fill this long-standing void and provide a detailed and comprehensive analysis of a prototypical Hund impurity model (specified below). We achieve this by advancing and combining two powerful complementary techniques that both arose in the very context of Kondo physics: An \textit{analytical} solution based on AL's celebrated CFT approach, generalized from a pure spin impurity to one with spin and orbital structure, and a quasiexact \textit{numerical} solution using a state-of-the-art implementation of Wilson's NRG, allowing studies of multiorbital systems by fully exploiting Abelian and non-Abelian symmetries.
This allows us to achieve a detailed understanding of the NFL behavior arising in this Hund impurity model.

\subsection{Minimal models for Hund metals}

We next describe the considerations motivating the specific choice of model studied below.

A minimal model for Hund metals has been proposed in Ref.~\cite{Yin2012}. It is a three-orbital Hubbard-Hund model, and it has been studied extensively in Refs.~\cite{Werner2008,Medici2011,Aron2015,Stadler2015,Dang2015,Stadler2018,Deng2019,
Stadler2019}. 
A treatment of this model by DMFT at 1/3 filling yields a self-consistent Hund impurity model. More specifically, one obtains a self-consistent three-orbital Anderson-Hund (\threeoAH) model, in which bath and impurity both have spin \textit{and} orbital degrees of freedom. The impurity hosts two electrons forming an antisymmetric orbital triplet and a symmetric spin triplet ($S=1$), reflecting Hund's rule. At energies so low that charge fluctuations can be treated by a Schrieffer-Wolff transformation \cite{Aron2015}, the \threeoAH\ model maps onto a three-channel spin-\-orbital Kondo (\threesoK) model whose impurity forms a $(3\times3)$-dimensional $\sutwo\times \suthree$ spin-orbital multiplet.

The \threeoAH\ model exhibits SOS
\cite{Yin2012,Aron2015,Stadler2015,Stadler2018,Deng2019}.  Within the
SOS window $[\Tspin, \Torb]$, the imaginary part of the spin
susceptibility scales as $\chi_\spin^\imp \sim \omega^{-6/5}$
\cite{Stadler2015,Stadler2019}.  The fact that the exponent,
$\gamma=6/5$, is larger than 1 has been argued to lead to the
anomalous superconducting state of the iron pnictide Hund metals, as
mentioned above \cite{Lee2018a}.  However, the origin of this power
law has remained unclear. One impediment toward finding an
explanation is the fact that for the $\threeoAH$ model the orbital and
spin screening scales cannot be tuned independently. The SOS window
turns out to be rather small, masking the NFL behavior expected to
occur within it.

In this paper, we sidestep this limitation by instead studying the
\threesoK\ model and treating its exchange couplings as independent
parameters, freed from the shackles of their \threeoAH\ origin.  We
tune these such that the regime of SOS is very wide, with $\Tspin \ll
\Torb$.  This enables us to characterize the NFL fixed point obtained
for $\Tspin = 0$, which also governs the intermediate NFL window if
$\Tspin \ll \Torb$.  We compute fixed-point spectra and the scaling
behavior of dynamical spin and orbital susceptibilities using both NRG
and CFT, with mutually consistent results. In particular, we find an
analytical explanation for the peculiar power law $\chi_\spin^\imp
\sim \omega^{-6/5}$: It turns out to be governed (albeit somewhat indirectly) by the NFL fixed
point mentioned above. Finally, we demonstrate the relevance of
  these \threesoK\ results for the low-energy behavior of the
  \threeoAH\ model by employing a hybrid Anderson-Kondo model which
  smoothly interpolates between the physics of the \threesoK\ and
  \threeoAH\ models. This interpolation shows that our new results
  also shed light on previous DMFT results for a self-consistent
  \threeoAH\ model \cite{Stadler2015,Stadler2019}.
 
Our CFT analysis builds on that devised by AL
\cite{Affleck1990,Affleck1991a,Affleck1991,Affleck1993,Ludwig1994a} for the $k$-channel Kondo
model, describing $k$ spinful channels exchange coupled to an impurity
with spin $S$, but no orbital degrees of freedom. If $k>2S$, the
impurity spin is overscreened. AL described the corresponding NFL fixed
point using a charge-spin-orbital $\uone \times \sutwo_k \times \suk_2$ Kac-Moody (KM) decomposition of the bath states, and
fusing the spin degrees of freedom of impurity and bath using
$\sutwo_k$ fusion rules. Here we generalize this strategy to our
situation, where the impurity has spin \textit{and} orbital
``isospin'' degrees of freedom: the NFL fixed point  at $\Tspin=0$
can be understood by applying 
$\suthree_2$ fusion rules in the orbital sector, leading to orbital
overscreening.  If $\Tspin$ is nonzero (but $\ll \Torb$), the
overscreened orbital degrees of freedom couple weakly to the impurity
spin, driving the system to a FL fixed point. There both spin and
orbital degrees of freedom are fully screened, in a manner governed by
$\susix_1$ fusion rules.

The paper is structured as follows. Section~\ref{sec:model} defines
the \threesoK\ model and discusses its weak-coupling renormalization group (RG) flow. Section~\ref{sec:NRG} presents our NRG results. Section~\ref{sec:CFT_synopsis}
gives a synopsis of our CFT results,  summarizing all essential insights and arguments, while Sec.\ \ref{sec:CFT_analysis} elaborates the
corresponding CFT arguments in more detail. 
Section~\ref{sec:3OAH-model}  
discusses a hybrid Anderson-Kondo model
which interpolates between the \threesoK\ model and the \threeoAH\ model.
Section~\ref{sec:conclusion} summarizes our conclusions. 
The Appendix revisits a two-channel
spin-orbital Kondo model studied by
Ye in 1997 \cite{Ye1997}, pointing out the similarities and differences
between his work and ours.

\section{Model, perturbative RG flow}
\label{sec:model} 

We study the \threesoK\ model proposed in Ref.\ \cite{Aron2015}.
$H_\bath = \sum_{pm\sigma} \varepsilon_p \psi_{pm\sigma}^\dagger
\psi_{pm\sigma}$
describes a symmetric, flat-band bath, where
$\psi_{pm\sigma}^\dagger$ creates an electron with momentum $p$
and spin $\sigma$ in orbital $m \in \{1,2,3 \}$. 
The bath couples to the impurity spin $\bS$ and orbital isospin  
$\bT$ via 
\begin{align}
H_\interaction = J_0 \, \bS \cdot \bJ_\spin + K_0 \, \bT \cdot \bJ_\orb + I_0  \bS \cdot \bJ_\spinorb \cdot \bT . 
\label{eq:Hamiltonian_AronKotliar}
\end{align}
Here $\bS$
are \sutwo\ generators in the $S=1$
representation, normalized as $\mathrm{Tr}(S^\alpha
S^\beta)=\tfrac12 \delta^{\alpha\beta}$, and
$\bT$
are \suthree\ generators in the representation with Young diagram
\raisebox{0.5mm}{\tiny $\yng(1,1)\,$},
and $\mathrm{Tr}(T^a
T^b)\!=\!\tfrac12 \delta^{ab}$.  $\bJ_\spin$,
$\bJ_\orb$ and $\bJ_\spinorb$ are the 
bath spin, orbital and spin-orbital densities at the impurity site,
with $J^\alpha_\spin
= \psi_{m\sigma}^\dagger \,\tfrac12
\sigma^\alpha_{\sigma\sigma^\prime}
\,\psi_{m\sigma^\prime}$, $J^a_\orb = \psi_{m\sigma}^\dagger
\,\tfrac12 \tau^a_{mm^\prime}
\,\psi_{m^\prime\sigma}$, $J^{\alpha,a}_\spinorb =
\psi_{m\sigma}^\dagger \,\tfrac12 \sigma^\alpha_{\sigma\sigma^\prime}
\tfrac12 \tau^a_{mm^\prime}
\,\psi_{m^\prime\sigma^\prime}$ 
(summation over repeated indices is implied),
where fields are evaluated at the impurity site, 
$\psi_{m\sigma}^\dagger(r=0)$,
and $\sigma^\alpha$ [$\tau^a$] are Pauli [Gell-Mann] matrices, with
normalization $\mathrm{Tr}(\sigma^\alpha \sigma^\beta) =
2\delta^{\alpha\beta}$ [$\mathrm{Tr}(\tau^a \tau^b) = 2\delta^{ab}$].
We use Young diagrams as labels for irreducible
representations (irreps) of the $\suthree$ group. An
  alternative notation, also frequently used, would be to label
  $\suthree$ irreps by their dimension: $\bullet=1$, ${\tiny \yng(1)}
  = 3$, ${\tiny \yng(1,1)} = \bar{3}$, where $\bar{3}$ refers to the
  conjugate represenation of $3$, ${\tiny \yng(2)} = 6$, ${\tiny
    \yng(2,2)} = \bar{6}$, ${\tiny \yng(2,1)} = 8$, etc. 

\begin{figure}[tb] 
\includegraphics[width=\columnwidth]{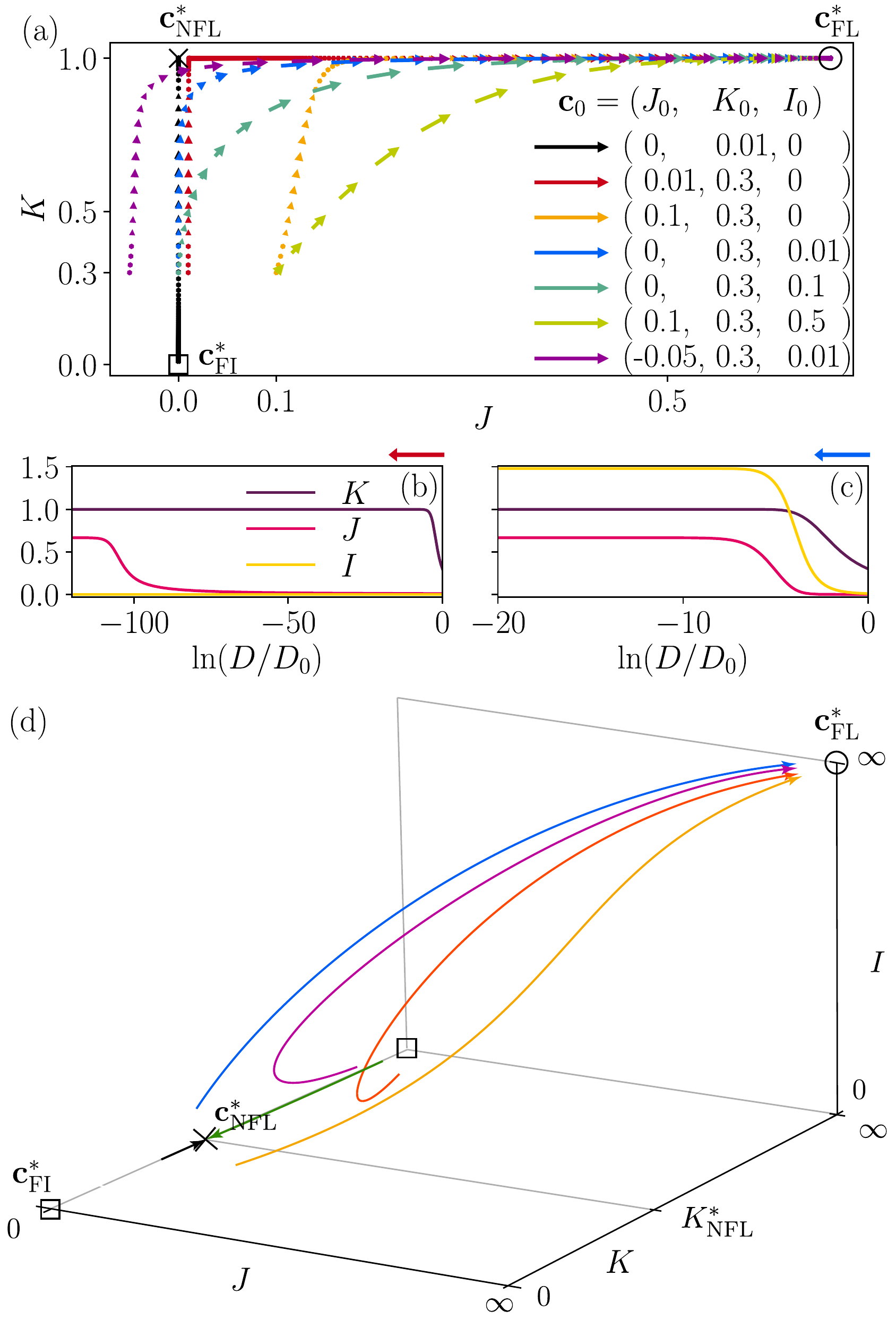}
\caption{(a) RG flow of the coupling vector $\bc = (J,K,I)$ (projected into
  the $J$-$K$ plane), obtained by solving the weak-coupling RG
  equations ~\eqref{eq:betaJKI} 
  [Eqs.~(8)-(10) of Ref.~\cite{Aron2015}]
for various initial values, $\bc_0 = (J_0,K_0,I_0)$. Arrows  
depict the gradient vector, 
$-\frac{\md}{\md \ln D} (J,K)$
at equal steps of $\ln D$.
(b),(c) Weak-coupling RG flow of $\bc(D)$ for   (b)
  $\bc_0 = (0.01,0.3,0)$ [red arrows in (a)] and (c)
  $(0,0.3,0.01)$ [blue arrows in (a)].
(d) Qualitative depiction of the conjectured RG flow in the 
full $J$-$K$-$I$ space, for all couplings non-negative. Fat, faint dashed lines show the solutions $\bc(D)$ of the weak-coupling equations \eqref{eq:betaJKI},  initialized at $K_0 \ll K^\ast_\nfl$ 
with $(J_0,I_0) = (0,0)$ (black), $(>\! 0, 0$) (yellow), or $(0,> \! 0)$ (blue), and plotted only in the weak-coupling regime (beyond the 
latter, Eqs.~\eqref{eq:betaJKI} lose validity).
Solid lines, drawn by hand, qualitatively show the flow expected
beyond the weak-coupling regime, including trajectories
initialized at $K_0 \gg K^\ast_\nfl$, with $(J_0,I_0) = (0,0)$ (green), 
$(>\! 0, 0$) (orange), or $(0,> \! 0)$  (purple).
The black squares, cross, and circle 
depict fixed points.}
\label{fig:RG_trajectories}
\end{figure}

The Hamiltonian has $\uonecharge \times \sutwospin \times \suthreeorb$
symmetry.  We label its symmetry multiplets by $Q=(q,S, \lambda)$,
with $q$ the \textit{bath} particle number relative to half filling
(the \threesoK\ impurity has no charge dynamics; hence we may choose
$q_\imp =0$), $S$ the total spin, and $\lambda$ a Young diagram
denoting an \suthree\ representation. The values of the spin, orbital,
and spin-orbital exchange couplings, $J_0$, $K_0$, $I_0$, can be
derived from the \threeoAH\ model by a Schrieffer-Wolff
  transformation \cite{Aron2015}.
When the \threeoAH\ model is studied in the regime relevant for Hund metals,
i.e., with a ferromagnetic on-site Hund coupling $J_{\rm H}$ favoring
maximization of the local spin, and with a local filling $n_d$ differing
by $\simeq 1$ from half filling, the resulting \threesoK\ exchange couplings
$J_0$, $K_0$, $I_0$ are typically all positive, i.e., antiferromagnetic.
[This can be inferred from Eqs.~(4)-(7) of Ref.~\cite{Aron2015}.] Furthermore, when
the weak-coupling RG flow of the \threesoK\ model is studied in the presence
of finite $K_0>0$ and $I_0>0$, one finds that $J_0$ flows toward
positive values regardless of whether its initial value is chosen
positive or negative [the latter case is illustrated by the purple
arrows in Fig.~\ref{fig:RG_trajectories}(a)].
Hence, we here focus on positive exchange couplings only. However, instead of
  using values obtained from a Schrieffer-Wolff transformation, here
  we take the liberty of choosing  $J_0$, $K_0$, $I_0$ to
  be independent, tuning them such that $\Tspin \ll \Torb$.
This is in extension of the \threeoAH\ model, 
in which $\Tspin$ is only
at most about an order of magnitude smaller than $\Torb$.

Aron and Kotliar \cite{Aron2015} have performed a perturbative
analysis of the RG flow of the \threesoK\ model.  Their Eqs.~(8)-(10)
describe the flow of the coupling vector, $\bc(D) = (J,K,I)$, upon
reducing the half-bandwidth $D$ starting from
$\bc_0 = (J_0,K_0,I_0)$ at $D_0$.
For the \threesoK\ model, these equations read
\begin{align}
\beta_J &= -(1-\tfrac32 J)(J^2 + \tfrac29 I^2) + \dots \,,
\nonumber 
\\
\label{eq:betaJKI}
\beta_K &= -\tfrac32 (1 - K)(K^2 + \tfrac12 I^2) + \dots \,,
\\ \nonumber 
\beta_I &= -\tfrac32 \left( (\tfrac43 J + 2K -J^2 -K^2) I -\tfrac{5}{18} I^2 - \tfrac{17}{36} I^3 \right) + \dots ,
\end{align}
where $\beta_J = \mathrm{d}J / \mathrm{d}\ln D$, etc., with energies
in units of $D_0$.
Figure~\ref{fig:RG_trajectories} illustrates the resulting RG flow.
There are several fixed points.
The free-impurity fixed point, $\bc^\ast_\FI=(0,0,0)$, is unstable:
for any nonzero $\bc_0$, one or more couplings flow toward strong
coupling, and the $D$ values where $J$ or $K$ become of order unity
yield estimates of $\Tspin$ and $\Torb$, respectively. For
$\bc_0 = (0,K_0\neq 0,0)$ [black arrows in
Fig.~\ref{fig:RG_trajectories}(a)], the system flows toward a NFL fixed
point, $ \bc^\ast_\nfl = (0,1,0)$.  This fixed point is unstable against nonzero
$J_0$ or $I_0$. For $I_0=0$, the flow equations for $J$ and $K$ are
decoupled, such that for a small but nonzero $J_0 \ll K_0$ (red
arrows) the flow first closely approaches \pnfl, until $J$
grows large, driving it toward a FL fixed point \pfl.
Figure.~\ref{fig:RG_trajectories}(b) shows that the NFL regime
($J \ll K$) governed by $\bc^\ast_\nfl$ can be large.  For
$I_0\neq 0$, the $J$ and $K$ flows are coupled, hence the growth of
$K$ triggers that of $J$, accelerating the  flow toward
$\bc^\ast_\fl$.
In this case, the NFL energy window is rather small [cf.\
Fig.~\ref{fig:RG_trajectories}(c)]. For example, for
$\bc_0 = (0.1, 0.3, 0.5)$ (light green arrows), typical for the values
obtained through a Schrieffer-Wolff \threeoAH\ to \threesoK\ mapping,
the RG flow does not approach $\bc^\ast_\nfl$  very closely;
thus fully developed NFL behavior is
not observed.

Figure~\ref{fig:RG_trajectories}(d) offers a qualitative depiction of the conjectured RG flow in the full $J$-$K$-$I$ space, for all couplings non-negative. Fat, faint dashed lines show the solutions $\bc(D)$ of the weak-coupling Eqs.~\eqref{eq:betaJKI}.  
However,
these equations lose validity once the couplings are no longer small
(and their above-mentioned predictions that $K^\ast_\nfl = K^\ast_\fl =1$
should not be trusted).
Solid lines, drawn by hand, qualitatively depict the flow expected
beyond the weak-coupling regime, based on the following considerations.
First, for $K_0 > 0$ and $J_0=I_0=0$, the NRG analysis of Sec.~\ref{sec:NRG} suggests that the flow proceeds along
a trajectory where $I$ and $J$ remain zero, reaching a NFL fixed point, \pnfl$ = (0,K^\ast_\nfl,0)$ at a finite value of $K^\ast_\nfl$. 
This fixed point is stable, approached by RG flow both from below and above. Correspondingly, the line $J_0=I_0=0$ contains another fixed point at $K_0 = \infty$, which is unstable. To understand the latter point heuristically, consider taking $K_0$ very large. Then the system
will attempt to screen its local orbital degree of freedom, with representation  \raisebox{0.5mm}{\tiny $\yng(1,1)\,$}, into an orbital singlet. Doing so by binding just a bath single electron, spin up or down, would break spin symmetry. Hence, it must bind two bath electrons, spin up and down, yielding a local orbital degree of freedom yet again, with representation ${\tiny \yng(1)}$. Thus, choosing $K_0$ very large is equivalent to initializing the model with local orbital representation ${\tiny \yng(1)}$ and small initial coupling (presumably $\sim 1/K_0$). This would grow under the RG flow; hence $K_0 = \infty$ is an unstable fixed point, just as $K_0=0$.
(This argumentation is entirely analogous to that familiar from the two-channel Kondo model \cite{Nozieres1980}; for the
present \threesoK\ model, it is further elaborated in Ref.\ \cite{Wang2020}.)

For $K_0 > 0$ and $J_0$, $I_0$ both non-negative but not both zero, the 
NRG analysis of Sec.~\ref{sec:FL-spectrum} suggests that the  
flow always ends up at a unique FL fixed point \pfl.
Hence \pnfl\ is unstable against turning on $J_0$ or $I_0$. 
The  fixed point \pfl\ features a fully screened spin and orbital singlet ground state
and an excitation spectrum with   $\susix$ symmetry.
This implies that as the flow approaches \pfl, all three couplings $J$, $K$, and $I$ tend to infinity, with relative values
such that the fixed-point Hamiltonian has \susix\ symmetry, 
i.e., $3 J = 2 K = I$ \cite{Aron2015}.

\section{NRG results}
\label{sec:NRG} 
 
\label{sec:finite_size_spectra}

To study the RG flow in a quantitatively reliable manner, we solve the
\threesoK\ model using NRG \cite{Wilson1975, Weichselbaum2012a,
  Weichselbaum2012b}, exploiting non-Abelian symmetries using QSpace
\cite{Weichselbaum2012a}.  The bath is discretized logarithmically and
mapped to a semi-infinite ``Wilson chain'' with exponentially decaying
hoppings, and the impurity coupled to site $0$. The chain is
diagonalized iteratively while discarding high-energy states, thereby
zooming in on low-energy properties: the (finite-size) level spacing
of a chain ending at site $k$ is of order
$\omega_k \!\propto\! \Lambda^{-k/2}$, where $\Lambda > 1$ is a
discretization parameter.  The RG flow can be visualized using NRG
eigenlevel spectra, showing how the chain's lowest-lying
eigenenergies $\cE$ evolve when $k$ is increased by plotting
the dimensionless rescaled energies
$E = (\cE-\cE_\reference)/\omega_k$ versus $\omega_k$ for odd  $k$.  The
$E$-level flow is stationary ($\omega_k$ independent) while $\omega_k$ traverses
an energy regime governed by one of the system's fixed points, but
changes during crossovers between fixed points.

\begin{figure}[tb!]
\includegraphics[width=.97\columnwidth]{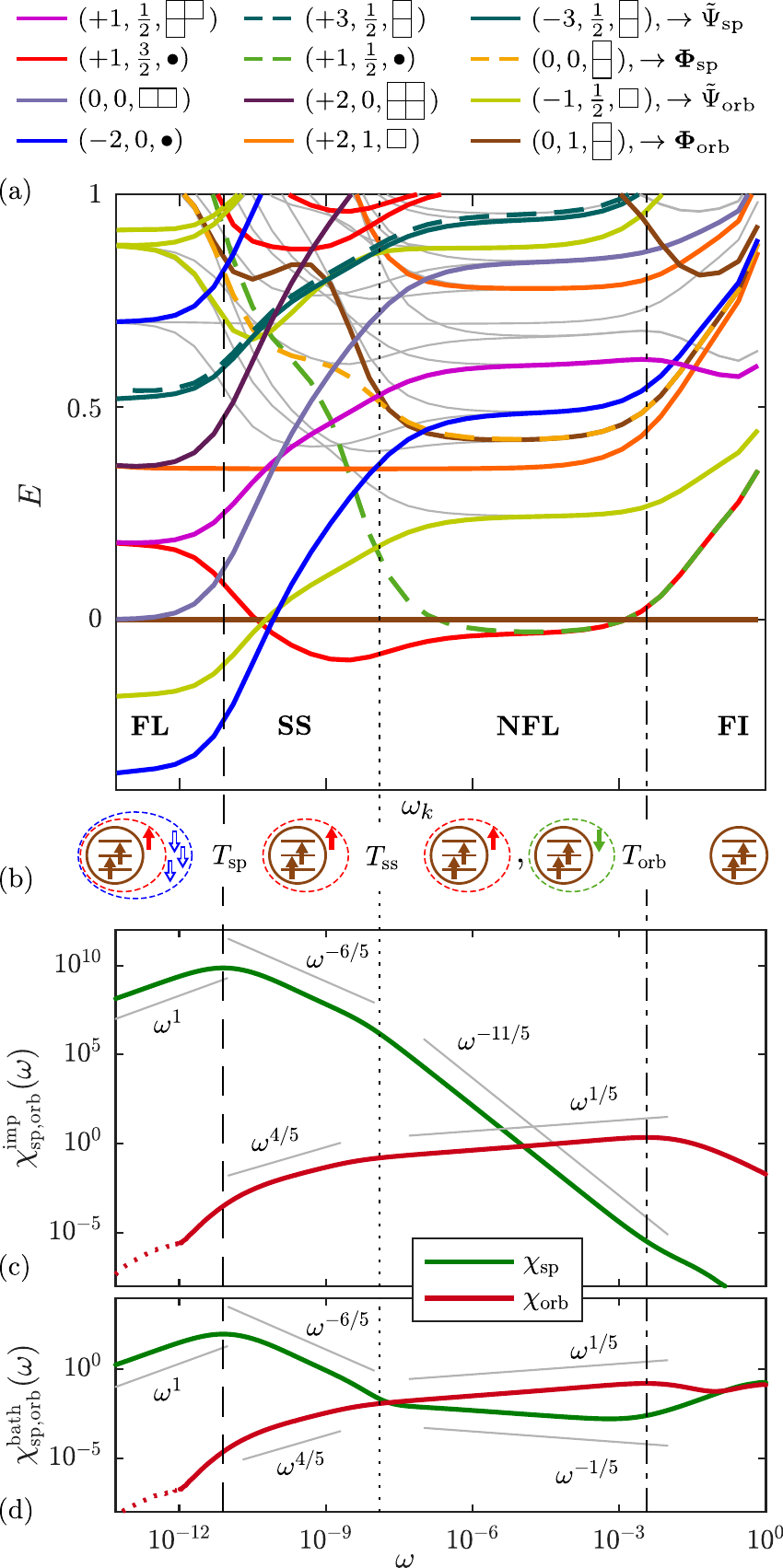}  
\caption{NRG results for $\bc_0 = (J_0,K_0,I_0) =(10^{-4},0.3,0)$.
  (a) Finite-size eigenlevel spectrum 
computed by NRG, with
$\cE_\reference = 
\cE(0,1, \protect\raisebox{0.6mm}{\tiny $\protect\yng(1,1)$}\,)$
as reference energy. 
Quantum numbers $Q=(q, S, \lambda)$ 
are  shown at the top, and $\to$ indicates boundary operators
  obtained via double fusion. (NRG parameters:
$\Lambda= 2.5$; number of kept multiplets, $N_\keep = 3000$; half-bandwidth of the bath, $D = 1$.)  (b)
  Illustrations of the ground states encountered during the flow.  (c),(d)
  Imaginary part of the spin and orbital susceptibilities of (c) the
  impurity and (d) the  bath site coupled to it (Wilson chain site $k=0$).
  Gray lines show power laws predicted by CFT.
Vertical lines show the crossover scales 
for orbital and spin screening, 
$\Torb$ and $\Tspin$, marking the maxima of $\chi^\imp_\orb$
and $\chi^\imp_\spin$,  and for spin splitting,
$\Tss$, marking kinks in  $\chi^{\imp,\bath}_{\spin,\orb}$.
}
\label{fig:spectra_susceptibilities_K0} 
\end{figure}

To analyze the NFL regime in detail, we choose $I_0=0$ and
$J_0 \ll K_0$, so that the SOS window becomes very large, with
$\Tspin \lll \Torb$.  Figure~\ref{fig:spectra_susceptibilities_K0}(a) shows
the NRG eigenlevel flow diagram for $\bc_0 = (10^{-4},0.3,0)$.  We discern
four distinct regimes, separated by three scales, $\Tspin$, $\Tss$,
$\Torb$.

\begin{enumerate}[label={(}\roman*{)}]

\item The \textit{free-impurity} (FI) regime, $\omega_k > \Torb$, involves
an unscreened impurity, with ground state multiplet
$Q=(0,1, \raisebox{0.5mm}{\tiny $\yng(1,1)\,$})$ (flat brown line).

\item In the NFL regime, $\Tss < \omega_k < \Torb$, two degenerate
multiplets, $(1,\frac{1}{2}, \bullet)$ and $(1,\frac{3}{2},\bullet)$
(dashed green and red lines) become the new ground state multiplets.
Below the scale $\Torb$, the impurity orbital isopin is thus screened
into an orbital singlet $\bullet \,$ by binding one bath electron, which couples to
the impurity spin 1 to yield a total spin of $\frac{1}{2}$ or
$\frac{3}{2}$.

\item In the \textit{spin-splitting} (SS) regime,
$\Tspin < \omega_k < \Tss$, the effects of nonzero $J_0$ become
noticeable, splitting apart $(1,\frac{1}{2}, \bullet)$ 
and $(1,\frac{3}{2},\bullet)$, the latter drifting down.

\item In the FL regime, $\omega_k < \Tspin$, $(-2,0, \bullet)$ becomes
the new ground state multiplet.  Below the scale $\Tspin$, the spin
3/2 is thus screened into a spin singlet by binding three
bath holes, yielding a \textit{fully} screened impurity.  Note the 
equidistant level spacing, characteristic of a FL.

\end{enumerate}

To further elucidate the
consequences of orbital and spin screening, 
we computed the impurity's zero-temperature orbital and spin susceptibilities, 
\begin{subequations}
\begin{align}
\chi_\orb^\imp (\omega) & = -\tfrac1{8 \pi}
\sum_a \mathrm{Im}
\langle T^a \| T^a \rangle_\omega , \\
\chi_\spin^\imp(\omega) & = -\tfrac1{3 \pi}
\sum_\alpha
\mathrm{Im} \langle S^\alpha \| S^\alpha \rangle_\omega , 
\end{align}
\end{subequations}
where $\langle X \| X \rangle_\omega$ refers to the Fourier-transformed retarded correlation functions $-i \Theta(t) \langle [X(t), X(0)] \rangle$ with frequency $\omega$,
and analogous susceptibilities, $\chi^\bath_\orb$, $\chi^\bath_\spin$ (involving
$\bJ_\orb$, $\bJ_\spin$) for the bath
site coupled to it.
To this end we used full-density-matrix (fdm) NRG  \cite{Weichselbaum2007} 
and adaptive broadening of the discrete NRG data \cite{Lee2016}.

Figures~\ref{fig:spectra_susceptibilities_K0}(c) and \ref{fig:spectra_susceptibilities_K0}(d) show these
susceptibilities on a log-log scale. $\chi^\imp_\orb$ and
$\chi^\imp_\spin$ each exhibit a maximum, at two widely different
scales, $\Torb$ and $\Tspin$, coinciding with the onset of the
stationary NFL or FL regimes in
Fig.~\ref{fig:spectra_susceptibilities_K0}(a), respectively.
Moreover, the four susceptibilities $\chi^{\imp,\bath}_{\orb,\spin}$
all exhibit kinks at a coinciding energy scale, $\Tss$, matching the
onset of the SS regime in
Fig.~\ref{fig:spectra_susceptibilities_K0}(a).  If $\omega$ lies
within
one of the regimes NFL, SS, or FL, the susceptibilities all show
behavior consistent with power laws (gray lines). These power laws
can all be explained by CFT, as discussed in Sec.~\ref{sec:CFT_synopsis}. 
Here we focus on their
qualitative features, which by themselves give striking clues
about the nature of orbital and spin screening.

In the NFL regime, where $\chi^\imp_\orb$ decreases with decreasing
$\omega$, it exhibits the \textit{same} power law as
$\chi^\bath_\orb$.  In this sense, the impurity's orbital isospin has
taken on the same character as that of the bath site it couples to,
indicative of orbital screening---in the parlance of AL's CFT
analysis, it has been ``absorbed'' by the bath.  This power law
$\omega^{1/5}$ is nontrivial, differing from the $\omega^1$ expected
for a fully screened local degree of freedom. This indicates that the
local orbital degree of freedom, even while being screened, is still
somehow affected by the spin sector. The converse is also true: the
onset of orbital screening at $\Torb$ is accompanied by a change in
behavior for both spin susceptibilities, $\chi_\spin^\imp$ and
$\chi_\spin^\bath$.  Both increase with decreasing $\omega$, with
\textit{different} powers, indicative of the absence of spin screening
in the NFL regime. The exponent for the impurity spin susceptibility,
$\chi^\imp_\spin \sim \omega^{-11/5}$, is remarkably large in
magnitude. (For comparison, for the standard spin-1/2,
  single-channel Kondo model, $\chi^\imp_\spin \sim \omega^{-1}$ for
  $\omega \gtrsim \Tspin$.) The highly singular $\omega^{-11/5}$
behavior---our perhaps most unexpected result---indicates that the
strength of spin fluctuations is strongly amplified by
the onset of orbital screening. Our CFT analysis below will reveal the
reason for this: orbital screening is accompanied by a renormalization
of the local bath spin density at the impurity site.

Upon entering the SS regime, all susceptibility lines show a kink,
i.e., change in power law, such that the impurity and bath exponents
match not only in the orbital sector, $\chi^\imp_\orb \sim
\chi^\bath_\orb$, but now also in the spin sector, $\chi^\imp_\spin
\sim \chi^\bath_\spin$. The latter fact indicates clearly that bath
and impurity spin degrees of freedom have begun to interact with each
other. However, this is only a precursor to spin screening, since the
spin susceptibilities still increase with decreasing $\omega$, albeit
with a smaller exponent, $\chi_\spin^{\imp,\bath} \sim \omega^{-6/5}$,
than in the NFL regime. However, since the exponent $\gamma =
  6/5$ is larger than 1, spin fluctuations are anomalously large also
  in this regime.  Importantly, this regime persists also for
  parameters corresponding to the more realistic \threeoAH\ model.
  Indeed, previous DMFT studies for a self-consistent \threeoAH\ model
  have yielded behavior for $\chi_\spin^\imp$ which in the SOS regime
  is consistent with an exponent of $\gamma=6/5$, as further discussed
  in Secs.\ \ref{sec:CFT_synopsis} and
  \ref{sec:3OAH-model}. Moreover, as mentioned in the Introduction,
  anomalously large spin fluctuations are of direct relevance for the
  superconducting state of the iron pnictide Hund metals: in
  Ref.~\cite{Lee2018a}, strong spin fluctuations with $\gamma > 1$
  were a key ingredient for a proposed explanation for the anomalously
  large ratio of $2\Delta_\text{max}/T_\text{c}$ observed
  experimentally.

Full spin screening eventually sets in in the FL regime, where the spin
susceptibilities $\chi^{\imp,\bath}_\spin$ show the $\omega^1$
behavior characteristic of a FL.  We expect this behavior also for the
orbital susceptibilities, but have not been able to observe it
directly, since our results for $\chi^{\imp,\bath}_\orb$ become
numerically unstable when dropping below $\simeq 10^{-5}$ [as
indicated by dotted lines in
Figs.~\ref{fig:spectra_susceptibilities_K0}(c) and (d)].

In the following two sections we explain 
how the above NRG results can be understood using
CFT arguments.

\renewcommand{\tabcolsep}{4pt} 
\begin{table*}[tb]   
\centering
\caption{Left: Five low-lying free-fermion multiplets ($|\mathrm{FS}\rangle$
  denotes the Fermi sea), with quantum numbers $(q,S,\lambda)$, multiplet dimensions $d$, and energies $E(q,S,\lambda)$. 
  Center: ``Single fusion'' with an impurity $Q_\mathrm{imp}=(0,1, \protect\raisebox{0.6mm}{\tiny $\protect\yng(1,1)$}\,)$ leads to multiplets  with quantum numbers $(q, S^\prime, \lambda^\prime)$, dimensions $d^\prime$, 
  eigenenergies $E^\prime = E(q,S,\lambda^\prime)$, and excitation
  energies $\delta E^\prime = E^\prime - E^\prime_\mathrm{min}$. 
  Right: ``Double fusion,'' which fuses multiplets from the middle column with an impurity in the conjugate representation $\Qbar_\imp = (0,1, 
  \protect\raisebox{0.7mm}{\tiny $\protect\yng(1)$}\,)$ [cf.~Sec.~\ref{sec:supp:fusion_SU(3)}, details on \step3], yields the 
multiplets  $(q, S^{\prime\prime}, \lambda^{\prime\prime})$. These characterize the CFT boundary operators $\hat{O}$, with scaling dimensions $\Delta=E(q,S,\lambda^{\prime\prime})$. $\bPhi_\orb$ and $\bPhi_\spin$ are the leading boundary operators in the orbital and spin sectors, respectively. 
  In the spin-splitting regime, their roles are taken by
  $\tPsi_\orb$ and $\tPsi_\spin$, respectively. ``Bare'' free-fermion versions of
  these boundary operators, having the same quantum numbers, 
  are listed on the very right.
  For clarity, not all possible multiplets arising from single and double fusion are shown. 
  A more comprehensive list is given in Table~\ref{tab:supp:fusion_NRG}.  
}
\renewcommand{\arraystretch}{1.4}
\begin{tabular}{cccc@{\hspace{8mm}}cc@{\hspace{2mm}}|@{\hspace{2mm}}ccc@{\hspace{8mm}}cc|ccc@{\hspace{8mm}}ccc} 
\hline \hline
\multicolumn{6}{c|@{\hspace{2mm}}}{Free fermions} & \multicolumn{5}{c|}{Single fusion} & \multicolumn{6}{c}{Double fusion}
\\ 
State & 
$q$ & $S$ & $\lambda$ & $d$ & $E$ &
$q$ & $S^\prime$ & $\lambda^\prime$ & $d^\prime$ & $\delta E^\prime$ & 
$q$ & $S^{\prime\prime}$ & $\lambda^{\prime\prime}$ & $\Delta$ & $\hat{O}$ 
& $\hat{O}_\mathrm{bare}$
\\ \hline 
$|\mathrm{FS}\rangle$ & 
0 & 0 & \y00 & 1 & 0 & 
0 & 1 & \raisebox{0.5mm}{\y01} & 9 & \f130 & 
0 & 0 & \raisebox{0.5mm}{\y11} & \fr35 & $\bPhi_\orb$ &
$\bT, \bJ_\orb$
\\ 
$\psi_{m\sigma}^\dagger |\mathrm{FS}\rangle$ &
1 & \fr12 & \raisebox{0.5mm}{\y10} & 6 & \fr12 &
1 & \big\{\fr12, \fr32\big\} & \y00 & 6 & 0 &
& $\cdots$ & & & & 
\\ 
$\psi_{m\sigma} |\mathrm{FS}\rangle$ &
$-1$ & \fr12 & \raisebox{0.5mm}{\y01} & 6 & \fr12 &
$-1$ & \fr12 & \raisebox{0.5mm}{\y10} & 6 & \f415 &
$-1$ & \fr12 & 
\raisebox{0.5mm}{\tiny$\yng(2)$} & \f910 & 
$\tPsi_\orb$ &
$(\psi_{l\sigma}^\dagger \psi_{l\sigma} \!-\! \psi_{m\sigma}^\dagger \psi_{m\sigma}) \psi_{n\sigma}$,
\\
& & & & & & & & & & & & & & & &
$\psi_{l\sigma}^\dagger \psi_{m\sigma} \psi_{n\sigma}$, {\footnotesize $l\!\neq\! m \!\neq\! n$}
\\ 
$\bJ_\spinorb |\mathrm{FS}\rangle$ &
0 & 1 & \raisebox{0.5mm}{\y11} & 24 & 1 &
0 & 0 & \raisebox{0.5mm}{\y01} & 3 & \fr{13}{30} &
0 & 1 & \y00 & \fr25 & $\bPhi_\spin$ &
$\bJ_\spin$
\\ 
$\cdots$ &
$-3$ & \fr12 & \raisebox{0.5mm}{\y11} & 16 & \fr32 &
$-3$ & \fr12 & \raisebox{0.5mm}{\y01} & 6 & \fr{14}{15} & 
$-3$ & \fr12 & \y00 & \f910 & $\tPsi_\spin$ &
$\psi_{1\sigma} \psi_{2\sigma} \psi_{3\bar{\sigma}}$
\\ \hline\hline
\end{tabular} 
\vspace{-4mm}
\label{tab:single_fusion_lowest}
\end{table*}

\section{CFT analysis: synopsis}
\label{sec:CFT_synopsis}

This section presents a synopsis of our 
CFT analysis.  It aims to be accessible also to readers
without in-depth knowledge of AL's CFT work on 
Kondo models. 
We begin by summarizing AL’s strategy for analyzing strong-coupling fixed points of quantum impurity models (Sec.~\ref{sec:keyconcepts}).
We then apply it to the NFL fixed point (Sec.~\ref{sec:NFLsynopsis})
and the FL fixed point (Sec.~\ref{sec:FLsynopsis}).
A more elaborate discussion of CFT details
follows in Sec.~\ref{sec:CFT_analysis}.

\subsection{General strategy}
\label{sec:keyconcepts}

AL's strategy for determining spectra and correlation functions from CFT involves three key concepts:

\begin{itemize}

\item[\step1]
\textit{Independent excitations.}---The starting assumption is that the low-energy spectrum of a multiorbital Kondo Hamiltonian at a conformally invariant fixed point can be constructed from combinations of \textit{independent} charge, spin, and orbital excitations. The excitation energies in each sector follow from the commutation relations of certain charge, spin, and orbital operators (these form a so-called Kac-Moody algebra); this is expressed in Eqs.~\eqref{eq:H_bosonized} and \eqref{eq:supp:E(qSlambda)-sup}.

\item[\step2]
\textit{Gluing conditions and fusion rules.}---The spectrum of excitations in each sector (charge, spin, orbital) is the same at the free and strong-coupling fixed points. However, the way in which these three types of excitations should be combined to obtain valid many-body excitations, specified by so-called \textit{gluing conditions}, differs for the free and strong-coupling fixed points. At the former, excitations are glued together in such a manner that a free-fermion spectrum is recovered. At the latter, the impurity has been absorbed by the bath, implying changes in the gluing conditions relative to those of the free fixed point.
These changes are governed by so-called \textit{fusion rules}, which specify how the impurity degrees of freedom should be ``added'' to those of the bath. This is conceptually similar to angular momentum addition, but with additional constraints to respect the Pauli principle. 

\item[\step3]
\textit{Scaling dimensions.}---Once the fusion rules and thus the spectrum of valid many-body excitations is known, the conformal scaling dimensions of operators living at the impurity site can be determined by using the same fusion rules once more (``double fusion''). Because of conformal invariance, the functional form of correlation functions is fully determined by the scaling dimensions of their operators.

\end{itemize}

In practice, analyzing a conformally invariant strong-coupling fixed point thus consists of three steps: \step1\ determine the independent excitations, \step2\ use ``single fusion'' to obtain the strong-coupling gluing conditions, and \step3\ use ``double fusion'' to obtain the scaling dimensions of operators living at the impurity site.
Even though AL’s justification of this strategy involved sophisticated CFT arguments, its application to an actual model is rather straightforward, once one has determined the appropriate fusion rules. For the \threesoK\ model, we present tables with the explicit fusion rules in the Supplemental Material (SM) \cite{suppmat}, and Table \ref{tab:supp:fusion_NRG} shows details on the fusion procedure. These tables are also meant to serve as a guide for future applications of AL's methodology.

\subsection{NFL regime}
\label{sec:NFLsynopsis}

In the following, we follow this strategy for the NFL fixed point of the \threesoK\ model.

\step1 
The \threesoK\ model, being spherically symmetric around the
  origin, describes an effectively one-dimensional system.  In the
  imaginary-time formalism, the field describing the conduction band,
  $\psi(\tau+ir)$, lives on the upper half of the complex plane, with
  time, $\tau$, on the real and the distance, $r$, from the impurity
  on the imaginary axis.  The impurity at $r=0$ constitutes a
  ``boundary'' at the real axis.  The fixed points of the model,
  assumed to be scale invariant, can thus be described using
  (1+1)-dimensional boundary CFT.

The bath of the \threesoK\ model trivially has $\uone \times \sutwo \times
  \suthree$ symmetry. Moreover, since we assumed a flat band, i.e., a linear dispersion,  it also has conformal symmetry. The  combination of both leads 
  to the symmetry  $\uone \times \sutwo_3 \times
  \suthree_2$, where $\sutwo_3$ and $\suthree_2$ refer to
  generalizations of the familiar $\sutwo$ and $\suthree$ algebras,
  known as Kac-Moody algebras \cite{Affleck1991,VonDelftPhD1995,Ludwig1994}.
  The subscript on $\sutwo_3$ states that only
  those spin representations are allowed which can be constructed from
  electrons living on 3 orbitals. In particular, spins larger than
  $3/2$ do not occur in this algebra. The subscript on 
$\suthree_2$ indicates analogous restrictions
for the allowed \suthree\ representations. (The 
consequences of these restrictions are made explicit in Tables~S3 and S2
of the Supplemental Material \cite{suppmat}.)

According to AL
  \cite{Affleck1991,Affleck1990,Affleck1991a,Affleck1993,Ludwig1994a},
  the fixed points can be analyzed as follows. First, standard $\uone \times
\sutwo_3 \times \suthree_2$ non-Abelian bosonization 
is used to decompose the bath Hamiltonian 
into charge, spin, and
orbital contributions, 
\begin{align} H_\mathrm{bath}  \sim
  \int \mathrm{d} r \left(  \tfrac1{12} \, J_\ch^2 (r)+ \tfrac15 \, \bJ_\spin^2 (r) +
    \tfrac15 \, \bJ_\orb^2 (r) \right) \,,
    \label{eq:H_bosonized}
\end{align}
with $J_\ch (r)= \psi_{m\sigma}^\dagger (r)\psi_{m\sigma}(r)$, etc. (We omitted overall prefactors; for a detailed discussion, see Refs.\
  \cite{Affleck1991,VonDelftPhD1995}.) Since $J_\ch$, $\bJ_\spin$,
  $\bJ_\orb$ are generators of the $\uone$, $\sutwo_3$, $\suthree_2$
  Kac-Moody algebras, respectively, the eigenstates of
  $H_\mathrm{bath}$ can be organized into multiplets forming irreps of
  the corresponding symmetry groups, labeled by quantum numbers
  $Q_\bath = (q, S, \lambda$). If the bath is put in a box of
finite size, the corresponding
  free-fermion excitation eigenenergies $E(q,S,\lambda)$
are discrete and simple functions
of the quantum numbers [see Eq.~\eqref{eq:supp:E(qSlambda)-sup}].

\step2
Next, we include the interaction with the impurity in the orbital
sector ($K_0>0$, $J_0=I_0=0$) to describe the properties of the NFL
fixed point $\bc_\nfl^\ast$.  The bosonized $H_\bath$ is quadratic in
$\bJ_\orb$, whereas the coupling term $H_\interaction = K_0 \, \bT
\cdot \bJ_\orb (r=0)$ is linear. The latter can thus
be absorbed into the former, 
 in the spirit of ``completing the square.'' AL conjectured
that at the
strong-coupling fixed point, this replacement takes the form
\begin{align}
\bJ_\mathrm{orb} (r) \mapsto \bcJ_\orb (r) = \bJ_\orb (r) + \delta(r) \, \bT \,,
\label{eq:absorbing_Jorb}
\end{align}
with $\bcJ_\orb$ satisfying the same Kac-Moody algebra as
$\bJ_\orb$. At the strong-coupling fixed point, the Hamiltonian can
thus be expressed as $H = H_\bath [\bJ_\orb] + H_\interaction =
H_\bath [\bcJ_\orb]$ (more details can be found in Sec.~\ref{sec:supp:fusion_SU(3)}
and Ref.\ \cite{Affleck1991a}). 

It follows immediately that at the
  fixed point, the spectrum of irreps of the full Hamiltonian can be
  obtained by combining the irreps of bath and impurity
  degrees of freedom, 
$Q_\bath \otimes Q_\imp = \sum_{\oplus} Q'$, 
and using ``fusion rules'' to deduce the resulting irreps $Q'$. This
is conceptually similar to coupling two $\sutwo$ spins, $\bS'' = \bS +
\bS'$, decomposing the direct product of their irreps as $S\otimes S'
= \sum_{\oplus} S''$, and deducing that $S''$ ranges from $|S - S'|$
to $S + S'$. However, in the present context, specific assumptions
must be made about which degrees of freedom are involved in the
screening processes and which are not, and for those which are, Kac-Moody
fusion rules have to be used when combining irreps.  For the present
situation, we have $Q_\bath = (q,S,\lambda)$ and $Q_\mathrm{imp} =
(0,1,\raisebox{0.5mm}{\tiny $\yng(1,1)$}\, )$, and place ourselves
\textit{at} the NFL fixed point, where bath and impurity couple only
in the orbital sector. 

To find the allowed irreps $Q' = (q', S',
\lambda')$, we therefore posit the following fusion strategy 
(inspired by and generalizing that of AL
\cite{Affleck1991,Affleck1990,Affleck1991a,Affleck1993,Ludwig1994a}).
In the charge sector, $q_\imp= 0$ trivially implies that $q'= q$.  In
the orbital sector, the impurity's orbital isospin is 
coupled to that of the bath [Eq.~\eqref{eq:Hamiltonian_AronKotliar}] and absorbed by it according to Eq.~\eqref{eq:absorbing_Jorb};
hence, $\lambda \otimes \lambda_\imp = \sum_\oplus
\lambda'$ is governed by the fusion rules of the $\suthree_2$
Kac-Moody algebra.  By contrast, in the spin sector the impurity spin
is a spectator, decoupled from the bath (we are \textit{at}
$\bc_\nfl^\ast$, where $J_0=I_0=0$); hence, $S \otimes S_\imp =
\sum_\oplus S'$ is governed by the fusion rules of the $\sutwo$ Lie
algebra [not the $\sutwo_3$ Kac-Moody algebra]. The set of
excitations $(q,S^\prime,\lambda^\prime)$ so obtained have energies
given by $E(q,S,\lambda^\prime)$, not $E(q,S^\prime,\lambda^\prime)$,
since $H_\interaction$ only acts in the orbital sector.  A more
complete discussion of our ``fusion hypothesis'' is given in
Sec.~\ref{sec:supp:fusion_SU(3)}.  The resulting spectrum reproduces
the NRG spectrum in the NFL fixed point regime (see
Table~\ref{tab:supp:fusion_NRG}).

Table \ref{tab:single_fusion_lowest} exemplifies a few many-body
states obtained via this fusion scheme (AL called
it  single fusion, in distinction from a second
fusion step, discussed below). In particular, the degenerate ground
state multiplets of $\bc_\nfl^\ast$, (1, \fr12, $\bullet$) and (1,
\fr32, $\bullet$)
[cf.\ Fig.~\ref{fig:spectra_susceptibilities_K0}(a)], arise via
fusion of a one-particle bath excitation, \mbox{(+1, \fr12, $
  \raisebox{0.7mm}{\tiny $\yng(1)$}$\,)}, with the impurity, $(0,1,
\raisebox{0.5mm}{\tiny $\yng(1,1)\,$})$, schematically depicted in
  Fig.~\ref{fig:spectra_susceptibilities_K0}(b).

\step3
Next, we want to compute the leading scaling behavior of spin and
  orbital correlation functions at the impurity site, i.e., on the
  boundary of the CFT. The absorption of the impurity into the
    bath (bulk) Hamiltonian translates, in CFT language, to a change
    in the boundary condition imposed on the theory at $r=0$. As a
    result, a new set of ``boundary operators,'' i.e., local operators
    living at the impurity site, appear in the theory. These fully
  characterize the strong-coupling fixed point. Each boundary
  operator can be viewed as the renormalized version, resulting
    from the screening process, of some bare local operator having
  the same quantum numbers. 

According to AL, the boundary
  operators can be obtained via a second fusion
step (double fusion) 
(cf.~Refs.~\cite{Affleck1991,Affleck1993,Ludwig1994a} and Appendix~C of
Ref.~\cite{VonDelftPhD1995}). Each multiplet
  $(q,S^{\prime\prime},\lambda^{\prime\prime})$ resulting from double
  fusion is associated with a boundary operator $\hat O$ with the
same quantum numbers, and a scaling dimension given by $\Delta =
E(q,S,\lambda^{\prime\prime})$ (cf.\ Table
\ref{tab:single_fusion_lowest}). The realization that the scaling
  dimensions of boundary operators are related to finite-size
  excitation energies is due to Cardy \cite{Cardy1984}.  Using a
conformal mapping, he mapped the complex upper half-plane to a strip of
infinite length and finite width, in such a way that the
nontrivial boundary condition of the half-plane is mapped to both boundaries of the
strip.  He then showed that the boundary operators of the half-plane and
their scaling dimensions can be associated with the finite-size
spectrum of a Hamiltonian defined along the width of this strip.
Since the strip has two nontrivial boundaries, one on each side, 
the finite-size spectrum can be found using a double-fusion
procedure. The scaling dimensions of the boundary operators
  fully determine their time- or frequency-dependent correlators,
$\langle \hat O(t)\hat O(0) \rangle \sim t^{-2 \Delta}$ and $\langle
\hat O || \hat O \rangle_\omega \simeq \omega^{2 \Delta -1}$.

To explain the power laws found in the NFL regime of 
Figs.~\ref{fig:spectra_susceptibilities_K0}(c) and \ref{fig:spectra_susceptibilities_K0}(d), 
and particularly the fact that there $\chi_\orb^\imp$
and $\chi_\orb^\bath$ exhibit the \textit{same} power law, 
while $\chi_\spin^\imp$ and $\chi_\spin^\bath$ do not, we posit that 
the local operators in the orbital and spin exchange terms of
Eq.~\eqref{eq:Hamiltonian_AronKotliar} are renormalized to 
\begin{align}
\label{eq:RG-map-1} 
\bJ_\orb \mapsto \bPhi_\orb , \quad \bT \mapsto \bPhi_\orb, \quad 
\bJ_\spin \mapsto \bPhi_\spin, \quad \bS \mapsto \bS. 
\end{align}
Here $\bPhi_\orb$ has quantum numbers $(0,0, \raisebox{0.5mm}{\tiny $\yng(2,1)$}\,)$ 
(same as $\bT$,  $\bJ_\orb$) and dimension $\Delta_\orb = \frac{3}{5}$,
while $\bPhi_\spin$ has quantum numbers $(0,1,\bullet)$ 
(same as $\bS$, $\bJ_\spin$)
and  $\Delta_\spin = \frac{2}{5}$, 
(cf.\ Table \ref{tab:single_fusion_lowest}).
The local impurity and bath  orbital susceptibilities
thus both scale as
\begin{align}
\label{eq:chi_orb^imp,bath}
\chi_\orb^{\imp,\bath} \sim 
\langle \bPhi_\orb || \bPhi_\orb\rangle_\omega
\sim \omega^{2\Delta_\orb-1} = \omega^{1/5},  
\end{align}
and the bath spin susceptibility as 
\begin{align}
\label{eq:chi_spin^bath}
\chi_\spin^\bath  \sim 
\langle \bPhi_\spin || \bPhi_\spin\rangle_\omega
\sim \omega^{2\Delta_\spin-1} = \omega^{-1/5}. 
\end{align}

By contrast, the impurity spin $\bS$ is not renormalized, because
\textit{at} the fixed point $\bc_\nfl^\ast$, where $J_0 = 0$, it is
decoupled from the bath. Thus its scaling dimension is zero. 
The leading behavior of $\chi_\spin^\imp$ is obtained
by now taking $J_0 \neq 0$ but very small $(\ll K_0)$, and doing
second-order perturbation theory in the renormalized
spin exchange interaction. Thus, $\chi_\spin^\imp$ is proportional
to the Fourier transform of 
$\langle \bS(t) \bS(0) (\int \text{d}t' J_0 \bS \cdot \bPhi_\spin)^2 \rangle$,
and  power  counting yields
\begin{align}
\label{eq:chi-spin-imp_NFL}
\chi_\spin^\imp \sim \omega^{2\Delta_\spin-3} = \omega^{-11/5}. 
\end{align}
The above predictions are all borne out in 
Figs.~\ref{fig:spectra_susceptibilities_K0}(c) and \ref{fig:spectra_susceptibilities_K0}(d).
 
The remarkably large negative exponent, $-\frac{11}{5}$, for
$\chi_\spin^\imp$ reflects the fact that the renormalized spin
exchange interaction $J_0 \bS \cdot \bPhi_\spin$, with scaling
dimension $\frac{2}{5} < 1$, is a relevant perturbation. Its strength,
though initially miniscule if $J_0 \ll 1$, grows under the RG flow,
causing a crossover away from $\bc_\nfl^\ast$ for
$\omega \lesssim \Tss$. This is reflected in the level crossings
around $\Tss$ in the NRG eigenlevel flow of
Fig.~\ref{fig:spectra_susceptibilities_K0}. In particular, the
double-fusion parent multiplets for $\bPhi_\orb$ and $\bPhi_\spin$,
namely $(0, 1, \raisebox{0.5mm}{\tiny $\yng(1,1)$}\,)$ and
$(0, 0, \raisebox{0.5mm}{\tiny $\yng(1,1)$}\,)$, undergo level
crossings with the downward-moving multiplets
$(-1, \frac{1}{2}, \raisebox{0.7mm}{\tiny $\yng(1)\,$})$ and
$(-3, \frac{1}{2}, \raisebox{0.5mm}{\tiny $\yng(1,1)$}\,)$,
respectively.
These in turn are double-fusion parent multiplets for
the boundary operators $\tPsi_\orb$ and $\tPsi_\spin$, with scaling
dimensions $\tDelta_\orb = \tDelta_\spin = \frac{9}{10}$
(Table~\ref{tab:single_fusion_lowest}). 
To explain the SS regime of 
Figs.~\ref{fig:spectra_susceptibilities_K0}(c) and \ref{fig:spectra_susceptibilities_K0}(d),
and  particularly that  there the power laws
for $\chi^\imp$ and $\chi^\bath$ match in both
the orbital \textit{and} spin sectors, 
we posit the RG replacements 
\begin{align}
\nonumber
\bJ_\orb \mapsto \tPsi_\orb , \; \bT \mapsto \tPsi_\orb , \; 
\bJ_\spin \mapsto \bS + \tPsi_\spin, \; \bS \mapsto \bS + \tPsi_\spin .
\end{align}
Here $\bS + \tPsi_\spin$ is symbolic notation for some linear
admixture of both operators, induced by the action of the renormalized
spin exchange interaction. We thus obtain
\begin{align}
\label{eq:tilde chi_orb^imp,bath}
\chi_\orb^{\imp,\bath} \sim 
\langle \tPsi_\orb || \tPsi_\orb\rangle_\omega
\sim \omega^{2\tDelta_\orb-1} = \omega^{4/5}, 
\end{align}
and the leading contribution to 
$\chi_\spin^\imp$ and $\chi_\spin^\bath$,
obtained by perturbing $\langle \bS(t) \bS(0)\rangle$
to second order in $ \bS \tPsi_\spin$ \cite{quantumnumbersdiffer},  
is 
\begin{align}
\label{eq:chi-spin-imp_SS}
\chi_\spin^{\imp,\bath} \sim 
  \omega^{2\tDelta_\spin-3} = \omega^{-6/5}. 
\end{align}
This reproduces the power laws found in
Figs.~\ref{fig:spectra_susceptibilities_K0}(c) and \ref{fig:spectra_susceptibilities_K0}(d). 

Remarkably, $\chi_\spin^\imp \sim \omega^{-6/5}$ behavior has
also been found in studies of the self-consistent \threeoAH\ model
arising in our DMFT investigations of the three-orbital Hubbard-Hund
model for Hund metals. For the \threeoAH\ model the spin-orbital
coupling $I_0$ in Eq.~\eqref{eq:Hamiltonian_AronKotliar} is always
nonzero, so that a fully fledged NFL does not emerge---instead,
$\Torb$ and $\Tss$ effectively coincide (as further discussed in
Sec.~\ref{sec:3OAH-model}). However, the SS regime between $\Tspin$
and $\Tss\simeq \Torb$ can be quite wide, typically at least an
order of magnitude.  In Fig.~3(c) of Ref.~\cite{Stadler2015}, the
behavior of $\chi^\imp_\spin$ in this regime (between the vertical
solid and black lines there) is consistent with $\omega^{-6/5}$
behavior. Though this fact was not noted in Ref.~\cite{Stadler2015},
it was subsequently pointed out in Ref.~\cite{Lee2018a} (see Fig.~S1
of their Supplemental Material).  Behavior consistent with
$\chi^\imp_\spin \sim \omega^{-6/5}$ can also be seen in
Figs.~5.1(c) and 5.1(d) of Ref.~\cite{Stadler2019}, as discussed on p.~152
therein. The explanation for this behavior presented here, via a CFT
analysis of the NFL and SS regimes, is one of the main results of
this work, and the justification for the first part of the title of
this paper.
    
\subsection{Fermi-liquid regime}
\label{sec:FLsynopsis}

As mentioned above, the low-energy regime below $\Tspin$ is a FL.
The fixed-point spectrum at $\bc_\fl^\ast$ can be obtained 
by fusing a free-fermion spectrum with an  impurity with
$Q_\imp = (1, \frac{3}{2}, \bullet$), representing the effective
local degree of freedom obtained after completion of orbital screening
(see Table~\ref{tab:supp:fusion_NRG_FL}). 
Since the ground state describes a fully screened orbital
and spin singlet, it actually is the singlet of a larger
symmetry group, $\uone \times \susix$. Indeed,
the fixed-point spectrum at $\bc_\fl^\ast$ matches that of the 
$\uone \times \susix$ symmetric Kondo model.
We demonstrate this, using both NRG and CFT with $\susix_1$ fusion rules, 
in Sec.~\ref{sec:FL-spectrum} (see Table~\ref{tab:fusion_SU(6)}).
The FL nature of
the ground state is also borne out by the $\omega^1$ scaling
of $\chi^{\imp,\bath}_\spin$ in the FL regime of 
Figs.~\ref{fig:spectra_susceptibilities_K0}(c) \ref{fig:spectra_susceptibilities_K0}(d).

\section{CFT  analysis: details} 
\label{sec:CFT_analysis} 

We now provide technical details for our CFT analysis of the NFL and
FL fixed points of the three-orbital Kondo (\threesoK) model discussed 
in Secs.~\ref{sec:NRG} and \ref{sec:CFT_synopsis}.
We closely follow the strategy devised by Affleck and
Ludwig for their pioneering treatment of the strong-coupling
fixed points of Kondo models
\cite{Affleck1991,Affleck1990,Affleck1991a,Affleck1993,Ludwig1994a}
(for pedagogical reviews, see Refs.~\cite{Ludwig1994,Affleck1995} and
Appendixes A--D of Ref.~\cite{VonDelftPhD1995}). In a series of works, they
considered a variety of Kondo models of increasing complexity.  These
include the standard one-channel, \sutwo\ spin Kondo model with 
a spin exchange interaction between bath and impurity 
with $\uone\times \sutwo_1$ symmetry; a spinful $k$-channel bath
coupled to an \sutwo\ impurity [$\uone\times \sutwo_k \times \suk_2$
symmetry], and  an \suN\ $k$-channel bath coupled to an \suN\
impurity [$\uone\times \suN_k \times \suk_N$ symmetry].

Our \threesoK\ model features a spinful three-channel bath and an
$\sutwo_\spin \times \suthree_\orb$ impurity
[$\uone \times \sutwo_3 \times \suthree_2$ symmetry]. The impurity
multiplet is a direct product of a spin triplet ($S=1$) and an orbital
triplet $(\lambda=\raisebox{-0.5mm}{\tiny\yng(1,1)\,})$.  Its
direct-product structure is more general than any of the cases
considered by AL. (A two-channel version of our model, with
$\uone \times \sutwo_2 \times \sutwo_2$ symmetry, has been studied by
Ye \cite{Ye1997}, which we discuss in the Appendix.)
However, at the NFL fixed point $\bc_\nfl^\ast$ of our model, where
$J_0=I_0=0$, the impurity's \sutwo\ spin is a decoupled, threefold
degenerate spectator degree of freedom. Hence AL's analysis 
\cite{Ludwig1994a}  can be employed, with $N=3$ and $k=2$ channels,
modulo some minor changes to account for the impurity spin.

By contrast, in the spin-splitting crossover regime 
the spin exchange interaction comes to life, so that the impurity's
\sutwo\ spin degrees of freedom cease to be mere spectators.  This
regime thus lies outside the realm of cases studied by AL; in
particular, it is not manifestly governed  by the NFL fixed point
$\bc_\nfl^\ast$, or any other well-defined fixed
point. Correspondingly, our discussion of this crossover regime in
Sec.~\ref{sec:SS-regime} is more speculative than that of the
NFL regime, though our heuristic arguments are guided by and
consistent with our NRG results.

Finally, for our model's FL fixed point $\bc_\fl^\ast$,  we are
again in well-chartered territory: it can be understood by applying
AL's strategy to an \susix\ one-channel bath coupled to an \susix\
impurity [$\uone \times \susix_1$ symmetry].

Below we assume the reader to be familiar with AL's work and just
focus on documenting the details of our
analysis. Section~\ref{sec:non-Abelian-bosonization} describes how the
free-fermion bath spectrum is decomposed into charge, spin, and orbital
excitations using $\uone \times \sutwo_3 \times \suthree_2$
non-Abelian bosonization.  Section~\ref{sec:supp:fusion_SU(3)} derives
the finite-size spectrum and boundary operators of the NFL fixed
point via single and double fusion, using  the fusion rules of
the $\suthree_2$ Kac-Moody algebra in the orbital sector and the
$\sutwo$ Lie algebra in the spin sector. 
Section~\ref{sec:CFT-susceptibilities} describes the computation of
the spin and orbital susceptibilities in the NFL and SS regimes,
linking AL's strategy for computing such quantities to the compact
scaling arguments used in Sec.~\ref{sec:CFT_synopsis}.
Section~\ref{sec:imp_spectral_function} presents our results for the
impurity spectral function in the NFL regime. 
Finally, Sec.~\ref{sec:FL-spectrum}, devoted to the
FL regime, shows how its spectrum can be derived using
either $\sutwo_3$ fusion rules in the spin sector or $\susix_1$ fusion rules
in the flavor (combined spin+orbital) sector.

\subsection{Non-Abelian $\uone \times \sutwo_3 \times \suthree_2$ bosonization} 
\label{sec:non-Abelian-bosonization} 

\step1
The first step of AL's CFT approach for multichannel Kondo models is
to use non-Abelian bosonization to decompose the bath degrees of
freedom into charge, spin, and orbital excitations in a manner
respecting the symmetry of the impurity-bath exchange interactions.
Our \threesoK\ model features a spinful three-channel bath, with
$H_\bath = \sum_{pm\sigma} \varepsilon_p \psi_{pm\sigma}^\dagger
\psi_{pm\sigma}$.
We assume a linear dispersion, $\varepsilon_p = \hbar v_\Fermi p$,
with $\hbar v_\Fermi = 1$.  Using non-Abelian bosonization with the
$\uone \times \sutwo_3 \times \suthree_2$ Kac-Moody (KM) current
algebra, the spectrum of bath excitations can be expressed as (see
Refs.~\cite{Affleck1990,Affleck1991}, or Appendix~A of
Ref.~\cite{VonDelftPhD1995})
\begin{subequations}
\label{eq:supp:E(qSlambda)-sup} 
\begin{align}
E(q,S,\lambda) & = 
\tfrac{1}{12}q^2 + \tfrac{1}{5}\kappa_2 (S) + \tfrac{1}{5} 
\kappa_3 (\lambda) + \ell \,,
\label{eq:supp:E(qSlambda)} 
\\
\label{eq:CasimirSU2}
\kappa_2 (S) &= S(S+1) \,,
\\
\label{eq:CasimirSU3}
\kappa_3(\lambda) &= 
\tfrac13 \left( \lambda_1^2 + \lambda_2^2 + \lambda_1\lambda_2 + 3\lambda_1 + 3\lambda_2  \right) \,.
\end{align}
\end{subequations}
Here $\kappa_2 (S)$ and $\kappa_3(\lambda)$ are the eigenvalues of the
quadratic Casimir operators of the $\sutwo$ and $\suthree$ Lie
algebras, respectively \cite{FuchsSchweigert}. $q \in \mathbb{Z}$ is
the \uone\ charge quantum number, $S \in \frac{1}{2} \mathbb{Z}$ the
\sutwo\ spin quantum number, and $\lambda = (\lambda_1,\lambda_2)$ the
\suthree\ orbital quantum number, denoting a Young diagram with
$\lambda_j$ $j$-row columns:
\[
\yng(5,2)
\hspace{-2.05cm}
\underbrace{\phantom{\yng(2,0)}}_{\lambda_2} \underbrace{\phantom{\yng(3,0)}}_{\lambda_1}
\qquad
\begin{matrix}
\text{$\lambda_1$ = number of one-row columns}
\\
\text{$\lambda_2$ = number of two-row columns}
\end{matrix}
\]
Finally, $\ell \in \mathbbm{Z}$ counts higher-lying
``descendent'' excitations; for present purposes it suffices to set
$\ell = 0$.

The free-fermion spectrum of $H_\bath$ is recovered from
Eq.~\eqref{eq:supp:E(qSlambda)} by imposing free-fermion ``gluing
conditions,'' allowing only those combinations of quantum numbers
$(q,S,\lambda)$ for which $E(q,S,\lambda)$ is an integer multiple of
1/2.  The
resulting multiplets are listed in the left-hand column (``Free fermions'')
of Table \ref{tab:supp:fusion_NRG}.

\begin{figure*}[tb]
\includegraphics[scale=0.9]{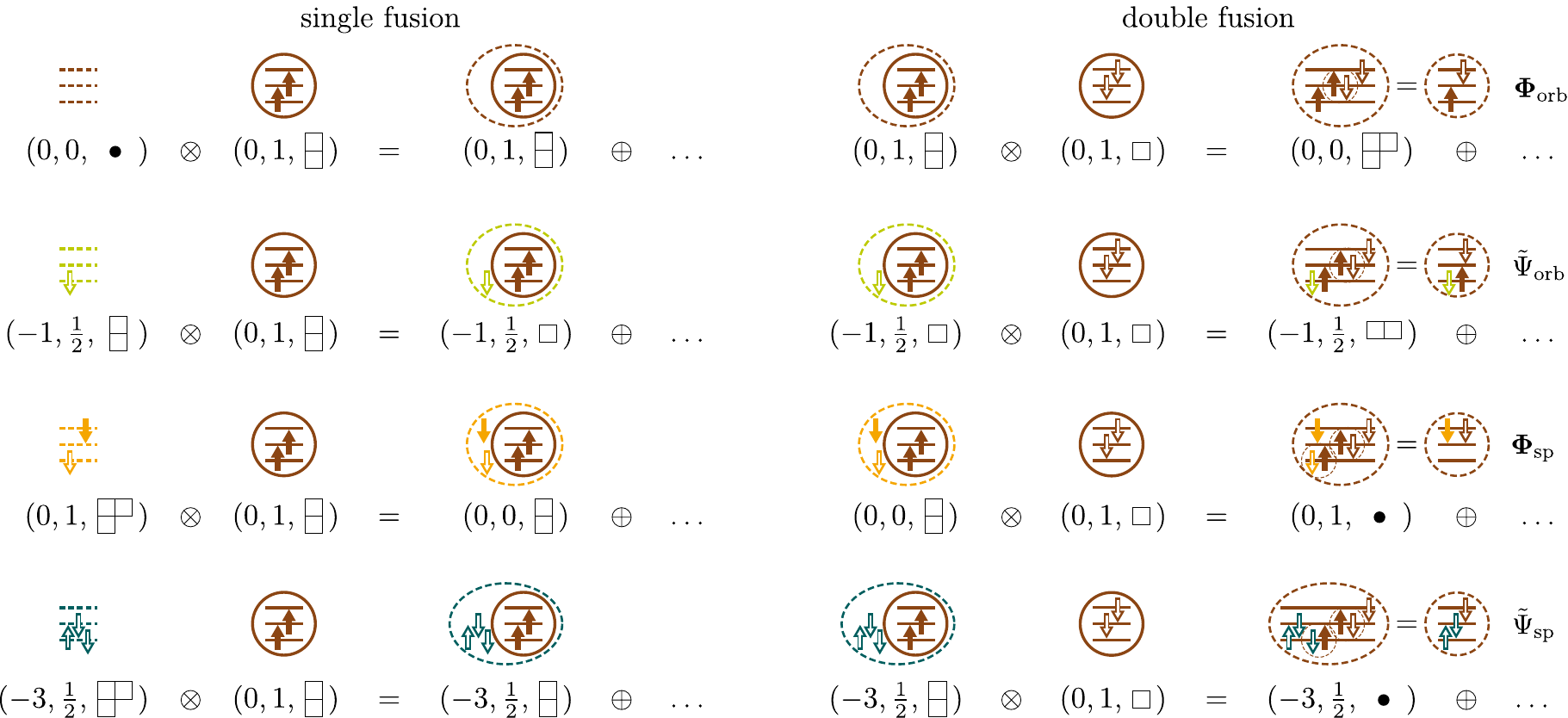}
\caption{Schematic depiction of single fusion (left) and double fusion
  (right), for the four multiplets giving rise to the boundary
  operators $\bPhi_\orb$, $\bPhi_\spin$, $\tPsi_\orb$, $\tPsi_\spin$
  discussed in Sec.~\ref{sec:CFT_synopsis} (corresponding to rows
  1,3,4,5 in Table \ref{tab:single_fusion_lowest}).  Filled arrows
  represent electrons, empty arrows represent holes.  An electron with
  spin $\uparrow$ and a hole with spin $\Downarrow$ (missing electron
  with spin $\uparrow$) can be combined to annihilate each other, as
  indicated by small dashed circles in the last column. Our illustrations
  depict the impurity using a fermionic representation, as would be
  appropriate for the \threeoAH\ model, even though the \threesoK\
  impurity has no charge dynamics.  In the ``single fusion'' column,
  excitations of the free bath are fused with the impurity,
  $Q_\imp=(0, 1, \,\protect\raisebox{-1mm}{\tiny
    \protect\yng(1,1)}\,)$,
  to obtain the eigenmultiplets of the full system at the NFL fixed
  point. In the ``double fusion'' column (right), the single-fusion
  results are fused with the conjugate impurity
  representation,
  $\Qbar_\imp = (0,1, \protect\raisebox{0.7mm}{\tiny
    $\protect\yng(1)$}\,)$. 
  Each of the resulting multiplets is associated with a boundary
  operator having the same quantum numbers. Colors relate the
  multiplets obtained after single fusion to the corresponding lines
  in Fig.~\ref{fig:spectra_susceptibilities_K0}. }
\label{fig:fusion_scheme}
\end{figure*}

\subsection{Non-Fermi-liquid fixed point}
\label{sec:supp:fusion_SU(3)} 
  
\step2 
We now focus on the NFL fixed point of the \threesoK\ model, at
$\bc_\nfl^\ast$, where $ (J_0,K_0,I_0) = (0,1,0)$.  According to AL's 
general strategy, the orbital isospin $T$ can be then ``absorbed'' by
the bath through the substitution
\begin{align}
\bJ_{\orb,n} \mapsto \bcJ_{\orb,n} = \bJ_{\orb,n} + \bT \,.
\label{eq:supp:absorb_T}
\end{align}
Here $\bJ_{\orb,n}$ and $\bcJ_{\orb,n}$
are Fourier components ($n$ being a Fourier index) of the bare and bulk orbital isospin currents,
respectively, defined for a bath in a finite-sized box.  (The local bath operator
$\bJ_\orb$ is proportional to $\sum_{n\in \mathbbm{Z}} \bJ_{\orb,n}$.)
The right-hand side of Eq.~\eqref{eq:supp:absorb_T} is reminiscent of 
 the addition of Lie algebra generators, 
$\bS' = \bS + \tilde \bS$, when performing a direct product 
decomposition, $S \otimes \tilde S = \sum_\oplus S'$, 
of \sutwo\ multiplets.  The terms added in Eq.~\eqref{eq:supp:absorb_T}, 
 however, generate two \textit{different} algebras: 
$\bJ_{\orb,n}$  are generators of the $\suthree_2$
KM algebra, $\bT$ of the \suthree\ Lie algebra.  AL proposed a remarkable
fusion hypothesis for dealing with such situations
(and confirmed its veracity by detailed comparisons to
 Bethe ansatz and NRG computations). For 
the present context their fusion hypothesis states: the
eigenstates of the combined bath+impurity system can be obtained by
combining (or ``fusing'') their orbital degrees of freedom, 
$\lambda \otimes \lambda_\imp = \sum_\oplus \lambda'$, using 
the fusion rules of the $\suthree_2$
KM algebra, as though the impurity's orbital multiplet were an $\suthree_2$,
not $\suthree$,
multiplet. The  $\suthree_2$ fusion rules 
 are depicted in Table \ref{tab:supp:fusion_SU(3)}
of the Supplemental Material \cite{suppmat}.

\onecolumngrid 

\renewcommand{\tabcolsep}{2.5pt}
\begin{table*}[ph!]
\centering 
\caption{Fusion table for orbital screening at the NFL fixed point $\bc_\nfl^\ast$ of the \threesoK\ model.
  Left: The 14 lowest low-lying free-fermion multiplets $(q,S,\lambda)$, with multiplet dimensions $d$ and energies $E(q,S,\lambda)$, 
  computed using Eqs.~\eqref{eq:supp:E(qSlambda)-sup} and Table \ref{tab:supp:E_SU(2)_SU(3)} of the SM \cite{suppmat}.
  Center: Single fusion with a $Q_\mathrm{imp}=(0,1, \protect\raisebox{0.6mm}{\tiny $\protect\yng(1,1)$}\,)$ impurity,
  using \sutwo\ fusion rules in the spin sector and 
  $\suthree_2$ fusion rules (listed in 
  Table~\ref{tab:supp:fusion_SU(3)} of the SM \cite{suppmat}) in the orbital sector. This yields multiplets  $(q, S^\prime, \lambda^\prime)$, with dimensions $d^\prime$, 
  energies $E^\prime = E(q,S,\lambda^\prime)$, and excitation
  energies $\delta E^\prime = E^\prime - E^\prime_\mathrm{min}$. These
  are compared to the  values, $E_\NRG$, computed by NRG  for 
  $(J_0,K_0,I_0) = (0,0.3,0)$. The NRG energies have been shifted and rescaled such that 
  the lowest energy is zero and the second-lowest values for $E_\NRG$ and $\delta E'$ 
  match. The single-fusion and NRG spectra agree well (deviations $\lesssim 10\%$).
  Right: Double fusion, which fuses multiplets from the middle column with an impurity in the conjugate
  representation $\Qbar_\imp = (0,1, 
  \protect\raisebox{0.7mm}{\tiny $\protect\yng(1)$}\,)$, yields the quantum numbers $(q, S^{\prime\prime}, \lambda^{\prime\prime})$. These characterize the CFT boundary operators $\hat{O}$, with scaling dimensions $\Delta=E(q,S,\lambda^{\prime\prime})$. 
}
\medskip
\renewcommand{\arraystretch}{1.7} 
\begin{tabular}{ccc@{\hspace{8mm}}cc@{\hspace{3mm}}|@{\hspace{2mm}}ccccc@{\hspace{8mm}}c@{\hspace{4mm}}c@{\hspace{4mm}}c|c|@{\hspace{2mm}}ccccc@{\hspace{8mm}}cc}
\hline \hline
\multicolumn{5}{c|@{\hspace{2mm}}}{Free fermions} 
& \multicolumn{8}{c|}{Single fusion, with $Q_\imp = (0, 1, {\tiny \yng(1,1)}\, )$} & NRG
& \multicolumn{7}{c}{Double fusion, with $\Qbar_\imp = (0,1, 
  \protect\raisebox{0.7mm}{\tiny $\protect\yng(1)$}\,)$}
\\ 
$q$ & $S$ & $\lambda$ & $d$ & $E$ 
& $q$ & & $S^\prime$ & & $\lambda^\prime$ & $d^\prime$ 
& $E^\prime$ & $\delta E^\prime$ & $E_\NRG$
& $\phantom{-}q$ & & $S^{\prime\prime}$ & & $\lambda^{\prime\prime}$ & $\Delta$ & $\hat{O}$
\\ \hhline{---|--|---------|-------}

\msix{\phantom{-}0} & \msix{\phantom{-}0} & \;\;\msix{$\bullet$}\;\; 
& \msix{1\phantom{4}} & \msix0
& \msix{\phantom{-}0} & \phantom{\lb} & \msix1 & \phantom{\lb} & \; \msix{\raisebox{0.5mm}{\y01}} \; & \msix{\phantom{4}9}
& \msix{\f415} & \msix{\f130 (0.033)} 
& \msix{0.033}
& \msix{\phantom{-}0} & \lbsix & \m0 & \lb & $\bullet$ & 0 & $\mathbbm{1}$
\\
& & & & 
& & & & & & 
& & &
& & & & & \raisebox{0.5mm}{\y11} & \fr35 & $\bPhi_\orb$
\\
& & & & 
& & & & & & 
& & &
& & & \m1 & \lb & \y00 & 0 & 
\\
& & & & 
& & & & & & 
& & &
& & & & & \raisebox{0.5mm}{\y11} & \fr35 & 
\\
& & & & 
& & & & & & 
& & &
& & & \m2 & \lb & \y00 & 0 & 
\\
& & & & 
& & & & & & 
& & &
& & & & & \raisebox{0.5mm}{\y11} & \fr35 & 
\\ 
\hhline{---|--|---------|-------}

\end{tabular}
\label{tab:supp:fusion_NRG}
\end{table*}

\clearpage

\renewcommand{\tabcolsep}{2.5pt}
\begin{table*}[ph!]
\centering
\renewcommand{\arraystretch}{1.5}
\begin{tabular}{ccc@{\hspace{8mm}}cc@{\hspace{3mm}}|@{\hspace{2mm}}ccccc@{\hspace{8mm}}c@{\hspace{4mm}}c@{\hspace{4mm}}c|c|@{\hspace{2mm}}ccccc@{\hspace{8mm}}cc}
\hline
\multicolumn{5}{c|@{\hspace{2mm}}}{Free fermions} 
& \multicolumn{8}{c|}{Single fusion, with $Q_\imp = (0, 1, {\tiny \yng(1,1)}\, )$} & NRG
& \multicolumn{7}{c}{Double fusion, with $\Qbar_\imp = (0,1, 
  \protect\raisebox{0.7mm}{\tiny $\protect\yng(1)$}\,)$}
\\ 
$q$ & $S$ & $\lambda$ & $d$ & $E$ 
& $q$ & & $S^\prime$ & & $\lambda^\prime$ & $d^\prime $ 
& $E^\prime$ & $\delta E^\prime$ & $E_\NRG$
& $q$ & & $S^{\prime\prime}$ & & $\lambda^{\prime\prime}$ & $\Delta$ & $\hat{O}$
\\ \hline 

\meight{+1} & \meight{\fr12} & \meight{\raisebox{0.5mm}{\y10}} & \meight{6} & \meight{\fr12}
& \meight{+1} & \lbeight & \msix{\fr12} & \lbsix & \m{\y00} & \m2
& \m{\f730} & \m0 & \m0
& \m{+1} & \lb & \fr12 & & \y10 & \fr12 &
\\
& & & & 
& & & & & &
& & &
& & & \fr32 & & \y10 & \fr12 &
\\
& & & & 
& & & & & \mfour{\y11} & \mfour{16}
& \mfour{\fr56} & \mfour{\fr35 (0.6)} & \mfour{0.64}
& \mfour{+1} & \lbfour & \m{\fr12} & \lb & \y10 & \fr12 &
\\
& & & & 
& & & & & & 
& & &
& & & & & \y02 & \f910 &
\\
& & & & 
& & & & & &
& & &
& & & \m{\fr32} & \lb & \y10 & \fr12 &
\\
& & & & 
& & & & & & 
& & &
& & & & & \y02 & \f910 &
\\
& & & & 
& & & \m{\fr32} & \lb & \y00 & 4
& \f730 & 0 & 0
& \multicolumn{5}{c}{\m{as above}} 
& &
\\
& & & & 
& & & & & \raisebox{0.5mm}{\y11} & 32
& \fr56 & \fr35 (0.6) & 0.64
& & & & & & &
\\ \hhline{---|--|---------|-------}

\meight{$-1$} & \meight{\fr12} & \meight{\raisebox{0.5mm}{\y01}} & \meight{6} & \meight{\fr12}
& \meight{$-1$} & \lbeight & \msix{\fr12} & \lbsix & \mfour{\y10} & \mfour6
& \mfour{\fr12} & \mfour{\f415 (0.27)} & \mfour{0.28}
& \mfour{$-1$} & \lbfour & \m{\fr12} & \lb & \y01 & \fr12 &
\\
& & & & 
& & & & & &
& & &
& & & & & \y20 & \f910 & $\tPsi_\orb$
\\
& & & & 
& & & & & & 
& & &
& & & \m{\fr32} & \lb & \y01 & \fr12 &
\\
& & & & 
& & & & & & 
& & &
& & & & & \y20 & \f910 &
\\
& & & & 
& & & & & \m{\y02} & \m{12}
& \m{\f910} & \m{\fr23 (0.67)} & \m{0.70}
& \m{$-1$} & \lb & \fr12 & & \y01 & \fr12 &
\\
& & & & 
& & & & & &
& & &
& & & \fr32 & & \y01 & \fr12 &
\\
& & & & 
& & & \m{\fr32} & \lb & \y10 & 12
& \fr12 & \f415 (0.27) & 0.28
& \multicolumn{5}{c}
{\m{as above}} 
& &
\\
& & & & 
& & & & & \raisebox{0.5mm}{\y02} & 24
& \f910 & \fr23 (0.67) & 0.70
& & & & & & & 
\\ 
\hhline{---|--|---------|-------}

\mseven0 & \mseven1 & \mseven{\raisebox{0.5mm}{\y11}} & \mseven{24} & \mseven1
& \mseven{0} & \lbseven & \mt0 & \lbt & \m{\y01} & \m3
& \m{\fr23} & \m{\fr{13}{30} (0.43)} & \m{0.46}
& \m0 & & \m1 & \lb & \y00 & \fr25 & $\bPhi_\spin$
\\
& & & & 
& & & & & &
& & &
& & & & & \y11 & 1 &
\\
& & & & 
& & & & & \y20 & 6
& \fr{16}{15} & \fr56 (0.83) & 0.88
& 0 & & 1 & & \y11 & 1 &
\\
& & & & 
& & & \m1 & \lb & \y01 & 9
& \fr23 & \fr{13}{30} (0.43) & 0.46
& \multicolumn{5}{c}
{\multirow{4}{3cm}{\centering as above, with $S^{\prime\prime} \in \{0,1,2 \}$}} & &
\\
& & & & 
& & & & & \y20 & 18
& \fr{16}{15} & \fr56 (0.83) & 0.88
& & & & & & &
\\
& & & & 
& & & \m2 & \lb & \y01 & 15
& \fr23 & \fr{13}{30} (0.43) & 0.46
& & & & & & & 
\\
& & & & 
& & & & & \y20 & 30
& \fr{16}{15} & \fr56 (0.83) & 0.88
& & & & & & & 
\\ \hline

\mfour{+2} & \mfour0 & 
\mfour{\raisebox{0.5mm}{{\tiny$\yng(2)$}}} 
& \mfour6 & \mfour1
& \mfour{+2} & & \mfour1 & & \mfour{\raisebox{0.5mm}{\y10}} & \mfour9
& \mfour{\fr35} & \mfour{\fr{11}{30} (0.37)} & \mfour{0.39}
& \mfour{+2} & \lbfour & \m0 & \lb & \y01 & \fr35 &
\\
& & & & 
& & & & & &
& & &
& & & & & \y20 & 1 &
\\
& & & & 
& & & & & & 
& & &
& & & \multicolumn{3}{c}
{\multirow{2}{3cm}{\centering as above, with $S^{\prime\prime} \in \{1,2 \}$}} & &
\\
& & & & 
& & & & & & 
& & &
& & & & & & & 
\\ \hline

\mfour{$-2$} & \mfour{0} & \mfour{\raisebox{1mm}{\y02}} & \mfour{6} & \mfour1
& \mfour{$-2$} & & \mfour1 & & \mfour{\raisebox{0.5mm}{\y11}} & \mfour{24}
& \mfour{\fr{14}{15}} & \mfour{\f710 (0.7)} & \mfour{0.74}
& \mfour{$-2$} & \lbfour & \m0 & \lb & \y10 & \fr35 &
\\
& & & & 
& & & & & & 
& & &
& & & & & \y02 & 1 &
\\
& & & & 
& & & & & & 
& & &
& & & \multicolumn{3}{c}
{\multirow{2}{3cm}{\centering as above, with $S^{\prime\prime} \in \{1,2 \}$}} & &
\\
& & & & 
& & & & & & 
& & &
& & & & & & & 
\\ \hline

\mseven{+2} & \mseven1 & \mseven{\raisebox{0.5mm}{\y01}} & \mseven9 & \mseven1
& \mseven{+2} & \lbseven & \mt0 & \lbt & \m{\raisebox{0.5mm}{\y10}} & \m3
& \m1 & \m{\fr{23}{30} (0.77)} & \m{0.82}
& \m{+2} & & \m1 & \lb & \y01 & 1 & 
\\
& & & & 
& & & & & & 
& & &
& & & & & \y20 & \fr75 &
\\
& & & & 
& & & & & \raisebox{0.5mm}{\y02} & 6
& \fr75 & \fr76 (1.17) & 1.24
& +2 & & 1 & & \y01 & 1 &
\\
& & & & 
& & & \m1 & \lb & \raisebox{0.5mm}{\y10} & 9
& 1 & \fr{23}{30} (0.77) & 0.82
& \multicolumn{5}{c}
{\multirow{4}{3cm}{\centering as above, with $S^{\prime\prime} \in \{0,1,2 \}$}} & &
\\
& & & & 
& & & &  & \raisebox{0.5mm}{\y02} & 18
& \fr75 & \fr76 (1.17) & 1.24
& & & & & & &
\\
& & & & 
& & & \m2 & \lb & \raisebox{0.5mm}{\y10} & 15
& 1 & \fr{23}{30} (0.77) & 0.82
& & & & & & & 
\\
& & & & 
& & & & & \raisebox{0.5mm}{\y02} & 30
& \fr75 & \fr76 (1.17) & 1.24
& & & & & & &
\\ \hline

\end{tabular}
\end{table*}

\clearpage

\renewcommand{\tabcolsep}{2.5pt}
\begin{table*}[h]
\centering
\renewcommand{\arraystretch}{1.5}
\begin{tabular}{ccc@{\hspace{8mm}}cc@{\hspace{3mm}}|@{\hspace{2mm}}ccccc@{\hspace{8mm}}c@{\hspace{4mm}}c@{\hspace{4mm}}c|c|@{\hspace{2mm}}ccccc@{\hspace{8mm}}cc}
\hline
\multicolumn{5}{c|@{\hspace{2mm}}}{Free fermions} 
& \multicolumn{8}{c|}{Single fusion, with $Q_\imp = (0, 1, {\tiny \yng(1,1)}\, )$} & NRG
& \multicolumn{7}{c}{Double fusion, with $\Qbar_\imp = (0,1, 
  \protect\raisebox{0.7mm}{\tiny $\protect\yng(1)$}\,)$}
\\ 
$q$ & $S$ & $\lambda$ & $d$ & $E$ 
& $q$ & & $S^\prime$ & & $\lambda^\prime$ & $d^\prime $ 
& $E^\prime$ & $\delta E^\prime$ & $E_\NRG$
& $q$ & & $S^{\prime\prime}$ & & $\lambda^{\prime\prime}$ & $\Delta$ & $\hat{O}$
\\ \hline

\mseven{$-2$} & \mseven1 & \mseven{\raisebox{0.8mm}{\y10}} & \mseven9 & \mseven1
& \mseven{$-2$} & \lbseven & \mt0 & \lbt & \y00 & 1
& \fr{11}{15} & \fr12 (0.5) & 0.52
& $-2$ & & 1 & & \y10 & 1 &
\\
& & & & 
& & & & & \m{\raisebox{0.5mm}{\y11}} & \m8
& \m{\fr43} & \m{\fr{11}{10} (1.1)} & \m{1.16}
& \m{$-2$} & & \m1 & \lb & \y10 & 1 &
\\
& & & & 
& & & & & & 
& & & 
& & & & & \y02 & \fr75 &
\\
& & & & 
& & & \m1 & \lb & \y00 & 3
& \fr{11}{15} & \fr12 (0.5) & 0.52
& \multicolumn{5}{c}
{\multirow{4}{3cm}{\centering as above, with $S^{\prime\prime} \in \{0,1,2 \}$}} & &
\\
& & & & 
& & & & & \raisebox{0.5mm}{\y11} & 24
& \fr43 & \fr{11}{10} (1.1) & 1.16
& & & & & & &
\\
& & & & 
& & & \m2 & \lb & \y00 & 5
& \fr{11}{15} & \fr12 (0.5) & 0.52
& & & & & & & 
\\
& & & & 
& & & & & \raisebox{0.5mm}{\y11} & 40
& \fr43 & \fr{11}{10} (1.1) & 1.16
& & & & & & &
\\ \hline

\msix{+1} & \msix{\fr32} & \msix{\raisebox{1mm}{\y02}} & \msix{24} & \msix{\fr32}
& \msix{+1} & \lbsix & \mfour{\fr12} & & \mfour{\y11} & \mfour{16}
& \mfour{\fr{43}{30}} & \mfour{\fr65 (1.2)} & \mfour{1.28}
& \mfour{+1} & \lbfour & \m{\fr12} & \lb & \y10 & \fr{11}{10} &
\\
& & & & 
& & & & & & 
& & & 
& & & & & \y02 & \fr32 & 
\\
& & & & 
& & & & & & 
& & & 
& & & \m{\fr32} & \lb & \y10 & \fr{11}{10} &
\\
& & & & 
& & & & & & 
& & & 
& & & & & \y02 & \fr32 & 
\\ 
& & & & 
& & & \fr32 & & \raisebox{0.5mm}{\y11} & 32
& \fr{43}{30} & \fr65 (1.2) & 1.28
& \multicolumn{5}{c}
{\m{as above}} & &
\\ 
& & & & 
& & & \fr52 & & \raisebox{0.5mm}{\y11} & 48
& \fr{43}{30} & \fr65 (1.2) & 1.28
& & & & & & & 
\\ \hline

\msix{$-1$} & \msix{\fr32} & 
\msix{\raisebox{0.5mm}{\tiny$\yng(2)$}} & \msix{24} & \msix{\fr32}
& \msix{$-1$} & \lbsix & \mfour{\fr12} & & \mfour{\y10} & \mfour6
& \mfour{\fr{11}{10}} & \mfour{\fr{13}{15} (0.87)} & \mfour{0.92}
& \mfour{$-1$} & \lbfour & \m{\fr12} & \lb & \y01 & \fr{11}{10} &
\\
& & & & 
& & & & & & 
& & & 
& & & & & \y20 & \fr32 & 
\\
& & & & 
& & & & & &
& & & 
& & & \m{\fr32} & \lb & \y01 & \fr{11}{10} &
\\
& & & & 
& & & & & & 
& & & 
& & & & & \y20 & \fr32 & 
\\ 
& & & & 
& & & \fr32 & & \y10 & 12
& \fr{11}{10} & \fr{13}{15} (0.87) & 0.92
& \multicolumn{5}{c}
{\m{as above}} & &
\\ 
& & & & 
& & & \fr52 & & \y10 & 18
& \fr{11}{10} & \fr{13}{15} (0.87) & 0.92
& & & & & & & 
\\ \hline

\meight{$\pm3$} & \meight{\fr12} & \meight{\raisebox{0.5mm}{\y11}} & \meight{16} & \meight{\fr32}
& \meight{$\pm3$} & \lbeight & \msix{\fr12} & \lbsix & \mfour{\y01} & \mfour6
& \mfour{\fr76} & \mfour{\fr{14}{15} (0.93)} & \mfour{0.98}
& \mfour{$\pm3$} & \lbfour & \m{\fr12} & \lb & \y00 & \f910 & $\tPsi_\spin$
\\
& & & & 
& & & & & &
& & & 
& & & & & \y11 & \fr32 & 
\\
& & & & 
& & & & & & 
& & & 
& & & \m{\fr32} & \lb & \y00 & \f910 & 
\\
& & & & 
& & & & & & 
& & & 
& & & & & \y11 & \fr32 & 
\\
& & & & 
& & & & & \y20 & 12
& \fr{47}{30} & \fr43 (1.33) & 1.41
& \m{$\pm3$} & \lb & \fr12 & & \y11 & \fr32 & 
\\
& & & & 
& & & & & &
& & & 
& & & \fr32 & & \y11 & \fr32 & 
\\
& & & & 
& & & \m{\fr32} & \lb & \y01 & 12
& \fr76 & \fr{14}{15} (0.93) & 0.98
& \multicolumn{5}{c}
{\m{as above}} & &
\\
& & & & 
& & & & & \y20 & 24
& \fr{47}{30} & \fr43 (1.33) & 1.41
& & & & & & & 
\\ \hline

\msix{$\pm3$} & \msix{\fr32} & \msix{\y00} & \msix4 & \msix{\fr32}
& \msix{$\pm3$} & \lbsix & \mfour{\fr12} & & \mfour{\y01} & \mfour6
& \mfour{\fr{53}{30}} & \mfour{\fr{23}{15} (1.53)} & \mfour{1.63}
& \mfour{$\pm3$} & \lbfour & \m{\fr12} & \lb & \y00 & \fr32 & 
\\
& & & & 
& & & & & &
& & & 
& & & & & \y11 & \fr{21}{10} & 
\\
& & & & 
& & & & & &
& & & 
& & & \m{\fr32} & \lb & \y00 & \fr32 & 
\\
& & & & 
& & & & & & 
& & & 
& & & & & \y11 & \fr{21}{10} & 
\\ 
& & & & 
& & & \fr32 & & \raisebox{0.5mm}{\y01} & 12
& \fr{53}{30} & \fr{23}{15} (1.53) & 1.63
& \multicolumn{5}{c}
{\m{as above}} & &
\\ 
& & & & 
& & & \fr52 & & \raisebox{0.5mm}{\y01} & 18
& \fr{53}{30} & \fr{23}{15} (1.53) & 1.63
& & & & & & & 
\\ \hline \hline 

\end{tabular}
\label{tab:supp:fusion_NRG_3}
\end{table*}

\twocolumngrid

Having discussed orbital fusion, we now turn to the spin sector---how
should the impurity's spectator spin be dealt with? This question goes
beyond the scope of AL's work, who did not consider impurities with
spectator degrees of freedom. We have explored several spin fusion
strategies and concluded that the following one yields spectra
consistent with NRG: In parallel to orbital fusion, the bath and impurity
spin degrees should be combined too, as
$S\otimes S_\imp = \sum_\oplus S'$, but using the fusion rules of the
$\sutwo$ Lie algebra, not the $\sutwo_3$ KM algebra.  Heuristically,
the difference---KM versus Lie---between the algebras governing
orbital and spin fusion reflects the fact that the bath and impurity
are \textit{coupled} in the orbital sector, where the bath ``absorbs''
the impurity orbital isospin, but \textit{decoupled} in the spin
sector, where the impurity spin remains a spectator.

The fusion of bath and impurity degrees of freedom, called single
fusion by AL, is illustrated schematically in the left-hand part of
Fig.~\ref{fig:fusion_scheme} for four selected multiplets. Table
\ref{tab:supp:fusion_NRG}  gives a comprehensive list of low-lying
multiplets obtained in this manner.  On the left it enumerates the 14
lowest-lying multiplets $(q,S, \lambda)$ of the free bath, with
dimensions $d$ and energies $E(q,S,\lambda)$.  Fusing these with a
$Q_\imp = (0,1, {\tiny \protect\yng(1,1)}\,)$ impurity yields the
 multiplets, $(q, S^\prime, \lambda^\prime)$, 
 listed in the center. Their energies are given by
$E^\prime = E(q,S,\lambda^\prime)$, not $E(q,S',\lambda^\prime)$,
since at the NFL fixed point, where $J_0=I_0 = 0$, the impurity spin
is decoupled from the bath.

The single-fusion excitation energies,
$\delta E^\prime = E^\prime - E^\prime_\mathrm{min}$, relative to the
lowest-lying multiplet ($E^\prime_\mathrm{min} = 7/30$) are in good
agreement (deviations $\lesssim 10\%$) with the values, $E_\NRG$, found by
NRG (for $K_0=0.3$, $J_0=I_0=0$) for multiplets with corresponding
quantum numbers.  The agreement improves upon decreasing the NRG
discretization parameter $\Lambda$ (here $\Lambda=2.5$ was used).
This remarkable agreement between CFT predictions and NRG confirms
the applicability of the $\sutwo\otimes \suthree_2$ 
fusion hypothesis proposed above.  

\step3
As mentioned in Sec.~\ref{sec:CFT_synopsis}, the fixed point 
$\bc_\nfl^\ast$ is characterized by a set of local operators,
called boundary operators by AL (since they live at the
impurity site, i.e., at the boundary of the two-dimensional
 space-time on which the CFT is defined). These
can be obtained by a second fusion step, called double fusion
by AL: the multiplets $(q, S^{\prime}, \lambda^{\prime})$ 
obtained from single fusion are fused
with the conjugate impurity representation, 
$\Qbar_\imp = (0,1, 
 \protect\raisebox{0.7mm}{\tiny $\protect\yng(1)$}\,)$,
to obtain another set of multiplets, $(q, S^{\prime\prime}, \lambda^{\prime\prime})$,
listed on the right-hand side of Fig.~\ref{fig:fusion_scheme}
 and Table~\ref{tab:supp:fusion_NRG}. 
 (The \textit{conjugate} impurity representation has to be used for double fusion to ensure
that the set of boundary operators contains the identity operator, $\bar \lambda \otimes \lambda = \mathbbm{1}$.)
 Each such multiplet is associated with 
 a  boundary operator $\hat{O}$ with the same quantum numbers
 and scaling dimension 
$\Delta = E^{\prime\prime} = 
 E(q,S,\lambda^{\prime\prime})$. The operators
called $\bPhi_\orb$ and $\bPhi_\spin$ are the 
leading boundary operators (with smallest scaling dimension)
 in the orbital and spin sectors, 
  respectively. They determine the behavior of the
orbital and spin susceptibilities in the NFL regime
(see Sec.~\ref{sec:CFT-susceptibilities}). 
In the spin-splitting regime, their role
is taken by the operators $\tPsi_\orb$ and $\tPsi_\spin$, respectively,
as discussed in Sec.~\ref{sec:CFT_synopsis}.

\subsection{Scaling behavior of the susceptibilities}
\label{sec:CFT-susceptibilities}

In this section, we compute the leading frequency dependence of the
dynamical spin and orbital susceptibilities. We begin with the NFL
regime, where we directly follow the strategy 
used by AL in Sec.~3.3
of Ref.~\cite{Affleck1991} and show how it reproduces the results presented
in Sec.~\ref{sec:CFT_synopsis}. Thereafter we discuss the SS regime, which has no
analog in AL's work, using somewhat more heuristic arguments.

\subsubsection{NFL regime}
\label{sec:susc-nfl-regime}

At the NFL fixed point, the impurity's orbital isospin $\bT$
 has been fully absorbed into the bath orbital current
$\bcJ_\orb$ [cf.\ Eq.~\eqref{eq:supp:absorb_T}]. From this perspective, 
the impurity orbital susceptibility $\chi_\orb^\imp$
is governed by the leading local perturbation
of the bulk orbital susceptibility, $\chi_\orb^\mathrm{bulk}
\sim \langle \bcJ^\bulk_\orb || \bcJ^\bulk_\orb \rangle_\omega$, where 
${\bcJ}^{\bulk}_\orb (t) = \int_{-\infty}^\infty \md x 
\bcJ_\orb (t,x) \sim \bcJ_{\orb,n=0}$ is the bulk orbital current.
The  leading local perturbations 
are  those combinations of  boundary operators 
(found via double fusion; see Table~\ref{tab:supp:fusion_NRG})
having the smallest scaling dimensions and the same
symmetry as the bare Hamiltonian \cite{Affleck1991,Affleck1991a,Affleck1993}.

In the orbital sector, the leading boundary operator is $\bPhi_\orb$,
with quantum numbers $(0,0,\raisebox{0mm}{\tiny \y11}\,)$ and scaling
dimension $\Delta_\orb = \frac{3}{5}$ (cf.\
Tables~\ref{tab:single_fusion_lowest} and
\ref{tab:supp:fusion_NRG}). The orbital current $\bcJ_\orb$ has the
same quantum numbers.  Its first descendant $\bcJ_{\orb, -1}$ can be
combined with $\bPhi_\orb$ to obtain an orbital \suthree\ singlet
boundary operator, $H_\orb^\prime = \bcJ_{\orb,-1} \cdot \bPhi_\orb$,
with scaling dimension $1+\Delta_\orb = 1+\frac35$. This is the
leading irrelevant (dimension $>1$) boundary perturbation to the
fixed-point Hamiltonian in the orbital sector. Its contribution to the
impurity orbital susceptibility,
$\chi_\orb^\imp \sim \chi_\orb^\bulk$, evaluated perturbatively to
second order, is
\begin{align}
\chi^\imp_\orb(\omega) 
&\sim 
\int_{-\infty}^{\infty} \md t_1
\int_{-\infty}^{\infty}  \md t_2
\int_{-\infty}^{\infty}  \md t_3
\, e^{i\omega t_1}
\nonumber 
\\ & 
\qquad 
\bigl\langle
\bcJ_\orb^\bulk (t_1) \cdot \bcJ_\orb^\bulk (0) 
H'_\orb (t_2)  H'_\orb (t_3) 
\bigr\rangle
\nonumber
\\
\label{eq:chi_orb_imp_supp}
 & \sim \omega^{2\Delta_\orb-1}
=  \, \omega^{1/5} \,.
\end{align}
The last line follows by power counting ($\bcJ_\orb^\bulk$ has dimension 0,
each time integral dimension $-1$). 

The local bath site coupled to the impurity will show the same
behavior, $\chi_\orb^\bath \sim \omega^{1/5}$, since the orbital
exchange interaction strongly couples its 
orbital isospin $\bJ_\orb$ to $\bT$---indeed,
$\bcJ_\orb$ is constructed from a linear
combination of both these operators [cf.\ Eq.~\eqref{eq:supp:absorb_T}].

The above results can be obtained in a more direct way by positing
that at the NFL fixed point, orbital screening causes both
$\bT$ and $\bJ_\orb$ to be renormalized into the same boundary
operator, $\bPhi_\orb$. We then obtain 
\begin{align}
\label{eq:chi_orb_direct}
\chi_\orb^\imp (\omega) \sim \chi_\orb^\bath (\omega) \sim 
\langle \bPhi_\orb ||\bPhi_\orb \rangle_\omega
\sim \omega^{2\Delta_\orb -1} , 
\end{align}
reproducing Eq.~\eqref{eq:chi_orb_imp_supp}.
This is the argument presented in Sec.~\ref{sec:CFT_synopsis}.

We next turn to the spin sector. Exactly \textit{at} the NFL fixed point,
where $J_0=I_0=0$, the impurity spin $\bS$ is decoupled from the bath.
\textit{At} $\bc_\nfl^\ast$ it hence has no dynamics, scaling
dimension 0, and a trivial spin susceptibility,
$\chi_\spin^\imp (\omega) \sim \delta(\omega)$.  By contrast, $\chi_\spin^\bath$, 
the susceptibility  of $\bJ_\spin$, the local bath spin
coupled to the impurity,  does show nontrivial dynamics at the fixed
point.  The reason is that some of the boundary operators induced by
orbital screening actually live in the spin sector (a highly nontrivial
consequence of non-Abelian bosonization and orbital fusion).  The
leading boundary operator in the spin sector is $\bPhi_\spin$, with
quantum numbers $(0,1,\bullet)$ and scaling dimension
$\Delta_\spin = \frac{2}{5}$ (cf.\
Tables~\ref{tab:single_fusion_lowest} and
\ref{tab:supp:fusion_NRG}). It can be combined with the first
descendant of the (bare, unshifted) spin current to obtain a spin
\sutwo\ singlet boundary operator,
$H_\spin^\prime = \bJ_{\spin,-1} \cdot \bPhi_\spin$, with scaling
dimension $1+\Delta_\spin = 1+\frac{2}{5}$. This is the leading irrelevant
boundary perturbation to the fixed-point Hamiltonian in the spin
sector. Its contribution to the local bath spin susceptibility,
$\chi_\spin^\bath \sim \chi_\spin^\mathrm{bulk}$, evaluated to second order, is
\begin{align}
\chi^\bath_\spin(\omega) 
&\sim 
\int_{-\infty}^{\infty} \md t_1
\int_{-\infty}^{\infty}  \md t_2
\int_{-\infty}^{\infty}  \md t_3
\, e^{i\omega t_1}
\nonumber 
\\ & 
\qquad 
\bigl\langle
\bJ_\spin^\bulk (t_1) \cdot \bJ_\spin^\bulk (0) 
H'_\spin (t_2)  H'_\spin (t_3) 
\bigr\rangle
\nonumber
\\
\label{eq:chi_spin_bath_supp}
 & \sim \omega^{2\Delta_\spin-1}
=  \, \omega^{-1/5} \,.
\end{align}
This result, too, can be obtained more directly, by positing that
$\bJ_\spin$ is renormalized to $\bPhi_\spin$, with 
\begin{align}
\label{eq:chi_sp_direct}
 \chi_\spin^\bath (\omega) \sim 
\langle \bPhi_\spin ||\bPhi_\spin \rangle_\omega
\sim \omega^{2\Delta_\spin -1} ,
\end{align}
as argued in Sec.~\ref{sec:CFT_synopsis}.

If the system is tuned very slightly away from the NFL fixed point,
$J_0 \ll 1$, $I_0 = 0$, the impurity spin does acquire nontrivial
dynamics, due to the action of the spin exchange interaction,
$J_0 \bS \cdot \bJ_\spin$. According to the above argument,
orbital screening renormalizes it to $J_0 \bS\cdot \bPhi_\spin$. 
Its second-order contribution to the impurity spin susceptibility
is 
\begin{align}
\chi^\imp_\spin(\omega) 
&\sim 
\int_{-\infty}^{\infty} \md t_1
\int_{-\infty}^{\infty}  \md t_2
\int_{-\infty}^{\infty}  \md t_3
\, e^{i\omega t_1}
\nonumber 
\\ & 
\qquad 
\bigl\langle
\bS (t_1) \cdot \bS (0) 
(\bS\cdot \bPhi_\spin) (t_2)  (\bS\cdot \bPhi_\spin) (t_3) 
\bigr\rangle
\nonumber
\\
\label{eq:chi_spin_imp_supp}
 & \sim \omega^{2\Delta_\spin-3}
=  \, \omega^{-11/5} \,.
\end{align}
The occurrence of such a large, negative exponent for the spin susceptibility
is very unusual. It reflects the fact that near (but not at) the NFL fixed
point the impurity spin is almost (but not fully) decoupled from the 
bath, and hence able to ``sense'' that orbital screening modifies
the bath spin current in a nontrivial manner.

\begin{figure*}[bth!]
\includegraphics[width=\textwidth]{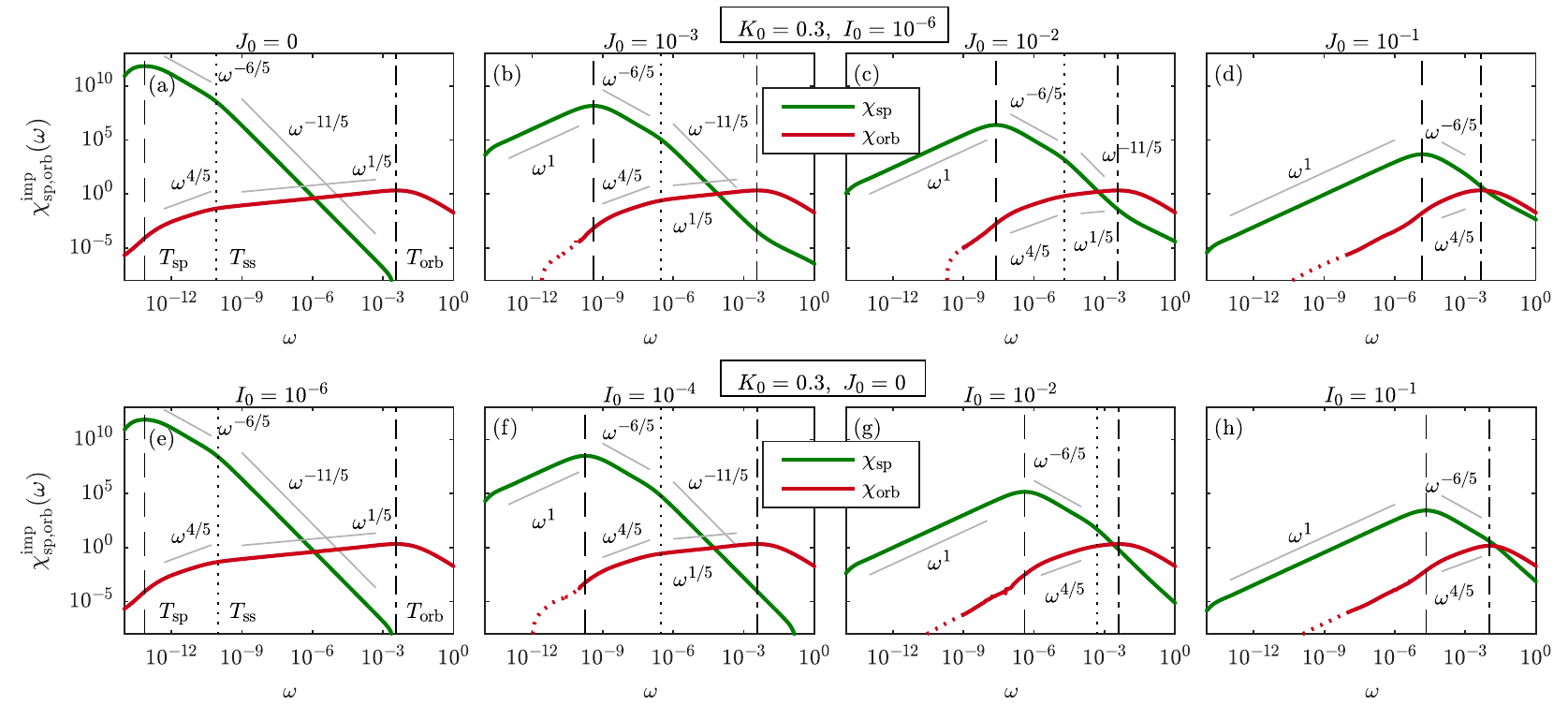}
\caption{Imaginary part of the zero-temperature impurity spin and orbital
  susceptibilities for the \threesoK\ model. 
We keep $K_0 = 0.3$ fixed throughout, and (a)-(d) 
vary $J_0$ at fixed $I_0 = 10^{-6}$,
or (e)-(h) vary $I_0$ at fixed $J_0=0$.
(a)-(d) As $J_0$ is increased from 
0 (left) to $10^{-1}$ (right), the width of the NFL regime
$[\Tss, \Torb]$ shrinks, while that of the SS regime
$[\Tspin,\Tss]$ remains constant. (e)-(h) We find the same
behavior when increasing $I_0$ with $K_0$ and $J_0$ kept fixed.}
\label{fig:SSwidth}
\end{figure*}

\subsubsection{Spin-slitting regime}
\label{sec:SS-regime}

The renormalized exchange interaction $J_0 \bS \cdot \bPhi_\spin$ is a
relevant perturbation, with scaling dimension $\frac{2}{5} < 1$.  It
grows under the RG flow, eventually driving the system away from the
NFL fixed point and into a crossover regime, 
$\Tspin < \omega < \Tss$, called the spin-splitting
regime in Sec.~\ref{sec:finite_size_spectra}. In the NRG flow diagram of
Fig.~\ref{fig:spectra_susceptibilities_K0}(a), this regime 
is characterized by level crossings, extending over several 
orders of magnitude in energy, rather than a 
stationary level structure. Hence the SS 
regime cannot be characterized by proximity to some
well-defined fixed point. (A stationary level structure, characteristic
of a FL fixed point, emerges only after another crossover,
setting in at the scale $\Tspin$.) 
Nevertheless, Figs.~\ref{fig:spectra_susceptibilities_K0}(c) and \ref{fig:spectra_susceptibilities_K0}(d)
show that the local orbital and spin susceptibilities \textit{do}
exhibit well-defined power-law behavior in the SS regime:
\begin{align}
\label{eq:SS-power-laws}
\chi_\orb^{\imp,\bath} (\omega) \sim \omega^{4/5}, \qquad
\chi_\spin^{\imp,\bath} (\omega) \sim \omega^{-6/5}. 
\end{align}
We define the width of the SS regime as
the energy range showing this behavior. 
It extends  over about 3 orders of magnitude,
independent of $J_0$ and $I_0$---increasing either of these
couplings rigidly shifts the SS regime to larger energies without 
changing its width (see Fig.~\ref{fig:SSwidth});
i.e., the ratio $\Tspin/\Tss$ is independent
of these couplings.

The latter fact leads us to conjecture that the NFL fixed point does,
after all, govern the SS regime too, though ``from afar'' rather than
from up close. In technical terms, we conjecture that 
the leading behavior in the SS regime is
governed by two different
boundary operators, $\tPsi_\orb$ and $\tPsi_\spin$, with scaling
dimensions $\tDelta_\orb = \tDelta_\spin = \frac{9}{10}$
(cf.\ Tables~\ref{tab:single_fusion_lowest} and \ref{tab:supp:fusion_NRG}) 
instead of the boundary operators $\bPhi_\orb$ and $\bPhi_\spin$ governing
the NFL regime. This conjecture is encoded in the
equation above Eq.~\eqref{eq:tilde chi_orb^imp,bath}.
It states that $\bJ_\orb$ and $\bT$  are both renormalized
to $\tPsi_\orb$, causing $\chi_\orb^\bath$ and $\chi_\orb^\imp$  to 
scale with the same power, 
\begin{align}
\label{eq:chi_orb_SS}
 \chi_\orb^{\bath,\imp}  \sim 
\langle \tPsi_\orb ||\tPsi_\orb \rangle_\omega
\sim \omega^{2\tDelta_\orb -1} = \omega^{4/5}, 
\end{align}
and  
that $\bJ_\spin$ and $\bS$  are both renormalized
to $\bS+\tPsi_\spin$, causing $\chi_\spin^\bath$ and $\chi_\spin^\imp$ 
to scale with the same power, 
\begin{align}
\label{eq:chi_spin_SS}
 \chi_\spin^{\bath,\imp}  \sim 
\langle \tPsi_\spin ||\tPsi_\spin \rangle_\omega
\sim \omega^{2\tDelta_\spin -3} = \omega^{-6/5}.  
\end{align}
The latter result is obtained in a manner analogous to
Eq.~\eqref{eq:chi_spin_imp_supp}, with $\bS \cdot \bPhi$ replaced by
$\bS  \tPsi_\spin$ \cite{quantumnumbersdiffer}.

\subsection{Impurity spectral function}
\label{sec:imp_spectral_function}

\begin{figure}[th]
\centering
\includegraphics[scale=0.7]{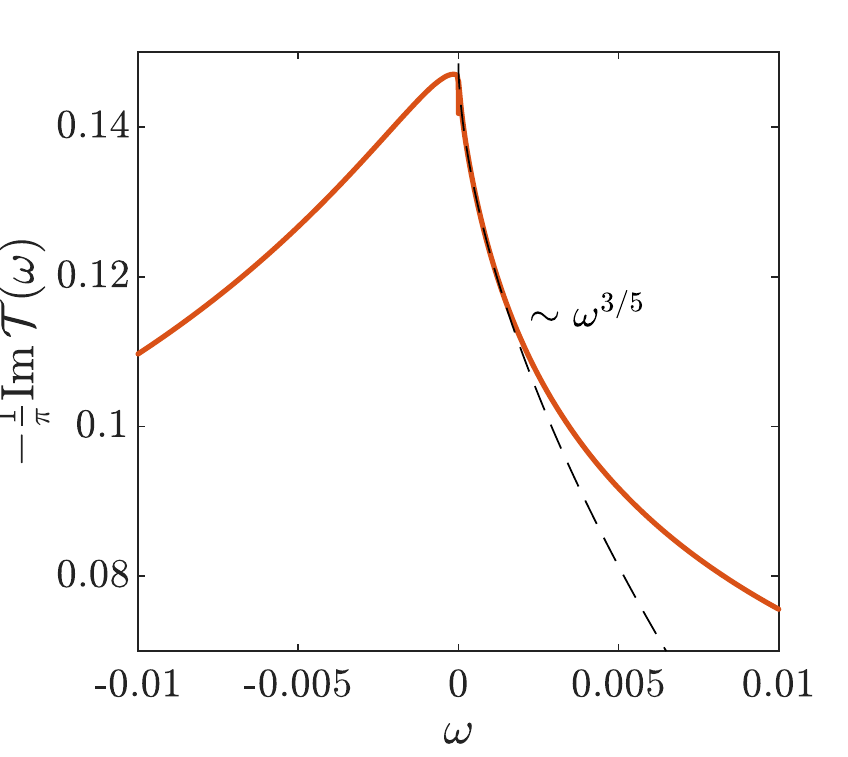}
\caption{Impurity spectral function, computed by fdm-NRG
  \cite{Weichselbaum2007}, for $(J_0,K_0,I_0) = (0,0.3,0)$. For
  $\omega>0$, the $\omega^{3/5}$ behavior is consistent with a
  boundary perturbation given by $H'_\orb$.  For $\omega<0$, clear
  power law cannot be determined.}
\label{fig:T-matrix}
\end{figure}

We next consider the leading frequency dependence of the
impurity spectral function in the NFL regime.
For  a Kondo-type impurity, this function  is given by
$-\tfrac{1}{\pi} \mathrm{Im} \mathcal{T}(\omega)$, where
$\mathcal{T}(\omega) = \langle O_{m\sigma} \| O_{m\sigma}^\dagger
\rangle_\omega$
is the impurity $\mathcal{T}$ matrix, with
$O_{m\sigma} = [\psi_{m\sigma}, H_\interaction]$ \cite{Costi2000}.

As discussed in Sec.~\ref{sec:susc-nfl-regime}, the leading irrelevant
boundary operators in the NFL regime are
$H_\orb^\prime = \bcJ_{\orb,-1} \cdot \bPhi_\orb$ and
$H_\spin^\prime = \bJ_{\spin,-1} \cdot \bPhi_\spin$, with scaling
dimensions $1+\Delta_\orb = 1+\frac{3}{5}$ and
$1+\Delta_\spin = 1+\frac{2}{5}$, respectively. AL have shown that a
boundary perturbation of this type, with dimension $1+ \Delta$, causes
the $\mathcal{T}$ matrix to acquire a leading frequency dependence of
$\mathrm{Im}\mathcal{T}\sim |\omega|^\Delta$ \cite{Affleck1993}.

For $\omega>0$ our NRG results are consistent with 
$\mathrm{Im}\mathcal{T}\sim \omega^{3/5}$ (cf.~Fig.~\ref{fig:T-matrix}).
This suggests that the prefactor of $H'_\orb$ is much larger
than that of $H'_\spin$, presumably because the computation 
was done for $J_0 = I_0 = 0$. For $\omega<0$, by contrast, 
our numerical results do not exhibit clear power-law behavior 
for small $|\omega|$, implying that $\mathrm{Im}\mathcal{T}$
does not have  particle-hole symmetry. This is not surprising: 
the \threesoK\ model itself breaks particle-hole symmetry,
since under a particle-hole transformation, the impurity's
orbital multiplet $\raisebox{-0.5mm}{\tiny \yng(1,1)}$ is mapped to $\raisebox{0.5mm}{\tiny \yng(1)}\,$.
We suspect that the prefactor of the
$|\omega|^{\Delta_\orb}$ contribution to $\mathrm{Im} \mathcal{T}$
vanishes for $\omega < 0$ for the impurity orbital representation
${\tiny \yng(1,1)}$\,, such that only subleading boundary operators,
with dimensions $\Delta \geq 9/10$ (cf.~Table \ref{tab:supp:fusion_NRG}),
determine the small-$\omega$ scaling behavior. However, a 
detailed understanding of this matter is still lacking.

\subsection{Fermi-liquid fixed point} 
\label{sec:FL-spectrum}

In this  section we show how the FL spectrum at the
fixed point $\bc_\fl^\ast$ can be derived analytically. 
This can be done in two complementary ways. The first
uses $\sutwo_3$ fusion in the spin sector, the second
$\susix_1$ fusion in the flavor (combined spin+orbital)
sector.

\subsubsection{Fermi-liquid spectrum via $ \sutwo_3$ fusion}
\label{eq:SU(2)3-FL}

\renewcommand{\tabcolsep}{2.5pt} 
\begin{table*}[t]
\centering
\caption{Fusion table for spin screening at the FL fixed point, $\bc_\fl^\ast$,   of the \threesoK\ model. It has the same structure
  as Table~\ref{tab:supp:fusion_NRG}, but 
  here single fusion of  bath and impurity multiplets 
  in the charge and spin sectors is performed using $\uone \times \sutwo_3$ fusion rules (listed in Table~\ref{tab:supp:fusion_SU(2)3} of the SM \cite{suppmat}).
Moreover, we choose   $Q_\imp = (1, \frac32, \bullet)$ for the impurity, representing the effective local degree of freedom obtained after the completion of orbital screening. 
  The resulting  multiplets $(q^\prime, S^\prime, \lambda)$
  have eigenenergies $E^\prime = E(q^\prime,S^\prime,\lambda)$ and excitation energies  $\delta E^\prime = E^\prime - E^\prime_\mathrm{min}$. 
  The NRG energies, computed for $(J_0,K_0,I_0) = (10^{-4},0.3,0)$,
  have been shifted and rescaled such that 
  the lowest energy is zero and the second-lowest values for $E_\NRG$ and $\delta E'$ 
  match. The single-fusion and NRG spectra agree very well
  (deviations $\lesssim 2\%$).}
\renewcommand{\arraystretch}{1.9}
\begin{tabular}{ccc@{\hspace{8mm}}c@{\hspace{4mm}}c@{\hspace{2mm}}|@{\hspace{2mm}}ccccc@{\hspace{8mm}}c@{\hspace{4mm}}cc|c|@{\hspace{2mm}}ccccc@{\hspace{8mm}}cc} 
\hline \hline
\multicolumn{5}{c|@{\hspace{2mm}}}{Free fermions} 
& \multicolumn{8}{c|}{Single fusion, with $Q_\imp = (1, \frac32, \bullet)$} & NRG
& \multicolumn{6}{c}{Double fusion, with $\Qbar_\imp = (-1, \frac32, \bullet)$}
\\ 
$q$ & $S$ & $\lambda$ & $d$ & $E$ 
& $q^\prime$ & & $S^\prime$ & & $\lambda^\prime$ & $d$ 
&   $E^\prime$   & $\delta E^\prime$ & $E_{\rm NRG}$
& $q^{\prime\prime}$ & & $S^{\prime\prime}$ & & $\lambda^{\prime\prime}$ & $\Delta$ 
\\ \hline 

0 & 0 & {\y00} & 1 & 0
& {+1} & & {\fr32} & & {\y00} & 4
& {\fr56} & \fr12 & 0.50
& 0 & & 0 & & \y00 & 0
\\ 

{+1} & {\fr12} & \raisebox{0.5mm}{\y10} & {6} & {\fr12}
& {+2} & & {1} & & \raisebox{0.5mm}{\y10} & 9
& 1 & \fr23 & 0.67
& +1 & & \fr12 & & \raisebox{0.5mm}{\y10} & \fr12
\\ 

{$-1$} & {\fr12} & \raisebox{0.5mm}{\y01} & {6} & {\fr12}
& 0 & & 1 & & \raisebox{0.5mm}{\y01} & 9
& \fr23 & \fr13 & 0.33
& $-1$ & & \fr12 & & \raisebox{0.5mm}{\y01} & \fr12 
\\ 

0 & 1 & \raisebox{0.5mm}{\y11} & {24} & 1
& {+1} & & \fr12 & & \raisebox{0.5mm}{\y11} & 16
& {\fr56} & \fr12 & 0.50
& 0 & & 1 & & \raisebox{0.5mm}{\y11} & 1 
\\ 

{+2} & 0 & 
\raisebox{0.5mm}{\tiny $\yng(2)$} & 6 & 1
& {+3} & & \fr32 & & 
\raisebox{0.5mm}{\tiny $\yng(2)$} & 24
& \fr{13}{6} & \fr{11}{6} & 1.87
& +2 & & 0 & & 
\raisebox{0.5mm}{\tiny $\yng(2)$} & 1
\\ 

{$-2$} & {0} & \raisebox{0.5mm}{\y02} & {6} & 1
& $-1$ & & \fr32 & & \raisebox{0.5mm}{\y02} & 24
& \fr32 & \fr76 & 1.16
& $-2$ & & 0 & & \raisebox{0.5mm}{\y02} & 1
\\ 

{+2} & 1 & \raisebox{0.5mm}{\y01} & 9 & 1
& {+3} & & \fr12 & & \raisebox{0.5mm}{\y01} & 6
& \fr76 & \fr56 & 0.84
& +2 & & 1 & & \raisebox{0.5mm}{\y01} & 1
\\ 

{$-2$} & 1 & \raisebox{0.5mm}{\y10} & 9 & 1
& $-1$ & & \fr12 & & \raisebox{0.5mm}{\y10} & 6
& \fr12 & \fr16 & 0.17
& $-2$ & & 1 & & \raisebox{0.5mm}{\y10} & 1
\\ 

{+1} & {\fr32} & \raisebox{0.5mm}{\y02} & {24} & {\fr32}
& {+2} & & 0 & & \raisebox{0.5mm}{\y02} & 6
& 1 & \fr23 & 0.68
& +1 & & \fr32 & & \raisebox{0.5mm}{\y02} & \fr32
\\ 

{$-1$} & {\fr32} & 
\raisebox{0.5mm}{\tiny $\yng(2)$} & {24} & {\fr32}
& 0 & & 0 & & 
\raisebox{0.5mm}{\tiny $\yng(2)$} & 6
& \fr23 & \fr13 & 0.34
& $-1$ & & \fr32 & & 
\raisebox{0.5mm}{\tiny $\yng(2)$} & \fr32
\\ 

{$\pm3$} & {\fr12} & \raisebox{0.5mm}{\y11} & {16} & {\fr32}
& $-2$ & & 1 & & \raisebox{0.5mm}{\y11} & 24
& \fr43 & 1 & 0.99
& $-3$ & & \fr12 & & \raisebox{0.5mm}{\y11} & \fr32
\\ 

{$\pm3$} & {\fr32} & {\y00} & 4 & {\fr32}
& $-2$ & & 0 & & {\y00} & 1
& \fr13 & 0 & 0.00
& $-3$ & & \fr32 & & \y00 & \fr32
\\ \hline \hline 

\end{tabular}
\label{tab:supp:fusion_NRG_FL}
\end{table*}

It is natural to ask whether the FL spectrum at $\bc_\fl^\ast$ can be
derived from the NFL spectrum of $\bc_\nfl^\ast$ via some type of
fusion in the spin sector, reflecting spin screening induced by the
spin exchange interaction. For example, we have tried the following
simple strategy (``naive spin fusion''): when setting up the fusion
table (Table \ref{tab:supp:fusion_NRG}), the bath and impurity spin degrees of
freedom are combined, $S \otimes S_\imp = \sum_\oplus S'$, using the
fusion rules of the $\sutwo_3$ KM algebra
(Table~\ref{tab:supp:fusion_SU(2)3} in the SM \cite{suppmat})
instead of the $\sutwo$ Lie algebra, 
and the orbital degrees of freedom,
$\lambda\otimes \lambda_\imp = \sum_\oplus \lambda'$, using
$\suthree_2$ KM fusion rules (as before; see Table~\ref{tab:supp:fusion_SU(3)} in the SM \cite{suppmat}). 
Moreover, the energies of
the resulting multiplets are computed as
$E(q,S^\prime,\lambda^\prime)$, not $E(q,S,\lambda^\prime)$.  However,
this naive spin fusion strategy completely fails to reproduce the FL
fixed point spectrum obtained by NRG, yielding completely different
excitation energies and degeneracies. 

We suspect that this failure is due to the fact that 
the RG flow does not directly pass from the NFL regime into the
FL regime, but first traverses the intermediate SS regime.
In the latter, the degeneracy between the two degenerate
ground state multiplets of the NFL regime, $(1, \frac{1}{2},\bullet)$
and $(1, \frac{3}{2},\bullet)$, is lifted, in a manner that
seems to elude a simple description via a modified spin fusion rule. 

Instead, the FL spectrum can be obtained via the following
arguments. The ground state multiplet of the SS regime,
$(1, \frac{3}{2},\bullet)$, describes an
effective local degree of freedom coupled to a bath in such a manner
that one bath electron fully screens the impurity orbital isospin,
while their spins add to a total spin of $\frac{1}{2}+1 = \frac{3}{2}$
[see Fig.~\ref{fig:spectra_susceptibilities_K0}(b)]. Let us view this
as an effective impurity with $Q_\imp = (1, \frac{3}{2},\bullet)$. 
If we combine its charge and spin degrees of freedom with
those of a \textit{free} bath, using
$q +q_\imp = q^\prime$ and  
$S\oplus S_\imp = \sum_\oplus S^\prime$, fused according
to the $\sutwo_3$ KM algebra, the resulting
single-fusion spectrum fully
reproduces the FL spectrum found by NRG,
as shown in Table~\ref{tab:supp:fusion_NRG_FL}.

\subsubsection{Fermi-liquid spectrum via $\susix_1$ fusion}
\label{eq:SU6-FL}

The FL ground state of the \threesoK\ model is a fully screened spin
and orbital singlet. It is therefore natural to expect
that the FL spectrum has a higher symmetry, namely that of the group
$\uone \times \susix$, which treats spin and orbital excitations
on an equal footing. 

This is indeed the case:
we now show that the FL spectrum of the \threesoK\ model
discussed above matches that of an  \susix\ Kondo model
which does not distinguish between spin and
orbital degrees of freedom. 
We consider  a bath with six flavors of electrons, 
$H_\bath = \sum_{p} \sum_{\nu=1}^6 \varepsilon_p \psi_{p\nu}^\dagger
\psi_{p\nu}$ 
and an  impurity-bath coupling of the form 
$J_U \bU \cdot \bJ_\flavor$. Here $\bJ_\flavor$ is the flavor density at the impurity
site, with $J^a_\flavor = \psi_{\nu}^\dagger \,\tfrac12
  \lambda^a_{\nu\nu^\prime} \,\psi_{\nu^\prime}$,
where  $\lambda^a$ are  \susix\ matrices in 
the fundamental representation, 
and $\bU$ describes the impurity's 
$\susix$ flavor isospin, chosen in the fully antisymmetric
representation ${\tiny \yng(1,1)}$\,. The latter has
dimension 15, representing the ${6 \choose 2}$ 
ways of placing two identical particles into six available 
states.

\begin{figure}[t]
\includegraphics[width=0.97\linewidth]{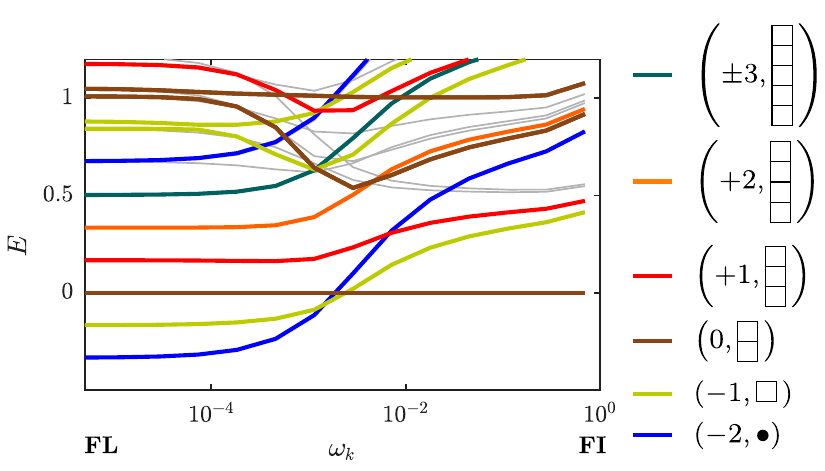}
\caption{NRG results for the \susix\ Kondo model with 
  $J_U=0.1$, shown using 
  $\cE_\reference = \cE(0,\protect\raisebox{0.6mm}{\tiny
    $\protect\yng(1,1)$}\,)$
  as reference energy.  The
  computation was performed using \mbox{QSpace}
  \cite{Weichselbaum2012a} to exploit the model's full
  $\uone \times \susix$ symmetry. (NRG parameters:
$\Lambda= 2.5$, $N_\keep = 2000$, $D = 1$.)}
\label{fig:supp:NRG_SU6}
\end{figure}

\renewcommand{\tabcolsep}{6pt}
\begin{table}[t]
\centering
\caption{Fusion table for flavor screening at the FL fixed point
  of the \susix\ Kondo model. The table  has the same structure
    as the left and center parts of Table~\ref{tab:supp:fusion_NRG}, 
    but here the free bath excitations are labeled $(q,\lambda)$,
    their energies are computed using Eqs.~\eqref{eq:E(qlambda)-sup}
    and Table~\ref{tab:supp:E_SU(6)} of the SM \cite{suppmat},  and  flavor fusion  with  $Q_\imp = (0,   \protect\raisebox{0.5mm}{\tiny $\protect\yng(1,1)$}\,)$ 
    is performed using $\susix_1$ fusion rules (listed in Table~\ref{tab:supp:fusion_SU(6)} of the SM \cite{suppmat}). The resulting 
  multiplets $(q,\lambda')$ have eigenenergies  $E'= E(q,\lambda')$,
  degeneracies $d'$ and excitation energies, $\delta E^\prime = E^\prime - E^\prime_\minimum$. The  FL spectrum, 
  obtained by $\uone \times \susix$  NRG calculations (Fig.~\ref{fig:supp:NRG_SU6}) for
  $J_U=0.1$,   is shown on the right. It has been shifted and rescaled such that the lowest energy is zero and the second-lowest values for $E_\NRG$ and $\delta E'$  match. The single-fusion and NRG spectra agree very well
  (deviations $\lesssim 1\%$).}
\renewcommand{\arraystretch}{2} 
\begin{tabular}{ScSc@{\hspace{8mm}}c@{\hspace{5pt}}c|ScSc@{\hspace{8mm}}Sc@{\hspace{10pt}}c@{\hspace{3pt}}Sc|Sc}
\hline \hline
\multicolumn{4}{c|}{Free fermions} & \multicolumn{5}{c|}{Single fusion, 
 $Q_\imp = (0, {\tiny \yng(1,1)}\,)$} & NRG
\\ 
$q$ & $\lambda$ & $d$ & $E$
& $q$ & $\lambda^\prime$ & $d^\prime$ & $E^\prime$ & $\delta E^\prime$ & $E_\NRG$
\\ \hline
\phantom{$-$}0 & $\bullet$ & 1 & 0
& \phantom{$-$}0 & \raisebox{0.5mm}{\tiny $\yng(1,1)$} & 15 & \fr23 & \fr13
& 0.33
\\ 


+1 & \raisebox{0.5mm}{\tiny $\yng(1)$} & 6 & \fr12
& +1 & \raisebox{0.5mm}{\tiny $\yng(1,1,1)$} & 20 & \fr56 & \fr12
& 0.50
\\ 

$-1$ & \raisebox{0.5mm}{\tiny $\yng(1,1,1,1,1)$} & 6 & \fr12
& $-1$ & \raisebox{0.5mm}{\tiny $\yng(1)$} & 6 & \fr12 & \fr16
& 0.17
\\ 


+2 & \raisebox{0.5mm}{\tiny $\yng(1,1)$} & 15 & 1
& +2 & \raisebox{0.5mm}{\tiny $\yng(1,1,1,1)$} & 15 & 1 & \fr23
& 0.67
\\ 

$-2$ & \raisebox{0.5mm}{\tiny $\yng(1,1,1,1)$} & 15 & 1
& $-2$ & $\bullet$ & 1 & \fr13 & 0 
& 0
\\ 


$\pm3$ & \raisebox{0.5mm}{\tiny $\yng(1,1,1)$} & 20 & \fr32
& $\pm3$ & \raisebox{0.5mm}{\tiny $\yng(1,1,1,1,1)$} & 6 & \fr76 & \fr56
& 0.84
\\ \hline \hline

\end{tabular}
\label{tab:fusion_SU(6)}
\end{table}

 Figure~\ref{fig:supp:NRG_SU6} shows the NRG finite-size eigenlevel flow
for this model. It exhibits a single crossover from a free-impurity
fixed point, with ground state multiplet
$(q,\lambda) = (0,{\tiny \yng(1,1)}\,)$, to a FL fixed point, whose
ground state multiplet $(-2, \bullet)$ involves complete screening of
the impurity's flavor isospin degree of freedom.

This crossover can be described analytically by using non-Abelian
bosonization followed by flavor fusion.
We begin by using non-Abelian bosonization
with the  $\uone \times \susix_1$ KM current
algebra to express the bath excitation spectrum in the  form
\begin{subequations}
\label{eq:E(qlambda)-sup}
\begin{align}
E(q,\lambda)  = &  \tfrac{1}{12}q^2 + \tfrac{1}{7}\kappa_6 
(\lambda) + \ell \,,
\label{eq:E(qlambda)}
\\
\label{eq:kappa_6}
\kappa_6(\lambda)
= & 
\tfrac{1}{12} ( 
5 \lambda_1^2 
+8 \lambda_2^2
+9 \lambda_3^2
+8 \lambda_4^2
+5 \lambda_5^2
)
\\ 
&+ 
\tfrac{1}{2} (
5 \lambda_1
+8 \lambda_2
+9 \lambda_3
+8 \lambda_4
+5 \lambda_5
)
\nonumber \\
&+
\tfrac{1}{6} (
6 \lambda_2 \lambda_3
+6 \lambda_3 \lambda_4
+4 \lambda_1 \lambda_2
+4 \lambda_2 \lambda_4
+4 \lambda_4 \lambda_5
\nonumber 
\\
\nonumber
& +3 \lambda_1 \lambda_3
+3 \lambda_3 \lambda_5
+2 \lambda_1 \lambda_4
+2 \lambda_2 \lambda_5
+ \lambda_1 \lambda_5) \, 
\end{align}
\end{subequations}
with $\ell \in \mathbbm{Z}$, where $\kappa_6(\lambda)$ is the
quadratic Casimir for the representation
$\lambda = (\lambda_1,\lambda_2,\lambda_3,\lambda_4,\lambda_5)$ of the
\susix\ Lie algebra \cite{FuchsSchweigert}.  [The contributions from
the two terms of Eq.~\eqref{eq:E(qlambda)} are listed in Table
\ref{tab:supp:E_SU(6)} in the SM \cite{suppmat} for all $q$ and $\lambda$
values needed in Table.~\ref{tab:fusion_SU(6)}.] The few lowest-lying $(q,\lambda)$ multiplets of
the free bath, having $E(q,\lambda) \in \frac{1}{2}\mathbbm{Z}$, are
listed on the left-hand side of Table~\ref{tab:fusion_SU(6)}. The
strong-coupling FL spectrum can be obtained by combining the bath and
impurity flavor degrees of freedom,
$\lambda \otimes \lambda_\imp = \sum_\oplus \lambda'$, using the
fusion rules of the $\susix_1$ KM algebra (see Table
\ref{tab:supp:fusion_SU(6)} in the SM \cite{suppmat}). The resulting
multiplets $(q,\lambda')$ are listed in the center of
Table~\ref{tab:fusion_SU(6)}. Their eigenenergies fully match
those from NRG.

\section{Three-orbital Anderson-Kondo model}
\label{sec:3OAH-model}

\begin{figure*}[th]
\includegraphics[width=\textwidth]{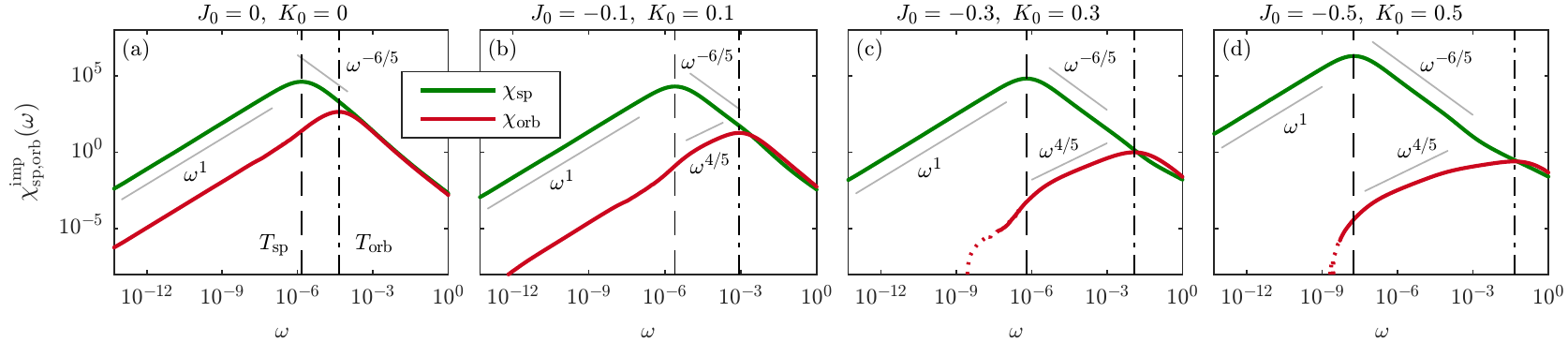}
\caption{Imaginary part of the impurity spin and orbital
    susceptibilities for the Anderson-Kondo model,
    with $U=5$, $J_{\rm H}=1$, $\Gamma=0.1$, $I_0=0$, and $J_0$, $K_0$
    varying from (a) $J_0=K_0=0$ (pure Anderson-Hund model) to (d)
    $-J_0= K_0=0.5$. The energy scales for spin and orbital screening,
    $\Tspin$ and $\Torb$ are pushed apart with increasing
    $-J_0 = K_0$.}
\label{fig:susceptibilities_AHM_AKM}
\end{figure*}

The \threesoK\ model, which we study in detail above, 
is derived from the more realistic \threeoAH\ model by a Schrieffer-Wolff
transformation. In the following, we explore another route for
smoothly connecting the physics of the two models, namely starting
from the \threeoAH\ model and then additionally turning on the spin and
orbital exchange couplings of the \threesoK\ model.

The Hamiltonian of the \threeoAH\ model \cite{Stadler2015} has the following form:
$H_\bath + H_\mathrm{hyb} + H_\threeoAH$,
\begin{align}
H_\threeoAH 
&= \tfrac34 J_{\rm H} N_\imp 
\!+\! \tfrac12 \left( U \!-\! \tfrac12 J_{\rm H} \right) N_\imp (N_\imp \!-\! 1) 
\!-\! J_{\rm H} \bS^2 ,
\nonumber 
\\
H_\mathrm{hyb} &= \sum_{pm\sigma} V_p (f_{m\sigma}^\dagger \psi_{pm\sigma} + \mathrm{H.c.}) \,,
\nonumber
\end{align}
with the impurity occupation
$N_\imp = \sum_{m\sigma} f_{m\sigma}^\dagger f_{m\sigma}$, where
$f_{m\sigma}^\dagger$ creates an impurity electron with spin $\sigma$ in 
orbital $m$. A hybridization function,
$\Gamma(\varepsilon) = \pi \sum_p |V_p|^2 \delta
(\varepsilon-\varepsilon_p) \equiv \Gamma \Theta(D-|\varepsilon|)$,
controls the hopping between the impurity and the bath. $U$ is the
local Coulomb interaction  and $J_{\rm H}$ the  Hund's coupling, favoring
a large spin. To this Hamiltonian we add
$J_0 \, \bS \cdot \bJ_\spin + K_0 \, \bT \cdot
\bJ_\orb$,
the Kondo-type spin and orbital exchange couplings between impurity
and bath from Eq.~\eqref{eq:Hamiltonian_AronKotliar}, with
$S^\alpha = f_{m\sigma}^\dagger \,\tfrac12
\sigma^\alpha_{\sigma\sigma^\prime} \,f_{m\sigma^\prime}$,
$T^a = f_{m\sigma}^\dagger \,\tfrac12 \tau^a_{mm^\prime}
\,f_{m^\prime\sigma}$. We treat $J_0$ and $K_0$ as free parameters
and use them to ``deform'' the \threeoAH\ model in a way that widens
the SOS regime between $\Tspin$ and $\Torb$. 

Figures~\ref{fig:susceptibilities_AHM_AKM}(a)-\ref{fig:susceptibilities_AHM_AKM}(d) show how the spin and
orbital susceptibilities change upon increasing $|J_0|$ and $|K_0|$,
with $J_0<0$ and $K_0>0$.  A pure \threeoAH\ model, with
$(J_0,K_0)=(0,0)$, clearly shows spin-orbital separation, but
$\Tspin$ and $\Torb$ differ by less than two decades
[Fig.~\ref{fig:susceptibilities_AHM_AKM}(a); see also
Ref.~\cite{Stadler2015}].
Though the SOS window is too small to reveal a true power law
for $\chi_\spin^\imp$, the hints of $\omega^{-6/5}$ behavior
are already discernable.
Turning on the additional exchange coupling
terms, with $J_0<0$ and $K_0>0$, causes $\Tspin$ to decrease and
$\Torb$ to increase, respectively, widening the SOS regime
[Figs.~\ref{fig:susceptibilities_AHM_AKM}(b)-\ref{fig:susceptibilities_AHM_AKM}(d)]. For
$(J_0,K_0)=(-0.5,0.5)$ it spans more than 6 orders of magnitude, so
that clear power laws,
$\chi_\spin^\imp \sim \omega^{-6/5}$ and
$\chi_\orb^\imp \sim \omega^{4/5}$,
become accessible
[Fig.~\ref{fig:susceptibilities_AHM_AKM}(d)].
These power laws are consistent with our 
findings for the spin-splitting regime  in Secs.~\ref{sec:NRG} and \ref{sec:CFT_analysis}.
This scenario is
evidently smoothly connected to that of the pure \threesoK\ model
[Fig.~\ref{fig:spectra_susceptibilities_K0}(c)]. There the absence of
charge fluctuations makes it possible to fully turn off the
$I_0$ contribution implicitly present in the \threeoAH\ model, thereby
widening the SOS regime even further and allowing the true \nfl\
regime to be analyzed in detail.

\section{Conclusion}
\label{sec:conclusion}

While the main aim of this work was to understand NFL behavior in Hund metals, 
it has much wider implications, as already indicated in Sec.~\ref{sec:Introduction}.
Let us assess these from several perspectives of
increasing generality. 

(i) We have used NRG and CFT to elucidate the NFL regime of a
\threesoK\ model, fine-tuned such that spin screening sets in at very
much lower energies than orbital screening. We were able to
  analytically compute the scaling behavior of dynamical spin and
  orbital susceptibilities, finding $\chi^\imp_\orb \sim
  \omega^{1/5}$, $\chi^\imp_\spin \sim \omega^{-11/5}$ in the NFL
  regime and $\chi^\imp_\spin \sim \omega^{-6/5}$ in the
  spin-splitting regime. The main significance of these findings
  lies in the qualitative physical behavior which they imply. An
  orbital susceptibility decreasing with an exponent $<1$, rather than
  the Fermi-liquid exponent 1, indicates that the orbital degrees of
  freedom, though screened, are still affected by the unscreened spin
  degrees of freedom.  A spin susceptibility diverging as
  $\omega^{-\gamma}$, with $\gamma > 1$, indicates anomalously strong
  spin fluctuations. This seems to be a characteristic property of
  the incoherent regime of Hund metals. As pointed out in Sec.~\ref{sec:Introduction}, anomalously strong spin fluctuations have direct
  consequences for theories of the superconducting state of the iron
  pnictides \cite{Lee2018a}.

(ii) We have uncovered the origin of hints of NFL behavior found
previously for a \threeoAH\ model and related models
\cite{Georges2013,Haule2009,Yin2011a,Yin2011b,Yin2012,Stadler2015,Stadler2018,Deng2019}.
There the spin-orbital coupling $I_0$ is always nonzero, preventing RG
trajectories from closely approaching the NFL fixed
point. Nevertheless, even if they pass this fixed point ``at a
distance,'' it still leaves traces of NFL behavior for various
observables, such as $\chi^\imp_\spin \sim \omega^{-6/5}$ behavior for
the imaginary part of the impurity's dynamical spin susceptibility. We
showed in Sec.~\ref{sec:3OAH-model} how NFL behavior emerges if
the \threeoAH\ model is ``deformed'' by additionally turning on the
spin and orbital exchange couplings of the \threesoK\ model,
thereby adiabatically connecting the SS regime of the
  \threesoK\ model to the incoherent regime of the \threeoAH\ model.
Furthermore, it has been shown in Ref.~\cite{Stadler2015} that
  DMFT self-consistency does not significantly influence the behavior
  of the susceptibilities in the \threeoAH\ model. Thus our
  conclusions, in particular regarding the prevalence of strong spin
  fluctuations in the SOS regime, should also apply to DMFT
  calculations. Indeed, DMFT studies \cite{Stadler2015,Stadler2019} of
  a self-consistent \threeoAH\ model contain results for 
  $\chi_\spin^\imp$ which, in the SOS window, are consistent with the
  $\omega^{-6/5}$ scaling found and explained here for the SS regime.

(iii) Taking a broader perspective, we have provided an analytic
solution of a paradigmatic example of a ``Hund impurity problem.''
We were able to address this fundamental type of problem by combining state-of-the-art multiorbital NRG with a suitable generalization of Affleck and Ludwig's CFT approach
\cite{Affleck1991,Affleck1990,Affleck1991a,Affleck1993,Ludwig1994a}.

(iv) Regarding experimental relevance, Hund impurities 
are of central importance for understanding Hund metals, including almost all 4$d$ and 5$d$ materials,
and even in the 5$f$ actinides Hund's coupling is the main cause for
electronic correlations. Our work illustrates paradigmatically why
hints of NFL physics can generically be expected to arise in such
systems.  Moreover, \textit{tunable} Hund impurities can be realized
using magnetic molecules on substrates \cite{Khajetoorians2015} or
multilevel quantum dots, raising hopes of tuning Hund impurities in
such a way that truly well-developed NFL behavior can be observed
experimentally.       

\section*{Acknowledgments}

We thank I. Affleck, A. Georges, M. Goldstein, O. Parcollet, E. Sela
and A. Tsvelik for helpful advice and, in
particular, I. Brunner for technical advice regarding $\suthree_2$
fusion rules. E.~W., K.~M.~S., and J.~v.~D.\ are supported by the Deutsche
Forschungsgemeinschaft under Germany's Excellence
Strategy|EXC-2111|390814868, and S.-S.~B.~L.\ by Grant No.\ LE3883/2-1. 
A.~W.\ was supported by the U.S. Department of Energy,
Office of Basic Energy Sciences, under Contract No. DE-SC0012704.
G.~K.\ was supported by the National Science Foundation Grant No.~DMR-1733071.
Y.~W.\ was supported by the U.S. Department of Energy, 
Office of Science, Basic Energy Sciences as a part of the
Computational Materials Science Program through the 
Center for Computational Design of Functional Strongly
Correlated Materials and Theoretical Spectroscopy.

\textit{Note added.}---       
Recently, a paper closely related to ours appeared \cite{Horvat2019},
with similar goals, a complementary analysis (using NRG but not CFT), 
and conclusions consistent with ours.

\appendix*

\renewcommand{\theequation}{A\arabic{equation}}

\section{Ye's SU(2)$\boldsymbol{\times}$SU(2) spin-orbital Kondo model}
\label{app:Ye}

In this appendix, we revisit an $\sutwo\times \sutwo$ spin-orbital
Kondo (\twosoK) model studied in a pioneering paper by Ye in 1997
\cite{Ye1997}. It is a simpler cousin
of our \threesoK\ model, having a Hamiltonian of precisely the same
form, with the following differences: the orbital channel index takes
only two values, $m = 1,2$; the local orbital current $\bJ_\orb$ is
defined using Pauli (not Gell-Mann) matrices; and the impurity spin
and orbital isospin operators, $\bS$ and $\bT$, are both
\sutwo\ generators, in the representation $S = \lambda = \frac{1}{2}$.

In the context of the present study, Ye's paper is of interest because
his Kondo impurity likewise features both spin and orbital degrees of
freedom. From a conceptual perspective, his and our models differ only
in the symmetry group, \sutwo\ versus \suthree\ in the orbital sector,
and the choice of impurity multiplet, $Q_\imp =
(\frac{1}{2},\frac{1}{2})$ versus $(1,\raisebox{-0.5mm}{\tiny
  \yng(1,1)}\, )$. Moreover, he was able to obtain 
exact results for his model using Abelian bosonization.
Below, we verify that when 
the NRG and CFT methodology 
used in the main text is applied to Ye's  \twosoK\ model, 
the results are consistent with his conclusions.

Before
    proceeding, however, let us also briefly address some differences
    between Ye's work and ours. Since he uses Abelian bosonization,
    his approach does not readily generalize to the $\uone\times
    \sutwo \times \suthree$ impurity model considered here.  Ye does
    mention very briefly that some of his results can also be obtained
    using non-Abelian bosonization, employing \textit{simultaneous}
    fusion in the spin \text{and} orbital sectors.  However, we show
    below that that fusion scheme is applicable only when
    particle-hole symmetry is present. This is the case for Ye's
    application, but not for our \threesoK\ model.  When particle-hole
    symmetry is absent, the fusion schemes needed for the spin and
    orbital are subtly different---indeed, clarifying these
    differences was the conceptually most challenging aspect of our
    work. Note that the particle-hole asymmetry of our
    \threesoK\ model is not a mere technical complication, but an
    essential ingredient of the physics of Hund metals, which
    typically feature fillings one particle away from
    half filling. Finally, note that Ye's model, involving an impurity
    with spin $1/2$, is not relevant for Hund metals, where Hund's
    coupling favors local spins larger than $1/2$.

\subsection{$\boldsymbol{I_0 = 0}$: NFL fixed point}
\label{sec:Ye-NFL}
\label{section:SU2SU2}

\begin{figure*}[tb]
\includegraphics[width=0.74\linewidth]{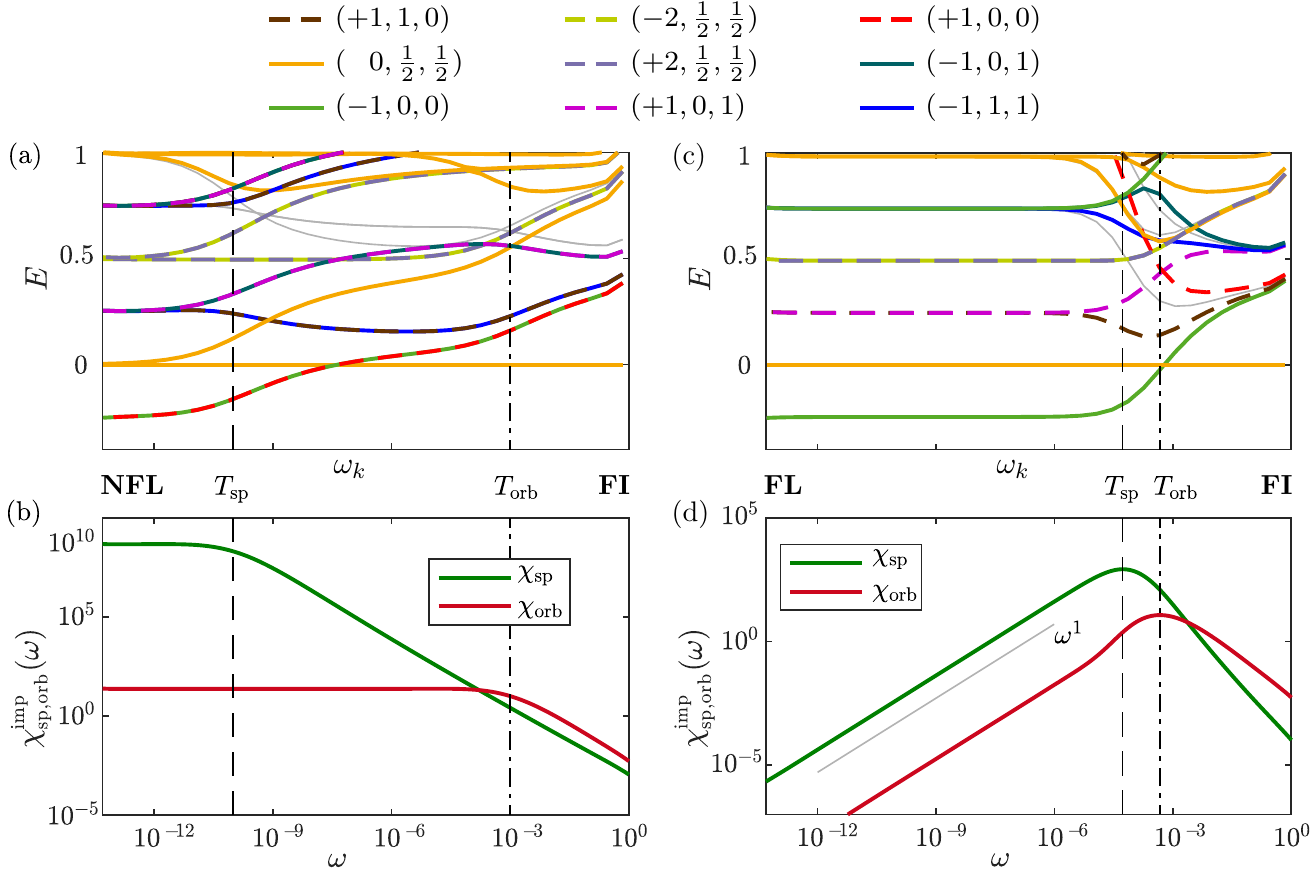}
\caption{
NRG results for Ye's  \twosoK\ model,
computed (a,b) for $(J_0,K_0,I_0) = (0.1, 0.3,0)$,
respecting particle-hole symmetry, 
and (c,d) for $ (0, 0.3,0.05)$,
breaking particle-hole symmetry.
(a,c) Eigenlevel flow diagrams, with quantum numbers $(q,S,\lambda)$
shown at the top.  The low-energy fixed points in (a) and (c)
 exhibit a NFL or FL spectrum, respectively,
reproduced analytically in Table~\ref{tab:supp:fusion_NRG_Ye_su2su2} or
\ref{tab:E_SU(4)}, respectively.
(b,d)   Imaginary part of the impurity's spin and orbital susceptibilities.
Vertical lines indicate the crossover scales 
for orbital screening, $\Torb$ (dash-dotted)
and spin screening, $\Tspin$ (dashed),
marking when $\chi^\imp_\orb$ and $\chi^\imp_\spin$ (b) bend over to constant behavior, or (d) reach their maxima, respectively.  (NRG parameters: $\Lambda= 2.5$, $N_\keep = 2000$, $D = 1$.)}
\label{fig:J}
\end{figure*}

For $I_0 = 0$, the \twosoK\ model obeys particle-hole symmetry.
Figure~\ref{fig:J}(a) shows the finite-size eigenlevel flow computed by
NRG for $\bc_0 = (J_0,K_0,I_0) = (0.1, 0.3,0)$.
 The low-energy fixed-point spectrum
features equidistant levels, but nevertheless has NFL properties, as
predicted by Ye, in that it cannot be understood in terms of
combinations of single-particle excitations.  Remarkably, 
this fixed-point spectrum can be reproduced by CFT arguments.  Using
non-Abelian bosonization according to the
$\uone \times \sutwo_2 \times \sutwo_2$ KM algebra, the  spectrum of
free bath excitations can be expressed as
\begin{subequations}
\label{eq:supp:E(qSlambda)su2level2-sup}
\begin{align}
  E(q,S,\lambda) & = \tfrac{1}{8}q^2 + \tfrac{1}{4}\kappa_2(S) + \tfrac{1}{4}\kappa_2(\lambda) + \ell \,,
\label{eq:supp:E(qSlambda)su2level2}
\\
\label{eq:eq:Su2Su2Casimir}
\kappa_2(S) &= S(S+1) \,, \quad \kappa_2(\lambda) = \lambda(\lambda+1) \, , 
\end{align}
\end{subequations}
with $\ell \in \mathbbm{Z}$, while $\kappa_2(S)$, 
$\kappa_2 (\lambda)$ are the quadratic $\sutwo$ Casimirs in the spin
and orbital sectors, respectively. We now combine  bath and
impurity degrees of freedom using \textit{simultaneous} fusion in the
spin and orbital sectors, $S \otimes S_\imp = \sum_\oplus S^\prime$
and $\lambda \otimes \lambda_\imp = \sum_\oplus \lambda^\prime$, employing the
fusion rules of the $\sutwo_2 \times \sutwo_2$ KM algebra
(Table~\ref{tab:supp:fusion_SU(2)2} in the SM \cite{suppmat}). 
This reproduces the NFL fixed-point spectrum
found by NRG, as shown  in Table~\ref{tab:supp:fusion_NRG_Ye_su2su2}.

By contrast, we recall that for the \threesoK\ model our attempts to
use simultaneous spin and orbital fusion to obtain the FL ground state
for $0 \neq J_0 \ll K_0$, $I_0 = 0$, were unsuccessful (cf.\
Sec.~\ref{eq:SU(2)3-FL}).  Thus the \twosoK\ and \threesoK\ models
provide an example and a counterexample for the success of
simultaneous spin and orbital fusion, succeeding or failing for a NFL
or FL fixed point spectrum, respectively.

\onecolumngrid

\renewcommand{\tabcolsep}{6pt}
\begin{table*}[tb!]
\centering
\caption{Fusion table for  the NFL fixed point
  of Ye's \twosoK\ model. It is 
  structured just as Table~\ref{tab:supp:fusion_NRG}
  for the \threesoK\ model, but here the 
  free bath excitations are computed using 
  Eqs.~\eqref{eq:supp:E(qSlambda)su2level2-sup}  
  and Table~\ref{tab:supp:E_SU(2)_SU(2)} of the SM \cite{suppmat},
  and single fusion of  bath and impurity degrees of freedom 
  is performed simultaneously in the spin 
  and orbital sectors, using $\sutwo_2 \times \sutwo_2$ fusion rules
  (listed in Table~\ref{tab:supp:fusion_SU(2)2} of the SM \cite{suppmat}).
  The resulting multiplets $(q, S^\prime, \lambda^\prime)$ have energies $E^\prime = E(q,S^\prime,\lambda^\prime)$ and excitation energies  $\delta E^\prime = E^\prime - E^\prime_\mathrm{min}$. 
  The NRG energies, computed for $(J_0,K_0,I_0)= (0.1,0.3,0)$ [Fig.~\ref{fig:J}(a)]
  have been shifted and rescaled such that 
  the lowest energy is zero and the second-lowest values for $E_\NRG$ and $\delta E'$ 
  match. The single-fusion and NRG spectra agree very well (deviations $\lesssim 1\%$).}

\renewcommand{\arraystretch}{1.5}
\begin{tabular}{ccc@{\hspace{10mm}}cc@{\hspace{3mm}}|@{\hspace{2mm}}ccccc@{\hspace{10mm}}c@{\hspace{5mm}}cc|c|@{\hspace{2mm}}ccccc@{\hspace{8mm}}cc}
\hline \hline
\multicolumn{5}{c|@{\hspace{2mm}}}{Free fermions} 
& \multicolumn{8}{c|}{Single fusion, with $Q_\imp =  (0,\tfrac12, \tfrac12)$} & NRG 
& \multicolumn{7}{c}{Double fusion, with $Q_\imp =  (0,\tfrac12, \tfrac12)$}
\\ 
$q$ & $S$ & $\lambda$ & $d$ & $E$ 
& $q$ & & $S^\prime$ & & $\lambda^\prime$ & $d^\prime $ 
& $E^\prime$ & $\delta E^\prime$ & $E_\NRG$
& $q$ & & $S^{\prime\prime}$ & & $\lambda^{\prime\prime}$ & $\Delta$ & $\hat{O}$
\\ \hline 

\mfour0 & \mfour0 & \mfour0 & \mfour1 & \mfour0
& \mfour{0} & & \mfour{\fr12} & & \mfour{\fr12} & \mfour4
& \mfour{\fr38} & \mfour{\fr14} & \mfour{0.25} 
& \mfour0 & \lbfour & \m0 & \lb & 0 & 0 & 
\\
& & & & 
& & & & & & 
& & 
& & & & & & 1 & \fr12 & $\mathbf{\Phi}_\mathrm{orb}$
\\
& & & & 
& & & & & & 
& & 
& & & & \m1 & \lb & 0 & \fr12 & $\mathbf{\Phi}_\mathrm{sp}$
\\
& & & & 
& & & & & & 
& & 
& & & & & & 1 & 1 & 
\\ \hline

\mfour{$\pm1$} & \mfour{\fr12} & \mfour{\fr12} & \mfour{4} & \mfour{\fr12}
& \mfour{$\pm1$} & \lbfour & \m{0} & \lb & 0 & 1
& \fr18 & 0 & 0
& & & & & & & 
\\
& & & & 
& & & & & 1 & 3
& \fr58 & \fr12 & 0.5
& & & & & & & 
\\
& & & & 
& & & \m1 & \lb & 0 & 3
& \fr58 & \fr12 & 0.5
& & & & & & & 
\\
& & & & 
& & & & & 1 & 9
& \fr98 & 1 & 0.99
& & & & & & & 
\\ \hline

0 & 1 & 0 & 3 & {\fr12}
& 0 & & \fr12 & & \fr12 & 4
& \fr38 & \fr14 & 0.25
& & & & & & & 
\\ 

0 & 0 & 1 & 3 & {\fr12}
& 0 & & \fr12 & & \fr12 & 4
& \fr38 & \fr14 & 0.25
& & & & & & & 
\\

$\pm2$ & 0 & 0 & 1 & {\fr12}
& $\pm2$ & & \fr12 & & \fr12 & 4
& \fr78 & \fr34 & 0.75
& & & & & & & 
\\ 

$\pm2$ & 1 & 0 & 1 & 1
& $\pm2$ & & \fr12 & & \fr12 & 4
& \fr78 & \fr34 & 0.75
& & & & & & & 
\\ 

$\pm2$ & 0 & 1 & 1 & 1
& $\pm2$ & & \fr12 & & \fr12 & 4
& \fr78 & \fr34 & 0.75
& & & & & & & 
\\ 

$\pm2$ & 1 & 1 & 1 & {\fr32}
& $\pm2$ & & \fr12 & & \fr12 & 4
& \fr78 & \fr34 & 0.75
& & & & & & & 
\\ \hline \hline 

\end{tabular}
\label{tab:supp:fusion_NRG_Ye_su2su2}
\end{table*}

\twocolumngrid

\renewcommand{\tabcolsep}{6pt}
\begin{table}[th!]
\centering
\caption{Fusion table for the FL fixed point of 
  the \sufour\ Kondo  model. It is  structured just as  Table~\ref{tab:fusion_SU(6)}
  for the \susix\ Kondo model, but here the free bath excitations are computed using   Eqs.~\eqref{eq:supp:E(qlambda)su41-sup} and Table~\ref{tab:supp:E_SU(4)}
of the SM \cite{suppmat}, and flavor fusion
is performed using $\sufour_1$ fusion rules (Table~\ref{tab:supp:fusion_SU(4)} of the SM\cite{suppmat}).  
  The NRG spectrum was computed for the \twosoK\ model with $(J_0,K_0,I_0) = (0,0.3,0.05)$ [Fig.~\ref{fig:J}(c)].
  The single-fusion and NRG spectra agree very well 
  (deviations $\lesssim 1\%$).}
\renewcommand{\arraystretch}{1.6}
\begin{tabular}{ScSc@{\hspace{7mm}}Sc@{\hspace{3mm}}Sc@{\hspace{2mm}}|@{\hspace{2mm}}ScSc@{\hspace{9mm}}Sc@{\hspace{5mm}}Sc@{\hspace{2mm}}Sc@{\hspace{1mm}}|Sc}
\hline \hline
\multicolumn{4}{c|@{\hspace{2mm}}}{Free fermions} & \multicolumn{5}{c|}{Single fusion, $Q_\imp = (0,\raisebox{0.3mm}{\tiny \yng(1)}\,)$} & NRG
\\ 
$q$ & $\lambda$ & $d$ & $E$
& $q$ & $\lambda^\prime$ & $d^\prime$ & $E^\prime$ & $\delta E^\prime$ & $E_\NRG$

\\ \hline


0 & $\bullet$ & 1 & 0
& 0 & {\tiny $\yng(1)$} & 4 & \fr38 & \fr14 & 0.25
\\ \hline


+1 & {\tiny $\yng(1)$} & 4 & \fr12
& +1 & {\tiny $\yng(1,1)$} & 6 & \fr58 & \fr12 & 0.50
\\ \hline

$-1$ & {\tiny $\yng(1,1,1)$} & 4 & \fr12
& $-1$ & $\bullet$ & 1 & \fr18 & 0 & 0
\\ \hline


$\pm2$ & {\tiny $\yng(1,1)$} & 6 & 1
& $\pm2$ & {\tiny $\yng(1,1,1)$} & 4 & \fr78 & \fr34 & 0.75
\\ \hline


$+3$ & {\tiny $\yng(1,1,1)$} & 4 & \fr32
& $+3$ & $\bullet$ & 1 & \fr98 & 1 & 1.00
\\ \hline

$-3$ & {\tiny $\yng(1)$} & 4 & \fr32
& $-3$ & {\tiny $\yng(1,1)$} & 6 & \fr{13}8 & \fr32 & 1.50
\\ \hline \hline 

\end{tabular}
\label{tab:E_SU(4)}
\end{table}

We have also computed the imaginary
parts of spin and orbital susceptibilities $\chi_{\text{sp,orb}}^\imp(\omega)$.
Figure~\ref{fig:J}(b) shows the results. Both functions approach constants in the 
zero-frequency limit, i.e., scale as $\omega^0$.
This can be understood in terms of the scaling dimensions of the
leading boundary operators in the spin and orbital sectors,
$\bPhi_\spin$ and $\bPhi_\orb$, which have dimensions
$\Delta_\spin = \Delta_\orb = \frac{1}{2}$
(Table~\ref{tab:supp:fusion_NRG_Ye_su2su2}). By  the arguments
of Sec.~\ref{sec:CFT-susceptibilities}, we thus obtain 
\begin{align*}
\chi^\imp_{\spin,\orb} \sim \omega^{2\Delta_{\spin,\orb}-1} \sim \omega^0 \, , 
\end{align*}
as predicted by Ye. This resembles the behavior observed for the
celebrated two-channel Kondo model, featuring a spin-$1/2$
impurity having no orbital isospin (obtained from Ye's model by using
$\lambda = \bullet$ for the impurity orbital pseudospin, and setting
$K_0 = I_0 = 0$).

\subsection{$\boldsymbol{I_0 \neq  0}$: FL fixed point}
\label{sec:Ye-FL}

For $I_0 \neq 0$, particle-hole symmetry is broken.
Figure~\ref{fig:J}(c) shows the eigenlevel flow computed by NRG
for $\bc_0 =  (0, 0.3,0.05)$.  The
low-energy fixed point is a FL, as predicted by Ye. Its spectrum shows
the same equidistant set of energies as the NFL spectrum of $I_0=0$
[Fig.~\ref{fig:J}(a)], but the
degeneracies are different. This
fixed point \textit{cannot} be understood by simultaneous fusion in
the spin and orbital sector. However, it agrees with the FL spectrum
of an \sufour\ Kondo model with the higher symmetry
$\mathrm{U}(1)_\mathrm{ch} \times \mathrm{SU}(4)_\mathrm{fl}$, defined
in analogy to the \susix\ Kondo model from Sec.~\ref{eq:SU6-FL}, with
a flavor index $\nu = 1, \dots, 4$ encoding both spin and orbital
degrees of freedom. Using non-Abelian bosonization according to the
$\uone \times \sufour_1$ KM algebra, the free bath spectrum can be
expressed as
\begin{subequations}
\label{eq:supp:E(qlambda)su41-sup}
\begin{align}
E(q,\lambda)  = &\tfrac{1}{8}q^2 + \tfrac{1}{5}\kappa_4 (\lambda) + \ell \,,
\label{eq:supp:E(qlambda)su41}
\\
\label{eq:kappa_4}
\kappa_4(\lambda)
= &
\tfrac{1}{8} (
3 \lambda_1^2
+4 \lambda_2^2
+3 \lambda_3^2
%
+ 4 \lambda_1 \lambda_2
+2 \lambda_2 \lambda_3
+4 \lambda_1 \lambda_3
\nonumber \\
&+12 \lambda_1
+16 \lambda_2
+12 \lambda_3
) \,. 
\end{align}
\end{subequations}
with $\ell \in \mathbbm{Z}$, where $\kappa_4(\lambda)$ is the
quadratic Casimir for the $\lambda = (\lambda_1,\lambda_2,\lambda_3)$
representation of the \sufour\ Lie algebra. [The contributions from the
two terms of Eq.\ \eqref{eq:supp:E(qlambda)su41} are listed
in Table \ref{tab:supp:E_SU(4)} of the Supplemental Material \cite{suppmat} for the lowest few $q$ and $\lambda$
values.]
Combining the flavor degrees of freedom of
bath and impurity, $\lambda \otimes \lambda_\imp = \sum_{\oplus}
\lambda^\prime$, using the fusion rules of the 
$\sufour_1$ KM algebra, we recover the FL fixed
point spectrum found by NRG. This is shown in 
Table~\ref{tab:E_SU(4)}. 
In the FL regime, the spin and orbital susceptibilities scale as
$\chi^\imp_{\spin,\orb} \sim \omega^1$ [Fig.~\ref{fig:J}(d)], as
expected for a Fermi liquid and predicted by Ye.

\newpage

\input{3soK-NRG-CFT.bbl}

\clearpage

\setcounter{equation}{0}
\setcounter{figure}{0}
\setcounter{page}{1}
\setcounter{table}{0}
\setcounter{section}{0}

\renewcommand{\theequation}{S.\arabic{equation}}
\renewcommand{\thefigure}{S\arabic{figure}}
\renewcommand{\thetable}{S\arabic{table}}
\renewcommand{\thepage}{S\arabic{page}}
%
% ...

\onecolumngrid

\begin{center}

{\bfseries\large Supplemental Material for ``Uncovering Non-Fermi-Liquid Behavior in Hund Metals:\\
\vspace{0.25em}
Conformal Field Theory Analysis of an $\sutwo \times \suthree$ Spin-Orbital 
Kondo Model''}

\vspace{1.2em}

E. Walter, K. M. Stadler, S.-S. B. Lee, Y. Wang, G. Kotliar, A. Weichselbaum, and J. von Delft

\vspace{.5em}

\date{\today}

\end{center}

Citations and equation numbers refer to references and equations given in the main text.

Below we provide a number of tables needed for various non-Abelian bosonization 
and Kac-Moody fusion schemes used in the main text: 
$\uone \times \sutwo_3 \times \suthree_2$, $\uone \times \susix_1$, 
$\uone \times \sutwo_2 \times \sutwo_2$, and $\uone \times \sufour_1$. 

The fusion rules for the $\suN_k$ Kac-Moody (KM) algebra differ from those
of the $\suN$ Lie algebra in that some Young diagrams arising for the
latter are forbidden for the former (such as Young diagrams with more
than $k$ columns, reflecting the fact that only two distinct spin
species are available when constructing $\suN_k$ representations). 
However, note that these fusion rules are in general more complicated than simply crossing out diagrams with more than $k$ columns. For example, 
in Table~\ref{tab:supp:fusion_SU(2)3} for $\sutwo_3$,
not all representations with $S''\leq 3/2$ are allowed. 
We constructed the KM fusion tables given below 
using a general recipe  due to Cummins 
\protect\cite{Cummins1991}, 
explained in pedagogical detail in Sec.\
16.2.4 of 
\protect\cite{DiFrancesco1997}.

\bigskip

\centering{\bfseries U(1)$\times$SU(2)${}_3\times$SU(3)${}_2$}

\vspace{.3em}

\twocolumngrid

\renewcommand{\tabcolsep}{3pt}
\begin{table}
\centering
\caption{The few lowest values of the quantum numbers  
  $q$, $S$ and $\lambda = (\lambda_1,\lambda_2)$ labeling \uone\ charge, 
$\sutwo_3$ spin and
$\suthree_2$  orbital multiplets, 
  their contributions to the energies $E(q,S,\lambda)$
 of Eq.~\eqref{eq:supp:E(qSlambda)},
  and the dimensions $d$ of the spin and orbital multiplets. 
$\kappa_2(S)$, $\kappa_3(\lambda) $ are given in Eqs.~\eqref{eq:CasimirSU2}, \eqref{eq:CasimirSU3}.}
\medskip 
\renewcommand{\arraystretch}{1.8}
\begin{tabular}{c|ccccccc}
\hline\hline 
$q$ & 0 & $\pm 1$ & $\pm 2$ & $\pm 3$ &  $\pm 4$ & $ \pm 5$
\\
\f112$q^2$ & 0 & \f112 & \fr13 & \fr34 & \fr43  & \fr{25}{12}  
\\ \hline 
$S$ & 0 & \fr12 & 1 & \fr32 & 2 & \fr52
\\ 
$\tfrac15 \kappa_2(S)$ & 0 & \f320 & \fr25 & \fr34 & \fr65 & \fr74
\\ 
$d(S)$ & 1 & 2 & 3 & 4 & 5 & 6
\\ \hline 
$(\lambda_1, \lambda_2)$ &   (0,0) & (1,0)  & (0,1)  & (2,0) & (0,2)  & (1,1) 
\\ 
$\lambda $ &  $\bullet$ 
&  \raisebox{0.5mm}{\tiny $\yng(1)$} 
&  \raisebox{0.5mm}{\tiny $\yng(1,1)$} 
& \raisebox{0.5mm}{\tiny $\yng(2)$} 
& \raisebox{0.5mm}{\tiny $\yng(2,2)$} 
&  \raisebox{0.5mm}{\tiny $\yng(2,1)$} 
\\ 
$\tfrac15 \kappa_3(\lambda) $ & 0 & \f415 & \f415 & \fr23 & \fr23 & \fr35
\\ 
$d(\lambda)$ & 1 & 3 & 3 & 6 & 6 & 8
\\ \hline \hline
\end{tabular} 
\vspace{-4mm}
\label{tab:supp:E_SU(2)_SU(3)}
\end{table}

\renewcommand{\tabcolsep}{3pt}
\begin{table}[h!]
\centering
\caption{$\suthree_2$ fusion rules, listing various 
direct product decompositions of the form
$\lambda \otimes \lambda' = \sum_\oplus \lambda''$. 
Crossed-out diagrams denote additional irreps occurring 
when considering  direct product decompositions for \suthree\
instead of $\suthree_2$.} 
\renewcommand{\arraystretch}{2.1}
\begin{tabular}{cccc|c|c}
\hline\hline 
\raisebox{-6pt}{$d(\lambda)$} & 
\raisebox{-6pt}{$\kappa_3(\lambda)$} & 
\raisebox{-6pt}{{$(\lambda_1, \lambda_2)$}} & 
\diagbox[width=40pt,height=30pt]{\hspace{10pt}${\lambda}$}
{{\color{white}.} \hspace{-20pt}${\lambda^\prime}$} 
& \raisebox{1mm}{\y10} & \raisebox{1mm}{\y01} 
\\ \hline 
3 & \fr43 & {(1,0)} & \hspace{-4mm}
\raisebox{0.5mm}{{\y10}} & 
\raisebox{1mm}{{\y01} $\soplus{5.5}$  {{\tiny $\yng(2)$}}} & 
\raisebox{1mm}{{\y00} $\soplus{3.5}$ {\y11}} 
\\
\hhline{----|--}%

3 & \fr43 & {(0,1)} & \hspace{-3.5mm}\raisebox{0.5mm}{{\y01}} 
& \raisebox{1mm}{{\y00} $\soplus{3.5}$ {\y11}} & 
\raisebox{1mm}{{\y10} $\soplus{5.5}$ {\y02}} 
\\
\hhline{----|--}

6 & \fr{10}{3} & {(2,0)} & \hspace{-4mm}
 \raisebox{0.5mm}{{{\tiny $\yng(2)$}}} & 
\raisebox{1mm}{{\y11} $\soplus{7.5}$ \ $\cancel{{\tiny \yng(3)}}$} & 
\raisebox{1mm}{{\y10} $\soplus{5.5}$  $\cancel{{\tiny \yng(3,1)}}$} 
\\ 
\hhline{----|--}

6 & \fr{10}{3} & {(0,2)} & \hspace{-4mm}
{\y02} & 
\raisebox{1mm}{{\y01} $\soplus{5.5}$ \ $\cancel{{\tiny \yng(3,2)}}$} 
& \raisebox{1mm}{{\y11} $\soplus{7.5}$  $\cancel{{\tiny \yng(3,3)}}$} 
\\ 
\hhline{----|--}

8 & 3 & {(1,1)} & \hspace{-4mm}
{\y11} & 
\raisebox{1mm}{{\y10} $\soplus{5.5}$ {\y02}}  & 
\raisebox{1mm}{{\y01} $\soplus{5.5}$  {{\tiny $\yng(2)$}}}  
\\
& & & & 
\raisebox{1mm}{$\soplus{5.5}$  $\cancel{{\tiny \yng(3,1)}}$} & 
\raisebox{1mm}{$\soplus{5.5}$  $\cancel{{\tiny \yng(3,2)}}$} 
\\  \hline \hline 
\end{tabular}
\label{tab:supp:fusion_SU(3)}
\end{table}

\renewcommand{\tabcolsep}{3pt}
\begin{table}[t]
\centering
\caption{$\sutwo_3$ fusion rules, listing various 
direct product decompositions of the form
$S \otimes S' = \sum_\oplus S''$. 
Crossed-out numbers denote additional irreps occurring 
when considering   direct product decompositions for \sutwo\
instead of $\sutwo_3$. 
} 
\renewcommand{\arraystretch}{2.1}
\begin{tabular}{ccc|c}
\hline\hline 
\raisebox{-6pt}{$d(S)$} & 
\raisebox{-6pt}{$\kappa_2(S)$} & 
\diagbox[width=40pt,height=30pt]{\hspace{10pt}${S}$}
{{\color{white}.} \hspace{-20pt}${S^\prime}$} 
& $\frac{3}{2}$
\\ \hline 
1 & 0 & 0 & $\frac{3}{2}$ \\
\hhline{---|-}%
2 & $\frac{3}{4}$ & $\frac{1}{2}$ & $1 \oplus \cancel{2}$ \\
\hhline{---|-}%
3 & 2 & 1 & $\frac{1}{2} \oplus \cancel{\frac{3}{2}} \oplus \cancel{\frac{5}{2}}$ \\
\hhline{---|-}%
4 & $\frac{15}{4}$ & $\frac{3}{2}$ & $0 \oplus \cancel{1} \oplus \cancel{2} \oplus \cancel{3}$ \\
\hline \hline 
\end{tabular}
\label{tab:supp:fusion_SU(2)3}
\end{table}

\clearpage

\onecolumngrid

\centering{\bfseries U(1)$\times$SU(6)${}_1$}

\vspace{.3em}

\twocolumngrid

\renewcommand{\tabcolsep}{7pt}
\begin{table}
\centering
\caption{The few lowest values of the quantum numbers 
  $q$ and $\lambda = (\lambda_1,\lambda_2,\lambda_3,\lambda_4,\lambda_5)$, 
labeling \uone\ charge and $\susix_1$ flavor 
multiplets,   their contributions to the eigenenergies $E(q,\lambda)$
of Eq.~\eqref{eq:E(qlambda)},  
  and the dimensions $d$ of the flavor multiplets.
Single-column Young diagrams with $i$ boxes have
$\lambda_j = \delta_{ij}$. $\kappa_6(\lambda) $ is given in 
Eq.~\eqref{eq:kappa_6}.}
\smallskip
\renewcommand{\arraystretch}{2}
\begin{tabular}{c|cccccSc}
\hline\hline 
$q$ & 0 & $\pm 1$ & $\pm 2$ & $\pm 3$  & $\pm 4$ & $ \pm 5$
\\ 
\f112$q^2$ & 0 & \f112 & \fr13 & \fr34 & \fr43  & \fr{25}{12}  
\\ \hline 
$\lambda$ & $\bullet$ 
& \raisebox{0.5mm}{\tiny $\yng(1)$} 
& {\tiny $\yng(1,1)$} 
& {\tiny $\yng(1,1,1)$} 
& {\tiny $\yng(1,1,1,1)$} 
& {\tiny $\yng(1,1,1,1,1)$} 
\\ 
$\tfrac17 \kappa_6(\lambda) $ & 0 & \f512 
& \fr23 
& \fr34 
& \fr23 
& \f512 
\\ 
$d(\lambda)$ & 1 & 6 
& 15 
& 20 
& 15 
& 6 
\\ \hline \hline
\end{tabular}
\label{tab:supp:E_SU(6)}
\end{table}

\renewcommand{\tabcolsep}{3pt}
\begin{table}
\centering
\caption{$\susix_1$ fusion rules, 
listing some 
direct product decompositions 
$\lambda \otimes \lambda' = \sum_\oplus \lambda''$, 
with $\lambda'= \tiny \protect\yng(1,1)\,$. 
Crossed-out diagrams denote additional irreps occurring 
when considering  direct product decompositions for \susix\
instead of $\susix_1$. 
}
\begin{tabular}{cccc|Sc}
\hhline{=====}
\raisebox{-6pt}{$d(\lambda)$} & 
\raisebox{-6pt}{$\kappa_6(\lambda)$} & 
\raisebox{-6pt}{{$(\lambda_1, \lambda_2, \lambda_3, \lambda_4, \lambda_5)$}}& 
\diagbox[width=30pt,height=25pt]{\hspace{5pt}${\lambda}$}
{{\color{white}.} \hspace{-20pt}\raisebox{-6pt}{\hspace{4pt}${\lambda^\prime}$}} & 
{\y01} 
\\ 
\hhline{-----}%

1 & 0 & {(0,0,0,0,0)} & \hspace{-4mm} {$\bullet$} & \raisebox{2pt}{{\y01}}
\\ 
\hhline{----|-}%

6 & \fr{35}{12} & {(1,0,0,0,0)} & \hspace{-4mm} 
{\raisebox{0.5mm}{\tiny $\yng(1)$}}
& 
{{\tiny $\yng(1,1,1)$}} \ $\oplus$ \ $\cancel{{\tiny \yng(2,1)}}$
\\ \hline

15 & \fr{14}{3} & {(0,1,0,0,0)} & \hspace{-4mm} 
{\raisebox{0.5mm}{\tiny $\yng(1,1)$}}
& {{\tiny $\yng(1,1,1,1)$}} \ $\oplus$ \ $\cancel{{\tiny \yng(2,1,1)}}$ \ $\oplus$ \ $\cancel{{\tiny \yng(2,2)}}$
\\ 
\hhline{----|-}

20 & \fr{21}{4} & {(0,0,1,0,0)} & \hspace{-4mm} 
{\raisebox{0.5mm}{\tiny $\yng(1,1,1)$}} & \;
{{\tiny $\yng(1,1,1,1,1)$}} \ $\oplus$ \ $\cancel{{\tiny \yng(2,1,1,1)}}$ \ $\oplus$ \ $\cancel{{\tiny \yng(2,2,1)}}$
\\ 
\hhline{----|-}%

  15 & \fr{14}{3} & {(0,0,0,1,0)} & \hspace{-4mm} 
{\raisebox{0.5mm}{\tiny $\yng(1,1,1,1)$}} & \; {\y00} \ 
$\soplus{3.5}$ \ $\cancel{{\tiny \yng(2,1,1,1,1)}}$ \ $\oplus$ \ $\cancel{{\tiny \yng(2,2,1,1)}}$
  \\ 
\hhline{----|-}%

6 & \fr{35}{12} & {(0,0,0,0,1)} & \hspace{-4mm} 
{\raisebox{0.5mm}{\tiny $\yng(1,1,1,1,1)$}} & {\raisebox{0.5mm}{\tiny $\yng(1)$}} \ $\oplus$ \ $\cancel{{\tiny \yng(2,2,1,1,1)}}$ 
\\ 
\hhline{=====}

\end{tabular}
\label{tab:supp:fusion_SU(6)}
\end{table}

\onecolumngrid

\centering{\bfseries U(1)$\times$SU(2)$_2\times$SU(2)$_2$}

\vspace{.3em}

\twocolumngrid

\renewcommand{\tabcolsep}{6pt}
\begin{table}
\centering
\caption{The few lowest values of the quantum numbers  
  $q$, $S$ and $\lambda$, 
labeling \uone\ charge, $\sutwo_2$ spin and $\sutwo_2$ orbital
 multiplets, respectively,
  their contributions to the eigenenergies $E(q,S,\lambda)$
of Eq.~\eqref{eq:supp:E(qSlambda)su2level2}, 
  and the dimensions $d$ of the spin and flavor multiplets.
$\kappa_2(S)$ and $\kappa_2(\lambda)$ are given in
Eq.~\eqref{eq:eq:Su2Su2Casimir}.} 
\renewcommand{\arraystretch}{1.8}
\begin{tabular}{c|ccccccc}
\hline \hline 
$q$ & 0 & $\pm 1$ & $\pm 2$ & $\pm 3$ 
\\ 
\fr18$q^2$ & 0 & \fr18 & \fr12 & \fr98
\\ \hline 
$S$, $\lambda$ & 0 & \fr12 & 1 & \fr32 
\\ 
$\tfrac14 \kappa_2(S)$, $\tfrac14 \kappa_2(\lambda)$ & 0 & \f316 & \fr12 & \fr{15}{16}
\\ 
$d(S)$, $d(\lambda)$ & 1 & 2 & 3 & 4 
\\ \hline \hline
\end{tabular}
\label{tab:supp:E_SU(2)_SU(2)}
\end{table}

\renewcommand{\tabcolsep}{3pt}
\begin{table}[t]
\centering
\caption{$\sutwo_2$ fusion rules, listing various 
direct product decompositions of the form
$S \otimes S' = \sum_\oplus S''$. 
Crossed-out numbers denote additional irreps occurring 
when considering direct product decompositions for \sutwo\
instead of $\sutwo_2$. 
} 
\renewcommand{\arraystretch}{2.1}
\begin{tabular}{ccc|c}
\hline\hline 
\raisebox{-6pt}{$d(S)$} & 
\raisebox{-6pt}{$\kappa_2(S)$} & 
\diagbox[width=40pt,height=30pt]{\hspace{10pt}${S}$}
{{\color{white}.} \hspace{-20pt}${S^\prime}$} 
& $\frac{1}{2}$
\\ \hline 
1 & 0 & 0 & $\frac{1}{2}$ \\
\hhline{---|-}%
2 & $\frac{3}{4}$ & $\frac{1}{2}$ & $0 \oplus 1$ \\
\hhline{---|-}%
3 & 2 & 1 & $\frac{1}{2} \oplus \cancel{\frac{3}{2}}$ \\
\hhline{---|-}%
\hline \hline 
\end{tabular}
\label{tab:supp:fusion_SU(2)2}
\end{table}

\clearpage

\onecolumngrid

\centering{\bfseries U(1)$\times$SU(4)$_1$}

\vspace{.3em}

\twocolumngrid

\renewcommand{\tabcolsep}{7pt}
\begin{table}
\centering
\caption{The few lowest values of the quantum numbers 
  $q$ and $\lambda = (\lambda_1,\lambda_2,\lambda_3)$,
labeling \uone\ charge and $\sufour_1$ flavor 
multiplets,   their contributions to the eigenenergies $E(q,\lambda)$
of Eq.~\eqref{eq:supp:E(qlambda)su41},
  and the dimensions $d$ of the flavor multiplets.
$\kappa_4(\lambda)$ is given in Eq.~\eqref{eq:kappa_4}.}
\renewcommand{\arraystretch}{1.8}
\begin{tabular}{c|cccSc}
\hline \hline
$q$ & 0 & $\pm 1$ & $\pm 2$ & $\pm 3$ 
\\ 
\fr18$q^2$ & 0 & \fr18 & \fr12 & \fr98 
\\ \hline 
$(\lambda_1, \lambda_2, \lambda_3)$ & (0,0,0) & (1,0,0) & (0,1,0) & (0,0,1) 
\\
$\lambda$ & $\bullet$ 
& {\tiny $\yng(1)$} 
& {\tiny $\yng(1,1)$} 
& {\tiny $\yng(1,1,1)$} 
\\ 
$\tfrac15 \kappa_4(\lambda) $ & 0 & \fr38 & \fr12 & \fr38
\\ 
$d(\lambda)$ & 1 & 4 & 6 & 4
\\ \hline \hline 
\end{tabular}
\label{tab:supp:E_SU(4)}
\end{table}

\renewcommand{\tabcolsep}{3pt}
\begin{table}[h!]
\centering
\caption{$\sufour_1$ fusion rules, 
listing some 
direct product decompositions 
$\lambda \otimes \lambda' = \sum_\oplus \lambda''$, 
with $\lambda'= \protect\raisebox{1mm}{\tiny $\protect\yng(1)$}\,$. 
Crossed-out diagrams denote additional irreps occurring 
when considering  direct product decompositions for \sufour\
instead of $\sufour_1$.}
\renewcommand{\arraystretch}{1.8}
\begin{tabular}{cccc|Sc}
\hhline{=====}
\raisebox{-6pt}{$d(\lambda)$} & 
\raisebox{-6pt}{$\kappa_4(\lambda)$} & 
\raisebox{-6pt}{{$(\lambda_1, \lambda_2, \lambda_3)$}} 
& 
\diagbox[width=30pt,height=25pt]{\hspace{5pt}\raisebox{-1.5pt}{${\lambda}$}}
{{\color{white}.} \hspace{-20pt}\raisebox{0pt}{\hspace{4pt}${\lambda^\prime}$}}
& {\y10} 
\\ \hline 

1 & 0 & {(0,0,0)} & $\bullet$  &  {\y10}
\\ \hline

4 & \fr{15}8 & {(1,0,0)} & {\tiny $\yng(1)$} 
& 
\; {{\tiny $\yng(1,1)$}} \ $\oplus$ \ $\cancel{{\tiny \yng(2)}}$
\\ \hline

6 & \fr52 & {(0,1,0)} & {\tiny $\yng(1,1)$} 
& 
\; {{\tiny $\yng(1,1,1)$}} \ $\oplus$ \ $\cancel{{\tiny \yng(2,1)}}$ 
\\ \hline

4 & \fr{15}8 &   {(0,0,1)} & {\tiny $\yng(1,1,1)$} 
& \; {$\bullet$} \ $\oplus$ \ $\cancel{{\tiny \yng(2,1,1)}}$
\\ \hhline{=====} 
\end{tabular}
\label{tab:supp:fusion_SU(4)}
\end{table}

\end{document}

%% file: 3soK-NRG-CFT.bbl
%